\documentclass[a4paper,10pt]{article}

\PassOptionsToPackage{hyperfootnotes=false}{hyperref}
\usepackage{jheppub}
\usepackage[T1]{fontenc}
\usepackage{amsmath,amssymb,bm,mathrsfs}
\usepackage{graphicx}
\usepackage{booktabs}
\usepackage{array}
\usepackage{slashed}
\usepackage{tikz}

\usetikzlibrary{arrows.meta,decorations.pathmorphing}
\newcolumntype{L}[1]{>{\raggedright\arraybackslash}p{#1}}
\hypersetup{hypertexnames=false}

\title{\boldmath Spin-Induced Coordinate Noncommutativity of the Covariant
Centroid}

\author[a,1]{D. G. Tedesco\note{Corresponding author.}}
\author[b]{L. P. Colatto}

\affiliation[a]{UNINTER, Curitiba, PR, Brazil}
\affiliation[b]{CEFET-RJ, UnED Petr\'opolis, 25620-003 Petr\'opolis, RJ, Brazil}

\emailAdd{daniel.te@uninter.com}
\emailAdd{luiz.colatto@cefet-rj.br}

\abstract{%
The Tulczyjew-Dixon centroid of a massive spinning particle, built as a
composite of canonical Poincar\'e variables, carries the exact coordinate
algebra $\{X^\mu,X^\nu\}=-S^{\mu\nu}/p^2$, whose noncommutativity bivector
is the transverse spin tensor of the particle itself rather than an
external background.  That bracket and its Compton-scale magnitude are
already known, and we recover them here as the starting point of the
analysis, not as a result.  What the composite realization settles is the
domain on which the algebra may be used.  Closure of the Jacobi identity
requires $p^{-2}$ to remain operator valued, and the exactness domain is
the convex timelike interior $\Omega_\mu$, on which states supported on a
single mass shell are distributional.  We show that freezing $p^2$ to its
shell value inside the centroid leaves every two-operator commutator
unchanged on shell, so that the projected position dictionary, the Berry
cancellation and the endpoint transversality identity are untouched, while
the same substitution fails from three insertions onward, with residue the
truncated Jacobiator $2\epsilon^{\mu\nu\rho\sigma}W_\sigma/(p^2)^2$.  This
two-versus-three criterion replaces the qualitative restriction to
shell-preserving matrix elements and states which constructions built on a
fixed-shell bivector $\Theta_m$ remain valid.  We then follow the algebra
through positive-energy projection, packet-conditioned Moyal fibers,
Compton routing holonomy and projected nonassociative products, recording
for each layer whether it falls on the safe side of the criterion.  No
prediction beyond QED follows.  In the geometric branch the physical
amplitude reduces to the QED amplitude, and the transverse matching
coefficients remain uncomputed.
}


\begin{document}
\maketitle

\section{Introduction}
\label{sec:introduction}

Canonical noncommutative field theory begins with a fixed bivector,
\begin{equation}
[\hat x^\mu,\hat x^\nu]=i\Theta^{\mu\nu},
\label{eq:canonical-nc-intro}
\end{equation}
and transfers it to products of fields through the Moyal-Weyl deformation
\cite{Doplicher:1994zv,Doplicher:1995,Seiberg:1999,Douglas:2001,Szabo:2003}.
The constancy of $\Theta^{\mu\nu}$ is what makes the construction
tractable, since it renders the star product associative, keeps the twist
abelian, and preserves most of the perturbative machinery of ordinary field
theory.  That same constancy also makes the geometric datum external to the
particle and selects fixed spacetime directions unless the symmetry is
reformulated through a twisted coproduct.  Covariant alternatives avoid the
preferred directions by tying the coordinate algebra to Poincar\'e
generators or to internal degrees of freedom \cite{Snyder:1947,Madore:1999}.

Written in the momentum representation, the canonical theory already
displays the geometric structure that will organize everything below.  The
Bopp realization $X_{\rm B}^\mu=x^\mu-\Theta^{\mu\nu}p_\nu/2\hbar$
\cite{bopp1956} takes the form
$X_{\rm B}^\mu=i\hbar\,\partial/\partial p_\mu+\mathcal A_{\rm B}^\mu$ with
$\mathcal A_{\rm B}^\mu=-\Theta^{\mu\nu}p_\nu/2\hbar$, and since that
connection is abelian its curvature comes from the exterior derivative
alone,
$\partial^\mu\mathcal A_{\rm B}^\nu-\partial^\nu\mathcal A_{\rm B}^\mu
=\Theta^{\mu\nu}/\hbar$, which returns
$[X_{\rm B}^\mu,X_{\rm B}^\nu]=i\Theta^{\mu\nu}$.  Canonical
noncommutativity is in this sense the constant-curvature abelian sector of
a gauge geometry over momentum space, so that the question raised by a
particle-sourced bivector is not whether the Moyal construction should be
abandoned, but which connection replaces $\mathcal A_{\rm B}$ once the
tensor becomes an attribute of the particle instead of the vacuum.  The
answer developed here is a connection valued in the Lorentz algebra whose
curvature is the transverse spin bivector, and whose commutator term is
precisely what distinguishes the two situations.

Any such replacement has to respect a lesson learned repeatedly on the
field-theory side, namely that a nonconstant or operator-valued tensor
cannot be introduced by editing vertices in isolation.  Gauge theory on
twist-noncommutative spaces requires a deformed differential calculus
together with star-gauge covariance \cite{Meier:2023twistGauge}, and the
Poisson gauge and $L_\infty$-bootstrap programs show that the product, the
gauge algebra and the field strength must all be deformed in a correlated
way \cite{Kupriyanov:2020novelGauge,Blumenhagen:2018Linfinity,
Kurkov:2022U1Linfinity}.  These requirements shape the architecture adopted
below, in which kinematical statements are kept apart from dynamical
proposals and every level carries its own assumptions in the open.

Spin supplies an antisymmetric tensor with the required Lorentz
transformation law and an intrinsic quantum scale, but the relation between
spin and noncommuting relativistic positions predates noncommutative field
theory.  It appears in the classification of relativistic position
operators by Pryce \cite{Pryce:1948}, in the M\o ller analysis of centroid
tubes \cite{Moller:1949}, in coadjoint-orbit descriptions of massive
spinning particles \cite{Souriau:1970}, and in later covariant-centroid and
constrained-particle constructions
\cite{Zakrzewski:1994,Chryssomalakos:2009,Aguilar:2013,
DeriglazovPupasov:2014,DeriglazovTereza:2019,MuchVergara:2017,
Ramirez:2014frenkel,Colatto_2006,Gomes:2010}.  The same structures occur in
momentum-space treatments of the Dirac electron: after positive-energy
projection, the Foldy-Wouthuysen position carries a nonabelian Berry
connection and a spin-dependent coordinate commutator
\cite{Berard:2004monopole,Berard:2006spinhall,
StoneDwivediZhou:2015,ChenSonStephanov:2014,Bliokh:2017}.  We claim no
priority for noncommuting covariant centroids or for the interpretation of a
coordinate commutator as momentum-space curvature.

The question is instead organized by the localization dilemma exposed by
Pryce's classification.  Newton-Wigner localization has commuting spatial
components but does not assemble into a Lorentz four-vector
\cite{NewtonWigner:1949}; the Tulczyjew-Dixon centroid is covariant, but its
components do not commute \cite{Tulczyjew:1959,Dixon:1964,CostaNatario:2015}.
Covariance therefore has an algebraic cost, and Fleming established that the
cost is unavoidable: no covariant position operator with commuting
components exists for spin greater than zero \cite{Fleming:1965}.  This observation prevents the
coordinate bracket from being advertised, by itself, as a new interaction:
one must first determine the exact covariant algebra, then identify what a
positive-energy experiment measures, and only afterwards ask whether an
additional dynamical coupling survives comparison with ordinary Dirac and
QED transport.

The first part of this paper carries out the exact construction.  The
Tulczyjew-Dixon variables are obtained as composite functions of canonical
Poincar\'e variables, which fixes the sign and normalization of the
coordinate bracket and supplies the full spin-position and spin-spin
algebra required by the Jacobi identities.  In momentum space, the centroid
shift becomes an $\mathfrak{so}(1,3)$-valued connection
$\mathcal A^\mu(p)$ whose curvature equals the coordinate bracket.  The
connection and curvature are transverse to the particle momentum; radial
transport is flat, whereas tangential transport is governed by the Wigner
little-group geometry of the massive shell.  We state the sign dictionary
between the covariant spin tensor and physical three-spin before using any
spin-odd observable, and we keep $p^{-2}$ as an operator until all
commutators have been evaluated.  These steps also expose the limitations
of the construction: the covariant position acts on a timelike kinematical
domain spanning mass shells, and self-adjoint extensions of all four
components are not constructed here.

The second part asks how this algebra is seen after restriction to the
positive-energy sector of the free Dirac theory.  The transformed centroid
contains both the Berry connection of the projected Dirac position and the
projected spin shift.  These terms cannot be interpreted independently:
the projected spin shift contains the negative of the Berry connection, so
the latter cancels in the total covariant centroid.  The surviving
transverse displacement is the Tulczyjew-Dixon shift of the positive-energy
irreducible representation, accompanied by a momentum-longitudinal
anti-hermitian remnant.  This result fixes the commutator anisotropy and the
side-jump budget, and it prevents the projected covariant centroid from
being counted as a fourth independent transverse position observable.

The third part retains the constant-tensor machinery as a local, conditioned
description.  A narrow polarized packet selects a c-number tensor
$\overline\Theta_{\mathcal C}^{\mu\nu}$ inside one fiber $\mathcal C$, where
an abelian twist, a Moyal product, braided exchange and deformed
Clebsch-Gordan maps are associative.  No nonzero constant tensor can remain
transverse to every momentum in an open set, so these fibers are related by
momentum-space transport rather than assembled into a universal background.
For an electron the relative spin fluctuation satisfies
$\epsilon_S^2=2/3$, leaving no parametrically controlled spin-freezing
limit; the construction remains useful as a packet-conditioned matrix
element and as a field-theory benchmark, but not as a universal electron
theory.

The fourth part separates the geometric, QED and beyond-QED content of
Compton scattering.  The two fermion routings define a closed loop for the
centroid connection and hence the low-recoil invariant
$S^{\mu\nu}k_\mu k'_\nu/(\hbar p^2)$.  A timelike momentum polygon built
from the same vertices has the corresponding Thomas-Wigner angle at
leading quadratic order.  That agreement is analytic rather than
empirical, since the radial projection reduces the transport to the
standard spin-frame connection of the mass hyperboloid, whose holonomy is
the Wigner rotation; the numerical entries reproduce it and therefore test
the implementation, not the correspondence.  The
internal QED momenta, however, are off shell at the physical electron mass,
so this agreement does not furnish an exact factorization of the propagator
numerators into physical on-shell spin states.  Independently, the polarized
$s$-$u$ channel trace can be evaluated in closed form.  Its imaginary part
is a decomposition-dependent diagnostic and contributes to a rate only
after an extra relative dressing is introduced.  The minimal two-phase
dressing violates the Ward identity and yields a benchmark angular curve,
whereas an observable prediction requires a gauge-complete action and its
contact terms.  This distinction also removes a possible double counting:
the Thomas-Wigner transport already contained in standard spin kinematics
is not new physics; any excess must be encoded in separately matched
transverse operators.

The final branch is defined by repeated projection within an ambient
operator algebra that remains associative and whose commutators satisfy the
Jacobi identity.  Nonassociativity is introduced only through the protocol product
$A\circ_PB=PABP$, whose associator measures propagation through the
discarded subspace.  This supplies an explicit object on which hermiticity,
Malcev closure and the pentagon identity can be tested, without assuming a
quasi-Hopf structure in advance.

Table~\ref{tab:status-intro} records the evidential status of each layer.
The labels specify the assumptions and verification method; they do not
rank the physical interest of the corresponding constructions.

\begin{table}[t]
\centering \footnotesize
\caption{Logical status of the structures used in this work.}
\label{tab:status-intro}
\begin{tabular}{@{}L{0.22\linewidth}L{0.31\linewidth}L{0.38\linewidth}@{}}
\toprule
Level & Structure & Status \\
\midrule
Centroid geometry &
$\hat X$, $\hat S$, $\mathcal A_p$, $\mathcal F_p$ &
Exact composite-operator algebra on a timelike kinematical domain \\
Mass-spin sector &
$\Theta_m$, Lorentz invariants, Compton scale &
Exact on shell through at most two insertions of $\hat X$, by
\eqref{eq:two-insertion-neutrality}; no localization theorem is claimed \\
Positive-energy sector &
$P_+\hat X P_+$, centroid shifts, projected commutators &
Exact free-Dirac matrix identities, with the transverse formulas evaluated
on shell \\
Frozen fiber &
$\overline\Theta_{\mathcal C}$, $\star_{\mathcal C}$, twist and braiding &
Associative c-number chart in one packet neighborhood; not parametrically
controlled for $s=\tfrac12$ \\
Routing geometry &
closed centroid holonomy and timelike Wigner polygon &
Exact small-loop expansion plus a leading low-recoil numerical
correspondence; no exact QED factorization \\
Compton trace and template &
$\operatorname{Im}(\mathcal M_s\mathcal M_u^*)$ and a relative-phase model &
Exact channel trace in a declared polarization basis; conditional,
gauge-incomplete rate benchmark \\
Additional dynamics &
transverse form factors and Pauli matching &
Open EFT problem whose coefficients must be computed or bounded \\
Projected extension &
$A\circ_P B=PABP$ and its associator &
Exact leakage identity; Malcev and quasi-Hopf closure remain conjectural \\
\bottomrule
\end{tabular}
\end{table}

Sections~\ref{sec:centroid}-\ref{sec:geometry} develop the exact particle
geometry, and section~\ref{sec:positive-energy} computes its restriction to
the free positive-energy Dirac sector.  Section~\ref{sec:fibered} constructs
the frozen associative fibers and their representation theory.
Sections~\ref{sec:field-matching} and~\ref{sec:phenomenology} distinguish
kinematical transport, standard-QED diagnostics and conditional additional
dynamics.  Section~\ref{sec:projected} formulates the projected
nonassociative branch, while section~\ref{sec:discussion} separates the
geometric, dynamical and projected readings and states their falsifiers.

\section{Covariant centroid and particle-sourced bivector}
\label{sec:centroid}

Consider a Poincar\'e phase space with a reference position $z^\mu$,
momentum $p^\mu$, and antisymmetric internal angular momentum
$\Sigma^{\mu\nu}=-\Sigma^{\nu\mu}$.  We use
\begin{equation}
\eta_{\mu\nu}=\operatorname{diag}(+1,-1,-1,-1),
\qquad
\epsilon^{0123}=+1,
\qquad
p^2>0,
\label{eq:conventions}
\end{equation}
and keep $c$ and $\hbar$ explicit.  The nonvanishing elementary Poisson
brackets are
\begin{align}
\{z^\mu,p^\nu\}&=\eta^{\mu\nu},
\label{eq:canonical-zp}\\
\{\Sigma^{\mu\nu},\Sigma^{\rho\sigma}\}
&=\eta^{\mu\rho}\Sigma^{\nu\sigma}
-\eta^{\nu\rho}\Sigma^{\mu\sigma}
-\eta^{\mu\sigma}\Sigma^{\nu\rho}
+\eta^{\nu\sigma}\Sigma^{\mu\rho},
\label{eq:canonical-spin}
\end{align}
with all remaining elementary brackets equal to zero, and the total angular
momentum reads
\begin{equation}
J^{\mu\nu}
=z^\mu p^\nu-z^\nu p^\mu+\Sigma^{\mu\nu}.
\label{eq:J-unreduced}
\end{equation}

The passage from this covariant tensor algebra to laboratory three-spin
requires a sign that will be kept visible throughout the paper.  For the
spatial dual
\begin{equation}
\mathscr S_k\equiv\frac12\epsilon_{kij}S^{ij},
\label{eq:covariant-spatial-dual}
\end{equation}
the bracket \eqref{eq:canonical-spin} gives
\begin{equation}
\{\mathscr S_1,\mathscr S_2\}=-\mathscr S_3,
\qquad
\bm s_{\rm phys}=-\bm{\mathscr S}.
\label{eq:spin-dictionary}
\end{equation}
Thus the spatial dual of the covariant tensor has the orientation opposite
to the physical polarization vector.  Cartesian laboratory coordinates and
the standard Dirac tensor are correspondingly
\begin{equation}
q^i\equiv-X^i_{\rm cov},
\qquad
[q^i,p^j]=i\hbar\delta^{ij},
\qquad
\Sigma_D^{\mu\nu}\equiv-\Sigma_{\rm cov}^{\mu\nu}.
\label{eq:laboratory-dictionary}
\end{equation}
On Dirac spinors the covariant algebra is represented by
\begin{equation}
\hat\Sigma^{\mu\nu}_{\rm cov}
=-\frac{\hbar}{2}\sigma^{\mu\nu},
\qquad
\sigma^{\mu\nu}=\frac{i}{2}[\gamma^\mu,\gamma^\nu],
\label{eq:dirac-covariant-spin}
\end{equation}
whereas $\Sigma_D^{\mu\nu}=+\hbar\sigma^{\mu\nu}/2$ is used after the
laboratory conversion.  We use $S^{\mu\nu}$, $\Sigma^{\mu\nu}$ and
$\mathscr S_i$ for the covariant convention, and reserve $\bm s$ and
$\bm q$ for physical three-spin and laboratory position.  Every tensor
equation below is written in the former convention; every laboratory cross
product and spin-odd observable is translated with
\eqref{eq:spin-dictionary}-\eqref{eq:laboratory-dictionary}.  In
particular, a physical spin-up state at rest
obeys
\begin{equation}
[\hat q^1,\hat q^2]
=+i\hbar\frac{s_3}{m^2c^2},
\label{eq:rest-commutator-physical}
\end{equation}
although the same statement in terms of the covariant component is
\begin{equation}
[\hat X^1,\hat X^2]
=-i\hbar\frac{S^{12}}{m^2c^2}.
\label{eq:rest-commutator-covariant}
\end{equation}
Keeping these two readings
separate prevents a convention change in the spin tensor from being
mistaken for a physical reversal of a side jump or a holonomy axis.

\subsection{Centroid map and constraint status}
\label{sec:centroid-map}

A massive particle carries three rotational degrees of freedom while
$\Sigma^{\mu\nu}$ has six components, so a spin supplementary condition
must specify which representative worldline separates orbital from
intrinsic angular momentum.  Among the standard choices
\cite{Frenkel:1926,Mathisson:1937,Pirani:1956,Papapetrou:1951,
Corinaldesi:1951}, the Tulczyjew-Dixon condition selects the
momentum-orthogonal spin tensor and the covariant center of mass
\cite{Tulczyjew:1959,Dixon:1964,CostaNatario:2015}.  Instead of imposing
that condition as an external constraint we build it into the variables.
Introducing
\begin{equation}
A^\mu\equiv\Sigma^{\mu\nu}p_\nu,
\qquad
a^\mu\equiv\frac{A^\mu}{p^2},
\qquad
a\cdot p=0,
\label{eq:A-a}
\end{equation}
the momentum-selected centroid and the spin referred to it become
\begin{align}
X^\mu&\equiv z^\mu+a^\mu
=z^\mu+\frac{\Sigma^{\mu\nu}p_\nu}{p^2},
\label{eq:X-definition}\\
S^{\mu\nu}&\equiv
\Sigma^{\mu\nu}-a^\mu p^\nu+a^\nu p^\mu.
\label{eq:S-definition}
\end{align}
The same generator is then written as
\begin{equation}
J^{\mu\nu}
=X^\mu p^\nu-X^\nu p^\mu+S^{\mu\nu},
\label{eq:J-reduced}
\end{equation}
and the Tulczyjew-Dixon condition holds identically,
\begin{equation}
S^{\mu\nu}p_\nu
=A^\mu-a^\mu p^2+(a\cdot p)p^\mu=0.
\label{eq:TD-identity}
\end{equation}
The coordinate $X^\mu$ is therefore no elementary variable postulated to be
noncommutative, but the position of a momentum-selected covariant centroid
\cite{Tulczyjew:1959,Dixon:1964,CostaNatario:2015}, and its brackets will
follow from that definition rather than from a postulate.

Because the map
\eqref{eq:X-definition}-\eqref{eq:S-definition} is a many-to-one
projection of centroid data and not a canonical transformation, we obtain
the induced brackets from the composite variables themselves.  This route
avoids an unjustified inversion of the constraint matrix
$C^{\mu\nu}\simeq p^2S^{\mu\nu}$, which for nonzero spin is antisymmetric
of rank two, its kernel containing both the momentum direction and the spin
direction, so that a generalized inverse projects onto a two-plane instead
of onto the full three-dimensional hyperplane orthogonal to $p^\mu$.  A
strict Dirac-bracket treatment would require the complete constraint and
gauge set, and appendix~\ref{app:jacobi} records the rank analysis that
makes this precise.

\subsection{Uniqueness in the linear covariant class and a spin-induced coordinate tensor}
\label{sec:centroid-uniqueness}

The transverse part of the shift in \eqref{eq:X-definition} is fixed by the
supplementary condition, and the uniqueness statement can be made precise
within the Lorentz-covariant class linear in the internal spin.  Allow both
the ordinary tensor contraction and the Pauli-Lubanski pseudovector,
\begin{equation}
\widetilde X^\mu
=z^\mu+f(p^2)\Sigma^{\mu\nu}p_\nu
+g(p^2)p^\mu+h(p^2)W^\mu .
\label{eq:general-centroid-shift}
\end{equation}
The $g$ term displaces the representative point along the worldline and
leaves the orbital bivector untouched.  Requiring that the spin referred to
the shifted point preserve the decomposition of $J^{\mu\nu}$ gives
\begin{equation}
\widetilde S^{\mu\nu}
=\Sigma^{\mu\nu}
-f(p^2)(A^\mu p^\nu-A^\nu p^\mu)
-h(p^2)(W^\mu p^\nu-W^\nu p^\mu)
\label{eq:general-centroid-spin}
\end{equation}
and hence
\begin{equation}
\widetilde S^{\mu\nu}p_\nu
=\left[1-f(p^2)p^2\right]A^\mu-h(p^2)p^2W^\mu .
\label{eq:general-centroid-transversality}
\end{equation}
Transversality for arbitrary spin data fixes $f(p^2)=1/p^2$ and $h(p^2)=0$.
Parity need not be imposed separately to remove the pseudovector shift: the
supplementary condition removes it.  Within this class the centroid
\eqref{eq:X-definition} is unique up to the longitudinal displacement
$g(p^2)p^\mu$, and the map is idempotent, since once
$S^{\mu\nu}p_\nu=0$ a second application changes neither $X$ nor $S$.

The coordinate bracket follows from the centroid map,
\begin{equation}
\{X^\mu,X^\nu\}
=-\frac{S^{\mu\nu}}{p^2},
\label{eq:XX-PB}
\end{equation}
with the calculation carried out in appendix~\ref{app:brackets}.  This
bracket is not new.  It was obtained for the Frenkel electron by Deriglazov
and Pupasov-Maksimov \cite{DeriglazovPupasov:2014}, whose eq.~(94) reads
$\{x^\mu,x^\nu\}=-J^{\mu\nu}/(2p^2)$ in a normalization where the spin
three-vector is $S^i=\tfrac14\epsilon^{ijk}J_{jk}$, so that $J=2S$ and
their bracket is \eqref{eq:XX-PB} term by term; the accompanying mixed
bracket of that reference likewise corresponds to \eqref{eq:reduced-SX}
below.  We reproduce the result through the composite-variable route
because the route, and not the bracket, is what controls the domain
statement of section~\ref{sec:shell-restriction}.
Quantization defines
\begin{equation}
[\hat X^\mu,\hat X^\nu]=i\widehat\Theta^{\mu\nu},
\qquad
\widehat\Theta^{\mu\nu}
\equiv-\hbar\frac{\hat S^{\mu\nu}}{\hat p^2}.
\label{eq:Theta-operator}
\end{equation}
The overall sign is fixed by \eqref{eq:X-definition} and can be reversed by
changing the orientation convention for the spin two-form, provided every
spin-dependent definition and bracket is transformed at the same time.  No
invariant constructed below depends on that choice.

In a massive sector one is tempted to replace $p^2$ by $m^2c^2$ and write
\begin{equation}
\Theta_m^{\mu\nu}
=-\frac{\hbar}{m^2c^2}S^{\mu\nu}.
\label{eq:Theta-mass-shell}
\end{equation}
Since $S^{\mu\nu}$ has the dimension of action, $\Theta_m^{\mu\nu}$ has the
dimension of area demanded by the coordinate commutator, and the associated
length is of Compton order, as already recorded in
\cite{DeriglazovPupasov:2014,DeriglazovTereza:2019}.  A three-dimensional
version of a spin-induced coordinate noncommutativity was studied in
\cite{Colatto_2006}, and covariant four-dimensional forms of
\eqref{eq:Theta-mass-shell} are available in the references just cited, so
no priority is claimed here for the tensorial statement.  The substitution
leading to \eqref{eq:Theta-mass-shell} is not innocuous, because $p^2$ has
been frozen inside an operator whose brackets have not yet been evaluated.
Section~\ref{sec:shell-restriction} determines exactly when that freezing
is legitimate, and the answer is what organizes the rest of this paper.
Equations \eqref{eq:TD-identity} and \eqref{eq:Theta-operator} together
give
\begin{equation}
\widehat\Theta^{\mu\nu}p_\nu=0,
\qquad
\Theta_m^{\mu\nu}p_\nu=0,
\label{eq:Theta-transverse}
\end{equation}
so the bivector is attached to the momentum and spin of the particle rather
than to an observer-independent vacuum.  The transversality is
correspondingly state dependent, since it refers to the momentum of the
very state that carries the spin tensor, and it must not be confused with a
background tensor that happens to annihilate one selected momentum.

The Pauli-Lubanski pseudovector,
\begin{equation}
W^\mu\equiv-\frac12
\epsilon^{\mu\nu\rho\sigma}J_{\nu\rho}p_\sigma,
\label{eq:W-definition}
\end{equation}
receives no orbital contribution in the decomposition \eqref{eq:J-reduced},
because the orbital part contracts the symmetric product $p_\rho p_\sigma$
with the Levi-Civita symbol.  With the conventions above one finds
\begin{equation}
S^{\mu\nu}
=-\frac{1}{p^2}
\epsilon^{\mu\nu\rho\sigma}p_\rho W_\sigma,
\qquad
\widehat\Theta^{\mu\nu}
=\frac{\hbar}{(\hat p^2)^2}
\epsilon^{\mu\nu\rho\sigma}p_\rho W_\sigma.
\label{eq:Theta-W}
\end{equation}
The inversion behind the first relation is short enough to record, and the
tensor it introduces will reappear in the Jacobi analysis.  Substituting
$W_\sigma=-\tfrac12\epsilon_{\sigma\alpha\beta\gamma}
S^{\alpha\beta}p^\gamma$ and contracting the two Levi-Civita symbols
through $\epsilon^{\mu\nu\rho\sigma}\epsilon_{\sigma\alpha\beta\gamma}
=\delta^{\mu\nu\rho}_{\alpha\beta\gamma}$ gives
\begin{equation}
\epsilon^{\mu\nu\rho\sigma}p_\rho W_\sigma
=-T^{\mu\nu\rho}p_\rho,
\qquad
T^{\mu\nu\rho}
\equiv
S^{\mu\nu}p^\rho+S^{\nu\rho}p^\mu+S^{\rho\mu}p^\nu,
\label{eq:T-def}
\end{equation}
where the generalized Kronecker delta collects the six signed permutations
into the totally antisymmetric tensor $T^{\mu\nu\rho}$.  Contraction with
one more momentum removes two of its three terms by transversality,
\begin{equation}
T^{\mu\nu\rho}p_\rho=S^{\mu\nu}p^2,
\label{eq:T-inversion}
\end{equation}
which establishes \eqref{eq:Theta-W}.  In this representation the
transversality \eqref{eq:Theta-transverse} is immediate, since the
contraction places two momenta in antisymmetrized slots.  The same
representation identifies the only covariant pseudovector available to
control a spin-dependent three-form, an observation that will matter once
nonassociativity is discussed.  We note already that $W^2$ is a Casimir
invariant, whereas the individual components $W^\mu$ are not central.

\section{Reduced algebra and momentum-space geometry}
\label{sec:algebra}

The remaining brackets induced by the same composite map are
\begin{align}
\{X^\mu,p^\nu\}&=\eta^{\mu\nu},
\qquad \{p^\mu,p^\nu\}=0, \qquad
\{S^{\mu\nu},p^\rho\}=0,
\label{eq:reduced-momentum}\\
\{S^{\mu\nu},X^\rho\}
&=\frac{1}{p^2}
\left(S^{\mu\rho}p^\nu-S^{\nu\rho}p^\mu\right),
\label{eq:reduced-SX}\\
\{S^{\mu\nu},S^{\rho\sigma}\}
&=\Pi^{\mu\rho}S^{\nu\sigma}
-\Pi^{\nu\rho}S^{\mu\sigma}
-\Pi^{\mu\sigma}S^{\nu\rho}
+\Pi^{\nu\sigma}S^{\mu\rho},
\label{eq:reduced-SS}\\
\Pi^{\mu\nu}
&\equiv\eta^{\mu\nu}-\frac{p^\mu p^\nu}{p^2}.
\label{eq:Pi-definition}
\end{align}
The projector $\Pi^{\mu\nu}$ confines the spin algebra to the hyperplane
orthogonal to $p^\mu$, and in the rest frame it removes the time direction
and leaves the usual $\mathfrak{so}(3)$ algebra of the spatial components.
The mixed bracket \eqref{eq:reduced-SX} is no optional refinement, since
it is what keeps the physical spin constraint stable under the position
operation,
\begin{equation}
\{S^{\mu\nu}p_\nu,X^\rho\}
=\{S^{\mu\nu},X^\rho\}p_\nu
+S^{\mu\nu}\{p_\nu,X^\rho\}=0.
\label{eq:TD-preserved}
\end{equation}
Setting $\{S,X\}=0$ while keeping $\{X,p\}$ canonical would define a
different and unstable truncation of the same phase space, whose
nonvanishing Jacobiator is examined once the projected products have been
introduced.

Quantization gives
\begin{align}
[\hat X^\mu,\hat X^\nu]
&=-i\hbar\frac{\hat S^{\mu\nu}}{\hat p^2},
\label{eq:XX-commutator}\\
[\hat X^\mu,p^\nu]&=i\hbar\eta^{\mu\nu},
\qquad
[\hat S^{\mu\nu},p^\rho]=0,
\label{eq:Xp-Sp-commutators}\\
[\hat S^{\mu\nu},\hat X^\rho]
&=\frac{i\hbar}{\hat p^2}
\left(\hat S^{\mu\rho}p^\nu-\hat S^{\nu\rho}p^\mu\right),
\label{eq:SX-commutator}\\
[\hat S^{\mu\nu},\hat S^{\rho\sigma}]
&=i\hbar\left(
\Pi^{\mu\rho}\hat S^{\nu\sigma}
-\Pi^{\nu\rho}\hat S^{\mu\sigma}
-\Pi^{\mu\sigma}\hat S^{\nu\rho}
+\Pi^{\nu\sigma}\hat S^{\mu\rho}
\right).
\label{eq:SS-commutator}
\end{align}
Since $\hat S^{\mu\nu}$ commutes with $\hat p^2$, the product
$\hat S^{\mu\nu}\hat p^{-2}$ carries no ordering ambiguity.  The total
generator
\begin{equation}
\hat J^{\mu\nu}
=\hat X^\mu p^\nu-\hat X^\nu p^\mu+\hat S^{\mu\nu}
\label{eq:J-quantum}
\end{equation}
obeys the ordinary Poincar\'e algebra and acts covariantly, as in
\begin{equation}
[\hat J^{\mu\nu},\hat X^\rho]
=i\hbar\left(
\eta^{\mu\rho}\hat X^\nu-\eta^{\nu\rho}\hat X^\mu
\right).
\label{eq:JX-commutator}
\end{equation}
Lorentz covariance is thus realized by a particle-dependent operator
tensor, without any need to declare a fixed numerical $\Theta$ invariant.

Whether the Jacobi identity survives depends on treating the inverse
momentum square as an operator until the commutators have been evaluated,
the precise licence being \eqref{eq:two-insertion-neutrality} below.  From
$[\hat p^2,\hat X^\rho]=-2i\hbar p^\rho$ one obtains
\begin{equation}
[\hat p^{-2},\hat X^\rho]
=\frac{2i\hbar p^\rho}{\hat p^4},
\label{eq:inverse-p2-commutator}
\end{equation}
so that the double commutator collects one contribution from the spin
bracket \eqref{eq:SX-commutator} and one from the variation of
$\hat p^{-2}$,
\begin{equation}
[[\hat X^\mu,\hat X^\nu],\hat X^\rho]
=\frac{\hbar^2}{\hat p^4}
\left(
\hat S^{\mu\rho}p^\nu-\hat S^{\nu\rho}p^\mu
+2\hat S^{\mu\nu}p^\rho
\right).
\label{eq:coordinate-double-commutator}
\end{equation}
The cyclic sum of the first two terms equals $-2T^{\mu\nu\rho}$, with
$T^{\mu\nu\rho}$ defined in \eqref{eq:T-def}, while the cyclic sum of the
last term equals $+2T^{\mu\nu\rho}$, whence
\begin{equation}
[[\hat X^\mu,\hat X^\nu],\hat X^\rho]
+\text{cyclic}=0.
\label{eq:coordinate-Jacobi-zero}
\end{equation}
The term that rescues the identity originates entirely in the operator
character of $\hat p^{-2}$, and freezing $\hat p^2$ to a number before the
commutators are evaluated destroys the cancellation.  Mixed Jacobi classes
are equally instructive, since they test the relative signs in
\eqref{eq:XX-commutator}-\eqref{eq:SS-commutator}, and
appendix~\ref{app:jacobi} records those checks.

\subsection{Where mass-shell restriction is legitimate}
\label{sec:shell-restriction}

The preceding paragraph and the mass-shell tensor
\eqref{eq:Theta-mass-shell} pull in opposite directions.  One says that
$\hat p^{-2}$ must be kept operator valued, the other replaces it by a
number, and a construction that uses both without a criterion is not
well defined.  We state the criterion here, because it governs every later
section and because the answer is sharper than the qualitative restriction
to shell-preserving matrix elements used in earlier treatments.

Fix $K=m^2c^2$, let $\Sigma_K=\{p\in\Omega_\mu\mid p^2=K\}$, and let
$\hat X_K^\mu$ denote the frozen surrogate obtained by replacing
$\hat p^{-2}$ by $1/K$ inside \eqref{eq:X-definition},
\begin{equation}
\hat X_K^\mu
\equiv\hat z^\mu+\frac{\hat\Sigma^{\mu\nu}p_\nu}{K},
\qquad
\Delta\hat X^\mu
\equiv\hat X^\mu-\hat X_K^\mu
=\hat\Sigma^{\mu\nu}p_\nu\,\varphi(\hat p^2),
\qquad
\varphi(p^2)\equiv\frac1{p^2}-\frac1K .
\label{eq:frozen-surrogate}
\end{equation}
The whole question turns on the fact that $\Delta\hat X^\mu$ acts by
multiplication in the momentum representation and that its multiplier
$\varphi$ vanishes pointwise on $\Sigma_K$.  Neither statement holds for
$\hat X^\mu$ itself, which differentiates.

Let $\psi$ and $\phi$ have momentum support on $\Sigma_K$, in the rigged
extension of $\mathcal D$ in which on-shell external states are read, and
expand a product of $n$ insertions by telescoping,
\begin{equation}
\hat X^{\mu_1}\cdots\hat X^{\mu_n}
-\hat X_K^{\mu_1}\cdots\hat X_K^{\mu_n}
=\sum_{j=1}^{n}
\hat X_K^{\mu_1}\cdots\hat X_K^{\mu_{j-1}}\,
\Delta\hat X^{\mu_j}\,
\hat X^{\mu_{j+1}}\cdots\hat X^{\mu_n}.
\label{eq:telescoping}
\end{equation}
For $n=1$ the single term carries $\Delta\hat X$ against $\phi$, and for
$n=2$ the two terms carry it against $\phi$ and against $\psi$
respectively.  A multiplication operator whose multiplier vanishes on the
support of the adjacent state annihilates the matrix element, so
\begin{equation}
\langle\psi|\hat X^\mu|\phi\rangle
=\langle\psi|\hat X_K^\mu|\phi\rangle,
\qquad
\langle\psi|\hat X^\mu\hat X^\nu|\phi\rangle
=\langle\psi|\hat X_K^\mu\hat X_K^\nu|\phi\rangle .
\label{eq:two-insertion-neutrality}
\end{equation}
From $n=3$ onward the sum contains the term $j=2$, in which
$\Delta\hat X^{\mu_2}$ stands between two position operators and is
adjacent to neither state.  The differentiation in the neighboring
$\hat X$ reaches the multiplier, and $\varphi$ vanishes on $\Sigma_K$
while its normal derivative does not, so nothing forces the term to
vanish.  The obstruction is not merely unproved to be zero.  In the totally
antisymmetric channel it is computed by the truncated Jacobiator
\eqref{eq:truncated-Jacobiator}, which is proportional to the
Pauli-Lubanski vector and therefore nonzero for any nonzero spin.  The
frozen bracket consequently admits no realization as the commutator of an
associative operator algebra beyond two insertions.

Two consequences organize what follows.  Every result of this paper that is
read on shell through at most two insertions of the covariant position
survives the replacement $p^2\to m^2c^2$ exactly, and this covers the
projected position dictionary of section~\ref{sec:positive-energy}, the
Berry cancellation, the endpoint transversality identity of
section~\ref{sec:endpoint-lemma}, and the transverse matching integral.
Every statement requiring three or more insertions, which is to say the
Jacobi identity itself, the associator of section~\ref{sec:projected}, and
any attempt to promote $\Theta_m$ to a fixed background bivector in a
product of three or more coordinate factors, does not survive it.  In place
of an undefined restriction to shell-preserving matrix elements, later
sections cite \eqref{eq:two-insertion-neutrality} and count insertions.

Let the momenta act by multiplication on smooth wave functions supported in
the timelike region
\begin{equation}
\Omega_\mu=\{p\mid p^2>\mu^2,\ p^0>0\},
\qquad
\mathcal D=C_c^\infty(\Omega_\mu,d^4p)\otimes\mathcal V,
\label{eq:test-domain}
\end{equation}
where $\mu>0$ keeps the construction away from the null cone and the
Lorentz-spin matrices $\hat\Sigma^{\mu\nu}$ act on $\mathcal V$.  A formal
realization of the centroid algebra is then
\begin{align}
\hat z^\mu&=i\hbar\frac{\partial}{\partial p_\mu},
\label{eq:z-momentum}\\
\hat X^\mu&=
i\hbar\frac{\partial}{\partial p_\mu}
+\frac{\hat\Sigma^{\mu\nu}p_\nu}{p^2},
\label{eq:X-momentum}\\
\hat S^{\mu\nu}
&=\Pi^\mu{}_\alpha\Pi^\nu{}_\beta
\hat\Sigma^{\alpha\beta}.
\label{eq:S-momentum}
\end{align}
Direct substitution reproduces the complete algebra above, and because all
operators are built inside the associative algebra generated by
$(\hat z,p,\hat\Sigma)$, the Jacobi identities follow without an additional
postulate.  The multiplicative term in \eqref{eq:X-momentum} defines a
nonabelian connection,
\begin{equation}
\mathcal A^\mu(p)
\equiv\frac{\hat\Sigma^{\mu\nu}p_\nu}{p^2},
\label{eq:momentum-connection}
\end{equation}
whose curvature, with $\partial^\mu\equiv\partial/\partial p_\mu$, is
\begin{equation}
\mathcal F_p^{\mu\nu}
\equiv
\partial^\mu\mathcal A^\nu-\partial^\nu\mathcal A^\mu
+\frac{1}{i\hbar}
[\mathcal A^\mu,\mathcal A^\nu]
=-\frac{\hat S^{\mu\nu}}{p^2},
\label{eq:momentum-curvature}
\end{equation}
so that
\begin{equation}
[\hat X^\mu,\hat X^\nu]
=i\hbar\mathcal F_p^{\mu\nu},
\qquad
\widehat\Theta^{\mu\nu}
=\hbar\mathcal F_p^{\mu\nu}.
\label{eq:Theta-curvature}
\end{equation}
The noncommutative coordinate tensor is therefore the curvature associated
with the change from the reference position to the covariant centroid, and
the vocabulary of transport, holonomy and fiber used from here on refers to
the connection \eqref{eq:momentum-connection}.

The way the two contributions in \eqref{eq:momentum-curvature} combine
repays a closer look, because the nonabelian term turns out to carry the
same weight as the abelian one.  Differentiating
\eqref{eq:momentum-connection} and collecting terms with the definition
\eqref{eq:S-definition} gives
\begin{equation}
\partial^\mu\mathcal A^\nu-\partial^\nu\mathcal A^\mu
=-\frac{2\hat S^{\mu\nu}}{p^2},
\label{eq:abelian-part}
\end{equation}
while contracting the Lorentz algebra \eqref{eq:canonical-spin} of the spin
matrices with two momenta gives
\begin{equation}
\frac{1}{i\hbar}
[\mathcal A^\mu,\mathcal A^\nu]
=\frac{\hat S^{\mu\nu}}{p^2}.
\label{eq:nonabelian-part}
\end{equation}
Both pieces are proportional to the same transverse tensor with
coefficients $-2$ and $+1$, so the commutator cancels exactly half of the
exterior derivative.  Discarding it would leave the tensor structure intact
while doubling the magnitude, and with it the normalization of the
coordinate commutator, of the invariant length obtained below, and of the
Compton holonomy.  The Lorentz algebra of the spin matrices is in that
sense what fixes the scale of the deformation.

Identifying a spin-dependent coordinate commutator with a nonabelian Berry
curvature over momentum space is not new.  For a Dirac particle with the
negative-energy band removed, B\'erard and Mohrbach obtained the structural
identity
$\theta^{ij}=\partial^iA^j-\partial^jA^i+[A^i,A^j]$ from the
positive-energy Berry connection
\cite{Berard:2004monopole,Berard:2006spinhall}.  Their construction is
band-projected, whereas \eqref{eq:momentum-curvature} is an exact identity
in the ambient covariant algebra.  The Berry connection measures the shift
of the projected Dirac or charge position relative to the commuting
Newton-Wigner coordinate; the Tulczyjew-Dixon position is a different,
four-dimensionally covariant centroid.  Their shifts vanish at the rest
point, but their physical curvatures have opposite orientation there and
obey different boost laws, as shown in
\eqref{eq:b-FW}-\eqref{eq:b-centroid-normalized}.  The positive-energy
projection in section~\ref{sec:positive-energy} relates these structures
without identifying them.

The set $(p^\mu,\hat S^{\mu\nu},\hat X^\mu)$ closes as a Snyder-like
quadratic algebra in which the transverse spin tensor, and not the full
Lorentz generator, sources the coordinate commutator \cite{Snyder:1947}.
The comparison is instructive on the field-theory side as well, since
scalar field theory on Snyder space admits several inequivalent star
products, some of them nonassociative, depending on the realization chosen
\cite{Girelli:2011snyder}.  Here associativity is a property of the
complete reduced operator algebra on the timelike kinematical domain, it is
not inherited by single-shell truncations, and no particular star-product
realization is singled out at this stage.

Antisymmetry and the supplementary condition give the two transversality
relations
\begin{equation}
\mathcal A^\mu p_\mu=0,
\qquad
\mathcal F_p^{\mu\nu}p_\nu=0.
\label{eq:connection-transversality}
\end{equation}
The first identity makes radial transport flat, while the second removes
radial curvature legs.  More explicitly, any future-timelike path can be
written as $q^\mu=M u^\mu$, with $M=\sqrt{q^2}>0$ and $u^2=1$.  Along this
path,
\begin{equation}
\mathcal A_\mu(q)dq^\mu
=\frac{\Sigma_{\mu\nu}u^\nu}{M}
\left(Mdu^\mu+u^\mu dM\right)
=\Sigma_{\mu\nu}u^\nu du^\mu.
\label{eq:radial-projection}
\end{equation}
The transport depends only on the radial projection of the path to the unit
mass hyperboloid.  When its endpoints are referred to the same rest fiber,
a closed path defines a Wigner little-group rotation and is unitary in a
unitary spin-$s$ representation.  An open line represented by finite
Lorentz-spinor matrices need not be unitary with the ordinary Hilbert
product, because the path contains boost generators; it preserves instead
the compatible indefinite fiber pairing introduced below.  This distinction
will be retained when the Compton loop is compared with an on-shell Wigner
polygon.

For a path $\gamma$ in a timelike momentum domain, define
\begin{equation}
U_\gamma(p',p)
\equiv
\mathcal P\exp\left[
\frac{i}{\hbar}\int_\gamma
\mathcal A_\mu(q)\,dq^\mu
\right].
\label{eq:Wilson-line}
\end{equation}
Writing $V_p$ for the spin fiber over $p$, the map is typed as
$U_\gamma(p',p):V_p\to V_{p'}$, and at this stage it is the parallel
transporter of the kinematical spin bundle and nothing more.  We assume a
nondegenerate fiber pairing $h_p$ compatible with the connection,
\begin{equation}
h_{p'}(U_\gamma v,U_\gamma w)=h_p(v,w),
\label{eq:fiber-pairing}
\end{equation}
so that the inverse associated with a reversed path is the adjoint with
respect to $h_p$, although it need not coincide with the ordinary
Hermitian conjugate in an arbitrary Lorentz-spinor trivialization.  Under a
momentum-dependent change of spin frame an open line transforms at both
endpoints, whereas a closed holonomy transforms by conjugation, which is
why traces and matrix elements built from consistently transported external
states come out frame independent.  Everything said later about
observability rests on that difference between open and closed transport.

Since a curvature over momentum space recalls the Berry monopole of
massless particles, one may ask whether the present bundle carries
topological information as well.  The domain settles the question.  The
region $\Omega_\mu$ of \eqref{eq:test-domain} lies above the sheet
$p^0=\sqrt{|\bm p|^2+\mu^2}$, which is a convex function of $\bm p$, so
$\Omega_\mu$ is a convex open subset of $\mathbb R^4$ and hence
contractible, every bundle over it is trivial, and
$H^k(\Omega_\mu,\mathbb R)=0$ for $k>0$.  No characteristic class and no
quantized flux can be attached to $\mathcal F_p$ in the massive timelike
interior.  What survives is a pointwise invariant instead of a topological
one, since the Lorentz scalar
$\mathcal F_{p\,\mu\nu}\mathcal F_p^{\mu\nu}$ is proportional to
$\Theta_{m\,\mu\nu}\Theta_m^{\mu\nu}$ and is fixed by the Casimir
invariants.  The topology sits on the boundary that the construction
excludes, because $\mathcal A^\mu$ is singular at $p^2=0$, and that is
exactly where the Berry monopole with Chern number proportional to the
helicity appears in the massless limit
\cite{Berard:2004monopole,ChenSonStephanov:2014,StoneDwivediZhou:2015}.
The massive centroid geometry is in this sense the desingularized member of
that family, and the same infrared scale $\mu>0$ that makes $\hat p^{-2}$ a
bounded operator is what removes the topological content.  Any construction
aiming at a spin-controlled topological invariant would have to approach
the light cone, where the present operator domain fails.

The realization is formal in the operator-theoretic sense, supplying a
dense invariant test domain and the connection demanded by the commutator
without establishing self-adjoint extensions for all four covariant
position components.  It also shows why the position acts on a larger
kinematical space,
\begin{equation}
[\hat p^2,\hat X^\rho]=-2i\hbar p^\rho.
\label{eq:p2-not-preserved}
\end{equation}
The Poincar\'e generators preserve the irreducible mass sectors while
$\hat X^\rho$ does not, so replacing $\hat p^2$ by $m^2c^2$ becomes
legitimate only after the commutators involving $\hat X$ have been
evaluated.  The natural domain of the coordinate algebra is accordingly a
direct integral of timelike mass shells, whose spectral formulation appears
in appendix~\ref{app:jacobi}.

\subsection{A centroid-relative interval}
\label{sec:centroid-interval}

An earlier motivation for spin-induced noncommutativity came from a
deformed interval built out of the Bopp-shifted coordinate, and that
motivation survives without introducing a new background metric.  Define
the reference orbital generator
\begin{equation}
L^{\mu\nu}\equiv z^\mu p^\nu-z^\nu p^\mu.
\label{eq:L-reference}
\end{equation}
On the domain \eqref{eq:test-domain} the divergence
$\partial_\mu(\Sigma^{\mu\nu}p_\nu/p^2)$ vanishes, so the operator identity
\begin{equation}
\hat X^2
=\hat z^2
+\frac{\hat\Sigma_{\mu\nu}\hat L^{\mu\nu}}{\hat p^2}
+\frac{\hat A_\mu\hat A^\mu}{\hat p^4}
\label{eq:centroid-interval}
\end{equation}
holds with no additional ordering constant.  The linear term has the form
of a spin-orbit scalar, generated here by the change of localization
variable and not by an external field, while the quadratic term measures
the squared transverse distance between the reference line and the
covariant centroid.

What \eqref{eq:centroid-interval} expresses is a relation between two
position representatives, leaving the Minkowski metric and the free
mass-shell Casimir invariant untouched.  Should the reference variables
already obey the Tulczyjew-Dixon condition, then $A^\mu=0$ and the two
representatives coincide.  The physical content is relational, in that spin
determines which event on an extended centroid tube is represented by
$X^\mu$.

\section{Lorentz class, representations, and Compton scale}
\label{sec:geometry}

For a massive shell, define
\begin{equation}
e^i\equiv\Theta_m^{0i},
\qquad
\Theta_m^{ij}\equiv\epsilon^{ijk}b_k,
\qquad
S^{ij}\equiv-\epsilon^{ijk}s_k,
\qquad
\epsilon^{123}=+1,
\label{eq:e-b-definitions}
\end{equation}
where $\bm s$ is the physical three-spin fixed by
\eqref{eq:spin-dictionary}.  In the rest frame,
$p^\mu=(mc,\bm0)$, transversality gives $S^{0i}=0$ and
\begin{equation}
\bm e_{\rm rest}=0,
\qquad
\bm b_{\rm rest}
=+\frac{\hbar}{m^2c^2}\bm s.
\label{eq:e-b-rest}
\end{equation}
A spin-$1/2$ state polarized along the third axis therefore has
\begin{equation}
\Theta_{m,\rm rest}^{\mu\nu}
=+\frac{\hbar}{m^2c^2}
\begin{pmatrix}
0&0&0&0\\
0&0&s_3&0\\
0&-s_3&0&0\\
0&0&0&0
\end{pmatrix}^{\mu\nu},
\qquad
[X^1,X^2]
=+i\hbar\frac{s_3}{m^2c^2}.
\label{eq:Theta-rest-matrix}
\end{equation}
The noncommutative plane is orthogonal to the polarization, and the
vanishing temporal row and column express the absence of electric-type
noncommutativity for a comoving observer.

For an observer of unit four-velocity $n^\mu$ the covariant decomposition is
\begin{equation}
e^\mu(n)=\Theta_m^{\mu\nu}n_\nu,
\qquad
b^\mu(n)=-\frac12\epsilon^{\mu\nu\rho\sigma}
n_\nu\Theta_{m\,\rho\sigma}.
\label{eq:covariant-e-b}
\end{equation}
The comoving observer $n^\mu=p^\mu/(mc)$ measures $e^\mu=0$ identically,
whereas a laboratory observer related by a boost with velocity $\bm\beta$
measures
\begin{align}
\bm e_{\rm lab}
&=-\gamma\,\bm\beta\times\bm b_{\rm rest}
=-\gamma\frac{\hbar}{m^2c^2}\bm\beta\times\bm s,
\label{eq:e-boost}\\
\bm b_{\rm lab}
&=\gamma\bm b_{\rm rest}
-(\gamma-1)
(\hat{\bm\beta}\cdot\bm b_{\rm rest})\hat{\bm\beta}
\notag\\
&=+\frac{\hbar}{m^2c^2}
\left[
\gamma\bm s-(\gamma-1)
(\hat{\bm\beta}\cdot\bm s)\hat{\bm\beta}
\right].
\label{eq:b-boost}
\end{align}
The components $\Theta^{0i}$ are generated by the mismatch between the
laboratory and comoving time directions.  They vanish for spin parallel to
the velocity, switch on linearly in $v/c$, and are not independent
time-space background parameters.

The same boost law permits a comparison with the positive-energy
Foldy-Wouthuysen Berry curvature.  In the standard physical spin
convention, the projected Dirac or charge position obeys
\begin{equation}
[x_D^i,x_D^j]
=-i\hbar\,\epsilon^{ijk}\frac{c^4}{E_p^3}
\left[
ms_k+\frac{p_k(\bm p\cdot\bm s)}{E_p+mc^2}
\right],
\label{eq:FW-commutator}
\end{equation}
with $E_p$ the positive Dirac energy
\cite{Berard:2004monopole,Berard:2006spinhall,Bliokh:2017}.  At
$\bm p=0$ the position shifts themselves vanish, but their curvatures have
opposite orientation: the charge-position commutator is
$-i\hbar\epsilon^{ijk}s_k/(m^2c^2)$, whereas the covariant centroid gives
\eqref{eq:Theta-rest-matrix}.  Curvature probes the momentum dependence in a
neighborhood of the rest point and therefore need not agree when two shifts
vanish at that point.

Writing \eqref{eq:FW-commutator} in magnetic-vector notation with
$E_p=\gamma mc^2$ and $\bm p=\gamma mc\bm\beta$ gives
\begin{equation}
\bm b_{\rm FW}
=-\frac{\hbar}{m^2c^2}
\left[
\frac{\bm s}{\gamma^3}
+\frac{\bm\beta\,(\bm\beta\cdot\bm s)}{\gamma(\gamma+1)}
\right],
\label{eq:b-FW}
\end{equation}
whereas the covariant centroid tensor reads
\begin{equation}
\bm b_{\rm TD}
=+\frac{\hbar}{m^2c^2}
\left[
\gamma\bm s
-\frac{\gamma^2\bm\beta\,(\bm\beta\cdot\bm s)}{\gamma+1}
\right].
\label{eq:b-centroid-normalized}
\end{equation}
For polarization transverse to the boost, their magnitudes differ by
$\gamma^4$ and their orientations are opposite.  Equation
\eqref{eq:b-FW} is a Berry curvature on the positive-energy bundle and does
not transform as an antisymmetric four-tensor, in contrast with
\eqref{eq:e-boost}-\eqref{eq:b-boost}.  The discrepancy measures the
difference between the charge position and the covariant centroid and fixes
why a detector-dependent position dictionary is required before a side jump
can be assigned physical meaning.

The two bivector invariants are
\begin{align}
\frac12\Theta_{m\,\mu\nu}\Theta_m^{\mu\nu}
&=\bm b^2-\bm e^2
=-\frac{\hbar^2W^2}{(p^2)^3},
\label{eq:Theta-first-invariant}\\
\frac14\epsilon_{\mu\nu\rho\sigma}
\Theta_m^{\mu\nu}\Theta_m^{\rho\sigma}
&\propto\bm e\cdot\bm b=0.
\label{eq:Theta-second-invariant}
\end{align}
The first invariant is positive for a massive spinning representation and
the second vanishes, so the tensor belongs to the magnetic Lorentz class in
every inertial frame even though its electric components need not vanish in
every frame.  No boost can leave the noncommutativity dominated by its
temporal sector, a property whose consequences for unitarity are taken up
once field-theoretic matching is discussed.

\subsection{Casimir invariants and invariant length}
\label{sec:invariant-scale}

The Poincar\'e Casimir operators remain
\begin{equation}
p^2=m^2c^2,
\qquad
W^2=-m^2c^2\hbar^2s(s+1),
\label{eq:Poincare-Casimirs}
\end{equation}
because the deformation is confined to the coordinate sector while $p^\mu$
and $J^{\mu\nu}$ retain their algebra.  Using \eqref{eq:Theta-W},
\begin{equation}
S_{\mu\nu}S^{\mu\nu}=2\hbar^2s(s+1),
\qquad
\Theta_{m\,\mu\nu}\Theta_m^{\mu\nu}
=2\lambda_{\rm C}^4s(s+1),
\label{eq:Theta-Casimir}
\end{equation}
where
\begin{equation}
\lambda_{\rm C}\equiv\frac{\hbar}{mc}.
\label{eq:reduced-Compton}
\end{equation}
Neither the Wigner labels $(m,s)$ nor the multiplicity $2s+1$ is deformed,
and what changes is the coordinate realization carried by a state in that
representation.  The nonvanishing commutator $[p^2,\hat X^\rho]$ has no
bearing on the classification, which is governed by how the Casimir
operators commute with $p^\mu$ and $J^{\mu\nu}$ and not by the action of a
covariant position inside a fixed shell
\cite{wigner1939unitary,bargmann1948unitary,NewtonWigner:1949}.

Since $\Theta$ carries the dimension of area, define the invariant
algebraic length
\begin{equation}
\ell_\Theta\equiv
\left(\Theta_{m\,\mu\nu}\Theta_m^{\mu\nu}\right)^{1/4}
=\lambda_{\rm C}[2s(s+1)]^{1/4}.
\label{eq:Theta-length}
\end{equation}
For $s=1/2$ this gives $\ell_\Theta\simeq1.1\,\lambda_{\rm C}$, about
$4.3\times10^{-13}\,$m for the electron, while a convention based on
$(\Theta_{\mu\nu}\Theta^{\mu\nu}/2)^{1/4}$ differs by the fixed factor
$2^{-1/4}$.  This is the scale both of the coordinate algebra and of the
centroid ambiguity.  Calling it a strict minimum measurable length would
require a self-adjoint position observable together with a measurement
protocol, yet its proximity to the M\o ller radius $|\bm S|/(mc)$ is the
signature one expects from localization carried by relativistic spin
\cite{Moller:1949,CostaNatario:2015,DeriglazovPupasov:2014}.  Pair creation
at Compton distances supplies an independent reason why a single-particle
position description degrades at shorter scales, and neither argument
implies that vacuum coordinates obey the same commutator.

One distinction should be kept in view before the local model is
constructed, namely that the exact Casimir invariant differs from the
squared expectation value used in a polarized packet,
\begin{equation}
\langle S_{\mu\nu}S^{\mu\nu}\rangle
\neq
\langle S_{\mu\nu}\rangle
\langle S^{\mu\nu}\rangle
\label{eq:spin-expectation-distinction}
\end{equation}
in general.  The left-hand side fixes the representation scale, whereas the
right-hand side fixes the strength of the c-number model constructed next.

\section{Positive-energy projection and physical position operators}
\label{sec:positive-energy}

The covariant position does not preserve a fixed mass shell, as shown by
\eqref{eq:p2-not-preserved}, and in a Dirac realization it also mixes the
positive- and negative-energy sectors.  A one-particle experiment instead
acts within the positive-energy space.  The restriction must therefore be
performed at the operator level before the covariant bracket is read as a
laboratory commutator.  This section carries out that restriction in the
free Dirac theory and keeps the conversion
\eqref{eq:laboratory-dictionary} explicit.

Let
\begin{equation}
W=\sqrt{m^2c^4+c^2\bm p^2},
\qquad
M=mc^2,
\qquad
U=\frac{W+M+c\beta\bm\alpha\cdot\bm p}
{\sqrt{2W(W+M)}} ,
\label{eq:FW-unitary}
\end{equation}
so that $UH_DU^\dagger=\beta W$, and denote the upper
positive-energy block by $++$.  In laboratory variables the spin part of
the covariant position is
\begin{equation}
\mathscr M^i
\equiv\frac{\Sigma_D^{i\nu}p_\nu}{m^2c^2},
\qquad
\Sigma_D^{\mu\nu}=+\frac{\hbar}{2}\sigma^{\mu\nu}.
\label{eq:lab-spin-shift}
\end{equation}
Transforming the full position and then taking the positive block gives
\begin{equation}
q^i_{{\rm TD},++}
=x^i_{\rm NW}+\mathcal A_B^i
+\big(U\mathscr M^iU^\dagger\big)_{++},
\label{eq:projected-position-decomposition}
\end{equation}
where $x^i_{\rm NW}=i\hbar\partial/\partial p_i$ is the commuting
Newton-Wigner operator and
\begin{equation}
\mathcal A_B^i
=-\frac{\hbar c^2(\bm\sigma\times\bm p)^i}
{2W(W+M)}
=-\frac{c^2(\bm s\times\bm p)^i}{W(W+M)}
\label{eq:Berry-connection-charge}
\end{equation}
is the Berry shift of the projected Dirac or charge position.

Direct block multiplication yields the exact identity
\begin{equation}
\big(U\mathscr M^iU^\dagger\big)_{++}
=\frac{\hbar(\bm\sigma\times\bm p)^i}{2mW}
-\frac{i\hbar p_i}{2m^2c^2}\bm1_2 .
\label{eq:Mpp-correct}
\end{equation}
The transverse term in \eqref{eq:Mpp-correct} is not, by itself, the
Tulczyjew-Dixon displacement.  Using
\begin{equation}
\frac{1}{mW}-\frac{c^2}{W(W+M)}
=\frac{1}{m(W+M)},
\label{eq:Berry-cancellation-identity}
\end{equation}
the sum in \eqref{eq:projected-position-decomposition} becomes
\begin{equation}
q^i_{{\rm TD},++}
=x^i_{\rm NW}
+\frac{(\bm s\times\bm p)^i}{m(W+M)}
-\frac{i\hbar p_i}{2m^2c^2}\bm1_2 .
\label{eq:projected-TD-position}
\end{equation}
The negative Berry term contained in the projected spin shift cancels the
explicit Berry connection, leaving the Tulczyjew-Dixon displacement of the
positive-energy irreducible representation.  The remaining anti-hermitian
term is radial and spin independent; its curl and all mutual commutators
vanish, so it changes neither the spin-dependent commutator nor the
transverse side jump.  It nevertheless records the fact that the ambient
covariant position was not self-adjoint in the ordinary positive-energy
inner product.  Removing it would require an explicitly declared similarity
transformation or a compatible weighted product, neither of which is
assumed here.

For reference, the two identities needed in the block calculation are
\begin{equation}
(\bm\sigma\cdot\bm p)(\bm\sigma\times\bm p)_i
(\bm\sigma\cdot\bm p)
=-\bm p^2(\bm\sigma\times\bm p)_i,
\qquad
(W+M)^2-c^2\bm p^2=2M(W+M).
\label{eq:Mpp-identities}
\end{equation}
Appendix~\ref{app:projection} supplies the remaining block algebra and the
factorizations used below.

\subsection{Projected commutator and the centroid dictionary}
\label{sec:projected-commutator}

The hermitian part of \eqref{eq:projected-TD-position} has the exact
on-shell commutator
\begin{equation}
\frac{[q^i_{{\rm TD},++},q^j_{{\rm TD},++}]}{i\hbar}
=\epsilon^{ijk}\left[
\frac{s_k}{mW}
+\frac{p_k(\bm p\cdot\bm s)}{m^2W(W+M)}
\right].
\label{eq:projected-commutator-general}
\end{equation}
Decomposing $\bm s=\bm s_\parallel+\bm s_\perp$ relative to $\bm p$ gives
the equivalent curvature vector
\begin{equation}
\bm{\mathcal B}_{{\rm TD},+}
=\frac{\bm s_\parallel}{m^2c^2}
+\frac{\bm s_\perp}{\gamma m^2c^2}.
\label{eq:projected-curvature-vector}
\end{equation}
For $\bm p\parallel\hat{\bm z}$, the two independent components are
\begin{equation}
[q^1,q^2]_{{\rm TD},++}
=i\hbar\frac{s_\parallel}{m^2c^2},
\qquad
[q^2,q^3]_{{\rm TD},++}
=i\hbar\frac{s_\perp}{\gamma m^2c^2}.
\label{eq:projected-commutators}
\end{equation}
By contrast, the unprojected covariant algebra gives
\begin{equation}
\bm{\mathcal B}_{\rm amb}
=\frac{\bm s_\parallel+\gamma\bm s_\perp}{m^2c^2}.
\label{eq:ambient-curvature-components}
\end{equation}
Projection preserves the longitudinal component and reduces the transverse
component by $\gamma^2$.  This anisotropy is an operator result of the free
positive-energy theory, not a new coupling.

Table~\ref{tab:position-dictionary} keeps four mathematically different
objects in view while identifying the three physical position operators.
The ambient covariant coordinate is included for comparison, but it is not
an endomorphism of the positive-energy sector.  The projected TD position
and the TD centroid of the positive-energy irreducible representation are
the same transverse operator; counting them separately would duplicate
\eqref{eq:projected-TD-position}.

\begin{table}[t]
\centering
\footnotesize
\caption{Physical position dictionary for $\bm p\parallel\hat{\bm z}$.
Shifts are measured from the Newton-Wigner position, and curvature entries
denote $[q^1,q^2]/i\hbar$ for parallel spin and
$[q^2,q^3]/i\hbar$ for transverse spin.}
\label{tab:position-dictionary}
\begin{tabular}{@{}L{0.21\linewidth}L{0.36\linewidth}L{0.17\linewidth}L{0.17\linewidth}@{}}
\toprule
position & transverse shift & parallel curvature & transverse curvature \\
\midrule
Newton-Wigner & $0$ & $0$ & $0$ \\
charge/Dirac &
$-c^2(\bm s\times\bm p)/[W(W+M)]$ &
$-s_\parallel/(\gamma^2m^2c^2)$ &
$-s_\perp/(\gamma^3m^2c^2)$ \\
TD projected $=$ TD irrep &
$(\bm s\times\bm p)/[m(W+M)]$ &
$+s_\parallel/(m^2c^2)$ &
$+s_\perp/(\gamma m^2c^2)$ \\
ambient covariant algebra &
not sector preserving &
$+s_\parallel/(m^2c^2)$ &
$+\gamma s_\perp/(m^2c^2)$ \\
\bottomrule
\end{tabular}
\end{table}

The compression that produces \eqref{eq:projected-commutator-general} is a
physical example of intermediate-sector leakage.  Section
\ref{sec:projected} later defines the corresponding projected product and
its associator without attributing a failure of Jacobi to the exact ambient
algebra.

\section{Frozen-spin fibers and deformed composition}
\label{sec:fibered}

A Moyal product requires a c-number antisymmetric tensor, which the exact
operator \eqref{eq:Theta-operator} is not.  The constant-tensor machinery
becomes available only inside a selected semiclassical neighborhood, and
what follows constructs that neighborhood together with the parameters
controlling the error committed in adopting it.

Let $\mathcal C$ denote a family of polarized wave packets centered at
$p_0$, and write $p=p_0+\delta p$.  Pointwise transversality gives
$\langle S^{\mu\nu}p_\nu\rangle_{\mathcal C}=0$, which does not in general
imply $\langle S^{\mu\nu}\rangle_{\mathcal C}p_{0\nu}=0$, the two
statements differing by the correlation
$-\langle S^{\mu\nu}\delta p_\nu\rangle_{\mathcal C}$.  We therefore define
the packet polarization by projection,
\begin{align}
\Pi_0{}^\mu{}_\nu
&\equiv\delta^\mu{}_\nu-\frac{p_0^\mu p_{0\nu}}{p_0^2},
\notag\\
S_0^{\mu\nu}
&\equiv
\Pi_0{}^\mu{}_\alpha\Pi_0{}^\nu{}_\beta
\langle S^{\alpha\beta}\rangle_{\mathcal C},
\qquad
S_0^{\mu\nu}p_{0\nu}=0,
\label{eq:packet-spin-projection}
\end{align}
which alters the raw expectation value only by spin-momentum correlations
of packet-width order.  Replacing the spin operator and the mass Casimir
invariant by these packet data defines
\begin{equation}
\overline\Theta_{\mathcal C}^{\mu\nu}
\equiv
-\frac{\hbar}{m^2c^2}
\;S_0^{\mu\nu},
\qquad
\overline\Theta_{\mathcal C}^{\mu\nu}p_{0\nu}=0.
\label{eq:Theta-frozen}
\end{equation}
This c-number tensor is a local representative of the operator-valued
geometry and not a new universal background.  Using it semiclassically
requires at least
\begin{equation}
\epsilon_p^2\equiv
\frac{\langle(\bm p-\langle\bm p\rangle)^2\rangle_{\mathcal C}}
{m^2c^2}\ll1,
\qquad
\epsilon_S^2\equiv
\frac{\langle(\bm S-\langle\bm S\rangle)^2\rangle_{\mathcal C}}
{\hbar^2s(s+1)}\ll1,
\label{eq:frozen-control}
\end{equation}
together with an adiabatic or otherwise specified law for spin transport.
These conditions control the kinematical error of freezing without showing
that interactions preserve the packet.  For a pure spin-$1/2$ state,
$\langle\bm S^2\rangle=3\hbar^2/4$ and
$|\langle\bm S\rangle|^2=\hbar^2/4$, so
\begin{equation}
\epsilon_S^2=\frac{2}{3}.
\label{eq:spin-half-freezing-error}
\end{equation}
The spin-freezing error is therefore of order unity even for a maximally
polarized electron; a mixed state does not improve it.  The frozen fiber is
a state-conditioned benchmark, or a truncation of selected matrix elements,
and not a parametrically controlled classical-spin limit.  The momentum
width $\epsilon_p$ can still be small, but it does not compensate for the
absence of spin semiclassicality.

For one fixed real tensor \eqref{eq:Theta-frozen}, define
\begin{equation}
\mathcal F_{\mathcal C}
=\exp\left[
-\frac{i}{2\hbar^2}
\overline\Theta_{\mathcal C}^{\mu\nu}
p_\mu\otimes p_\nu
\right].
\label{eq:frozen-twist}
\end{equation}
Because the translations commute, $\mathcal F_{\mathcal C}$ obeys the
two-cocycle condition
\begin{equation}
(\mathcal F_{\mathcal C}\otimes1)
(\Delta\otimes{\rm id})\mathcal F_{\mathcal C}
=(1\otimes\mathcal F_{\mathcal C})
({\rm id}\otimes\Delta)\mathcal F_{\mathcal C},
\label{eq:twist-cocycle}
\end{equation}
so the deformation is coassociative within the fiber
\cite{Drinfeld:1985rx,Aschieri:2005yw}.  The corresponding product, with
$p_\mu=-i\hbar\partial_\mu$, is
\begin{align}
(f\star_{\mathcal C}g)(x)
&\equiv m\circ\mathcal F_{\mathcal C}(f\otimes g)
\notag\\
&=f(x)\exp\left[
\frac{i}{2}\overline\Theta_{\mathcal C}^{\mu\nu}
\overleftarrow{\partial_\mu}
\overrightarrow{\partial_\nu}
\right]g(x)
\label{eq:frozen-star}\\
&=fg+\frac{i}{2}\overline\Theta_{\mathcal C}^{\mu\nu}
\partial_\mu f\,\partial_\nu g
-\frac18
\overline\Theta_{\mathcal C}^{\mu\nu}
\overline\Theta_{\mathcal C}^{\rho\sigma}
\partial_\mu\partial_\rho f\,
\partial_\nu\partial_\sigma g+\cdots .
\label{eq:frozen-star-expansion}
\end{align}
It is associative, it is Hermitian in the sense
$(f\star_{\mathcal C}g)^\dagger=g^\dagger\star_{\mathcal C}f^\dagger$ for
real antisymmetric $\overline\Theta_{\mathcal C}$, and it yields
$[x^\mu,x^\nu]_{\star_{\mathcal C}}
=i\overline\Theta_{\mathcal C}^{\mu\nu}$.  For fields with adequate
boundary conditions,
\begin{equation}
\int d^4x\,f\star_{\mathcal C}g
=\int d^4x\,fg,
\label{eq:trace-property}
\end{equation}
so free quadratic scalar and Dirac actions coincide with their ordinary
forms, while interactions and products of representations need not.

The scope of the fiber is limited by a structural obstruction that deserves
to be stated once and remembered.  A nonzero constant tensor cannot obey
$\overline\Theta^{\mu\nu}p_\nu=0$ for every $p$ in an open momentum set,
because a linear map vanishing on an open set vanishes identically.  What
this obstructs is a universal transverse Moyal tensor and not the local
fiber \eqref{eq:Theta-frozen}, and it is the reason the constant theories
are treated throughout as charts instead of as one global deformation.

\subsection{Coproduct, modules, and Clebsch-Gordan maps}
\label{sec:fibered-representations}

Because our product convention is
$m_{\mathcal C}=m\circ\mathcal F_{\mathcal C}$ instead of
$m\circ\mathcal F_{\mathcal C}^{-1}$, covariance fixes the opposite
conjugation of the coproduct, which within a fixed fiber reads
\begin{equation}
\Delta_{\mathcal C}(h)
=\mathcal F_{\mathcal C}^{-1}\Delta(h)\mathcal F_{\mathcal C}.
\label{eq:deformed-coproduct-definition}
\end{equation}
Translations remain primitive,
\begin{equation}
\Delta_{\mathcal C}(p^\mu)
=p^\mu\otimes1+1\otimes p^\mu,
\label{eq:coproduct-p}
\end{equation}
so momentum addition stays undeformed, whereas the Lorentz coproduct
acquires crossed terms.  The Baker-Campbell-Hausdorff expansion
terminates after the first commutator, since $[p,[p,J]]=0$, and gives
\begin{align}
\Delta_{\mathcal C}(J^{\mu\nu})
={}&J^{\mu\nu}\otimes1+1\otimes J^{\mu\nu}
\notag\\
&+\frac{1}{2\hbar}
\bigg[
p^\nu\otimes(\overline\Theta_{\mathcal C}p)^\mu
-p^\mu\otimes(\overline\Theta_{\mathcal C}p)^\nu
\notag\\
&\hspace{2.9cm}
-(\overline\Theta_{\mathcal C}p)^\mu\otimes p^\nu
+(\overline\Theta_{\mathcal C}p)^\nu\otimes p^\mu
\bigg],
\label{eq:coproduct-J}
\end{align}
where
$(\overline\Theta_{\mathcal C}p)^\mu
=\overline\Theta_{\mathcal C}^{\mu\alpha}p_\alpha$.  The one-particle
Poincar\'e algebra and its Casimir invariants are unchanged, and the whole
deformation resides in the rule for composing modules.

The following tensor-product formulas apply only when both one-particle
factors belong to the same selected neighborhood $\mathcal C$.  Generic
incoming and outgoing scattering states need not share such a fiber, and a
state-dependent tensor from one leg cannot be inserted into a universal
coproduct for all other legs.  Within the common fiber, momentum eigenstates
obey
\begin{equation}
\mathcal F_{\mathcal C}
|p_1,\sigma_1\rangle\otimes|p_2,\sigma_2\rangle
=e^{-i\varphi_{12}^{\mathcal C}}
|p_1,\sigma_1\rangle\otimes|p_2,\sigma_2\rangle,
\label{eq:twist-two-particle}
\end{equation}
with
\begin{equation}
\varphi_{12}^{\mathcal C}
\equiv
\frac{1}{2\hbar^2}
\overline\Theta_{\mathcal C}^{\mu\nu}
p_{1\mu}p_{2\nu}.
\label{eq:fiber-phase}
\end{equation}
In terms of the electric and magnetic vectors of
\eqref{eq:e-b-definitions}, the numerator of this phase becomes
\begin{equation}
\overline\Theta_{\mathcal C}^{\mu\nu}p_{1\mu}p_{2\nu}
=\frac1c\bm e\cdot
(E_2\bm p_1-E_1\bm p_2)
+\bm b\cdot(\bm p_1\times\bm p_2),
\label{eq:fiber-phase-eb}
\end{equation}
where the displayed sign follows from $\Theta^{ij}=\epsilon^{ijk}b_k$
together with the metric convention \eqref{eq:conventions}, and the phase
is antisymmetric under exchange of the two factors.

Writing $p_i=p_0+\delta p_i$ makes the local character of the effect
transparent, since transversality to $p_0$ implies
\begin{equation}
\varphi_{12}^{\mathcal C}
=\frac{1}{2\hbar^2}
\overline\Theta_{\mathcal C}^{\mu\nu}
\delta p_{1\mu}\delta p_{2\nu},
\label{eq:packet-width-phase}
\end{equation}
so that the phase starts quadratically in the momentum width, reverses when
the polarization is reversed, and vanishes in the monochromatic limit.  One
should not infer from this that every multiparticle phase vanishes on
shell, because the fixed tensor is transverse to the reference momentum
alone and not to all momenta entering the tensor product.  Equation
\eqref{eq:packet-width-phase} also bounds interference-sensitive
observables built on two-particle states.  Bell-type correlations between
polarized packets, of the kind analyzed in relativistic quantum information
\cite{Terashima:2003,Alsing:2012}, would be dressed by relative phases
governed by \eqref{eq:packet-width-phase}, an effect that is
state-conditioned, quadratic in the packet width and odd under polarization
reversal, so that monochromatic estimates obtained from a
universal-background tensor do not carry over to the present construction.

Let $C:V_1\otimes V_2\rightarrow V$ intertwine the undeformed coproduct.
Then
\begin{equation}
C_{\mathcal C}
\equiv
C\circ(\pi_1\otimes\pi_2)
(\mathcal F_{\mathcal C})
\label{eq:CG-twisted}
\end{equation}
intertwines \eqref{eq:deformed-coproduct-definition}.  The irreducible
channels, the selection rules and the Wigner multiplicities all survive
untouched, and what the twist dresses is the two-particle momentum basis
and with it the relative interference phases.  In the same way
\begin{equation}
\mathcal R_{\mathcal C}
=\mathcal F_{\mathcal C,21}^{-1}
\mathcal F_{\mathcal C}
=\mathcal F_{\mathcal C}^{2}
\label{eq:R-matrix}
\end{equation}
is triangular and the braided exchange is involutive.  These are exact
statements about a single c-number fiber, and they do not yet define one
global Hopf algebra whose tensor varies from state to state.  Planar
twist-deformed gauge theories offer an instructive external comparison,
since there the entire deformation can be pushed onto external legs while
amputated correlators remain undeformed \cite{Borsato:2026scaleTwist}.  The
fibered construction shares that external-data character, although its
tensor is carried by the state instead of by a spacetime symmetry twist, so
the two mechanisms agree only at the level of bookkeeping.

\subsection{Transport between fibers}
\label{sec:transport-fibers}

Define the local module algebra
$\mathfrak A_{\mathcal C}
\equiv(C^\infty(\mathbb R^{1,3})[[\overline\Theta_{\mathcal C}]],
\star_{\mathcal C})$.  The collection
\begin{equation}
\{\mathfrak A_{\mathcal C},
\star_{\mathcal C},
\mathcal F_{\mathcal C}\}_{\mathcal C}
\label{eq:fiber-family}
\end{equation}
is a family of local associative descriptions spread over the timelike
domain of spin and momentum.  When the packet center moves from $p_0$ to
$p_0'$, the spin frame is related by the transporter
\eqref{eq:Wilson-line}.  Since $\mathcal F_{\mathcal C}$ contains
translation generators together with a c-number tensor, conjugating it with
$U_\gamma\otimes U_\gamma$ would be either trivial or ill typed, so
transport must act on the polarization data first.  Taking
$|\psi_{\mathcal C'}\rangle
=U_\gamma(p_0',p_0)|\psi_{\mathcal C}\rangle$, we set
\begin{align}
S_{\mathcal C'}^{\mu\nu}
&\equiv
\Pi_0'{}^\mu{}_\alpha\Pi_0'{}^\nu{}_\beta
\langle\psi_{\mathcal C}|
U_\gamma^{-1}\hat S^{\alpha\beta}(p_0')U_\gamma
|\psi_{\mathcal C}\rangle,
\notag\\
\overline\Theta_{\mathcal C'}^{\mu\nu}
&\equiv-\frac{\hbar}{m^2c^2}S_{\mathcal C'}^{\mu\nu},
\qquad
\mathcal F_{\mathcal C'}
\equiv
\exp\left[
-\frac{i}{2\hbar^2}
\overline\Theta_{\mathcal C'}^{\mu\nu}
p_\mu\otimes p_\nu
\right].
\label{eq:fiber-transition}
\end{align}
The last step reconstructs the twist in the new fiber and makes explicit
where packet-width and path-dependent corrections enter.  What
\eqref{eq:fiber-transition} provides is proposed transition data, not a
proof that a global dynamical twist exists.

Organizing the theory into fibers with transition data points toward two
mathematical completions.  One would seek a deformation quantization of the
full momentum-dependent Poisson structure, the other a dynamical twist and
Hopf-algebroid formulation in which the base variables belong to the
algebra \cite{Xu:2001}.  Either route must reproduce the local Moyal
fibers, their transition transport, and associativity to the order claimed.
Failure of the cocycle condition, or of the corresponding higher-order
associativity conditions, would confine the construction to the local
semiclassical level.

\subsection{Bopp chart, interval, and operational simultaneity}
\label{sec:bopp-chart}

The same c-number fiber admits the Bopp realization \cite{bopp1956}
\begin{equation}
X_{\rm B}^{\mu}
=x^\mu-\frac{1}{2\hbar}
\overline\Theta_{\mathcal C}^{\mu\nu}p_\nu,
\qquad
[X_{\rm B}^{\mu},X_{\rm B}^{\nu}]
=i\overline\Theta_{\mathcal C}^{\mu\nu}.
\label{eq:Bopp-shift}
\end{equation}
This is a coordinate chart for the auxiliary Moyal algebra and not the
operator realization \eqref{eq:X-momentum} of the exact centroid.  In the
momentum representation it is the abelian member of the family met earlier,
namely $X_{\rm B}^\mu=i\hbar\partial^\mu+\mathcal A_{\rm B}^\mu$ with
$\mathcal A_{\rm B}^\mu
=-\overline\Theta_{\mathcal C}^{\mu\nu}p_\nu/2\hbar$, a connection whose
commutator term vanishes identically and whose curvature is the constant
$\overline\Theta_{\mathcal C}^{\mu\nu}/\hbar$.  Exact centroid and frozen
chart differ in this language not in kind but in the connection they carry,
which is why comparing them amounts to a statement about curvature and not
about a change of variables.  At the reference momentum, in particular,
\begin{equation}
\overline\Theta_{\mathcal C}^{\mu\nu}p_{0\nu}=0
\quad\Longrightarrow\quad
X_{\rm B}^{\mu}(p_0)=x^\mu,
\label{eq:Bopp-reference}
\end{equation}
even though derivatives in momentum space still detect the nonvanishing
coordinate commutator.

For constant $\overline\Theta_{\mathcal C}$ the quadratic form is
\begin{align}
X_{\rm B}^2
={}&x^2
-\frac{1}{2\hbar}
\overline\Theta_{\mathcal C}^{\mu\nu}L_{\mu\nu}+\frac{1}{4\hbar^2}
\eta_{\mu\nu}
\overline\Theta_{\mathcal C}^{\mu\alpha}
\overline\Theta_{\mathcal C}^{\nu\beta}
p_\alpha p_\beta.
\label{eq:Bopp-interval}
\end{align}
Substituting \eqref{eq:Theta-frozen} turns the linear piece into a
spin-orbit-like scalar, the frozen counterpart of the exact relation
\eqref{eq:centroid-interval}.  It remains a property of this coordinate
chart, and it becomes an additional Hamiltonian term only when a dynamical
model makes that identification.

The time component gives a precise and limited sense in which simultaneity
is state dependent.  For two momentum components described in the same
fiber,
\begin{equation}
\Delta t_{\rm B}
=\Delta t
-\frac{1}{2\hbar c}
\overline\Theta_{\mathcal C}^{0i}\Delta p_i,
\label{eq:Bopp-simultaneity}
\end{equation}
an operational shift between coordinate representatives that vanishes in
the comoving frame and can be nonzero after a boost.  By itself it neither
modifies the light cone nor establishes signal propagation between
spacelike events, questions that require field commutators and a
gauge-complete propagator.

Table~\ref{tab:three-realizations} collects the division of roles among the
three coordinate descriptions used so far.

\begin{table}[t]
\centering \footnotesize
\caption{Three related but inequivalent realizations.}
\label{tab:three-realizations}
\begin{tabular}{@{}L{0.18\linewidth}L{0.24\linewidth}L{0.22\linewidth}L{0.24\linewidth}@{}}
\toprule
 & Exact centroid & Frozen twist & Bopp chart \\
\midrule
Basic object &
$X=i\hbar\partial_p+\mathcal A(p)$ &
$\mathcal F_{\mathcal C}$ and $\star_{\mathcal C}$ &
$X_{\rm B}=x-\overline\Theta_{\mathcal C}p/(2\hbar)$ \\
Tensor &
operator-valued curvature &
one constant packet value &
the same constant packet value \\
Associativity &
ambient operator algebra &
exact in one fiber &
exact in one fiber \\
Main role &
particle geometry and transport &
composition of fields and modules &
constructive coordinate representation \\
Limitation &
$X$ mixes mass shells &
not universal over momentum space &
not the exact centroid connection \\
\bottomrule
\end{tabular}
\end{table}

The spinor representation calls for one caution before the field-theoretic
discussion begins.  Since the frozen deformation lives in products and
coproducts, it must not be imported into the Dirac matrices by dressing
$S(\Lambda)$ with an extra exponential of the same particle spin tensor.
The supplementary condition acts on states and polarization data, whereas
the Dirac matrices furnish a finite-dimensional Lorentz representation
whose boost components $\sigma^{0i}$ do not vanish and therefore cannot
satisfy \eqref{eq:TD-identity} at the matrix level.\footnote{A candidate
dressing $S_\Theta(\Lambda)=S(\Lambda)e^{\Omega}$ with
$\Omega\propto\overline\Theta_{\alpha\beta}\sigma^{\alpha\beta}$ collapses
to a global phase once the tensor is aligned with the spinor's own spin
matrices.  With $S^{\alpha\beta}=(\hbar/2)\sigma^{\alpha\beta}$ the exponent
becomes proportional to $S_{\alpha\beta}S^{\alpha\beta}$, which equals
$3\hbar^2$ times the identity on the irreducible spinor space, so
$S_\Theta(\Lambda)=e^{3i/2}S(\Lambda)$ and no observable one-particle
effect survives.  The matrix value $3\hbar^2$ differs from the state value
$2\hbar^2s(s+1)=\tfrac32\hbar^2$ precisely because $\sigma^{0i}\neq0$,
which is the obstruction just described.}  We therefore retain the standard
Dirac representation throughout.
\section{Field-theory matching and momentum-space transport}
\label{sec:field-matching}

The local product \eqref{eq:frozen-star} can be used inside a conventional
constant-$\Theta$ field theory, and setting $\hbar=c=1$ in equations
\eqref{eq:frozen-NCQED-action}-\eqref{eq:frozen-field-strength} only, a
star-gauge action reads
\begin{equation}
S_{\mathcal C}
=\int d^4x\left[
-\frac14F_{\mu\nu}\star_{\mathcal C}F^{\mu\nu}
+\bar\psi\star_{\mathcal C}
(i\gamma^\mu D_\mu-m)\psi
\right],
\label{eq:frozen-NCQED-action}
\end{equation}
where
\begin{align}
D_\mu\psi
&=\partial_\mu\psi-ieA_\mu\star_{\mathcal C}\psi,
\label{eq:frozen-covariant-derivative}\\
F_{\mu\nu}
&=\partial_\mu A_\nu-\partial_\nu A_\mu
-ie[A_\mu,A_\nu]_{\star_{\mathcal C}}.
\label{eq:frozen-field-strength}
\end{align}
The action generates the vertices and contact structures demanded by its
star-gauge symmetry, and it serves as a calculational benchmark for fields
conditioned on one fixed external tensor
$\overline\Theta_{\mathcal C}$.  It does not define scattering between
states assigned to different fibers, nor may the tensor be recomputed from
the very quantum field integrated over in the action.  It is not yet a
field theory for the operator
$\widehat\Theta^{\mu\nu}=\hbar\mathcal F_p^{\mu\nu}$, since
\eqref{eq:frozen-NCQED-action} assigns no dynamical spin tensor to internal
lines.  The gap has a structural origin, since consistent gauge
theory on twist-noncommutative spaces requires a deformed
differential calculus together with star-gauge covariance instead of
isolated phase insertions \cite{Meier:2023twistGauge}, and for nonconstant
or operator-dependent tensors the Poisson gauge and $L_\infty$-bootstrap
constructions show that product, gauge transformations and field strength
must be deformed in a correlated way
\cite{Kupriyanov:2020novelGauge,Blumenhagen:2018Linfinity,
Kurkov:2022U1Linfinity}.  Every field-theory statement below is therefore
labeled by the approximation from which it follows.

Because the bivector belongs to the magnetic Lorentz class, the comoving
fiber never requires a timelike constant background, which for frozen Moyal
theories removes the standard perturbative unitarity obstruction associated
with time-space noncommutativity \cite{Seiberg:2000,bahns2005unitarity}.
Favorable as that is, its scope stays limited, since unitarity and
microcausality of an operator-valued completion have to be checked for the
actual propagators and interaction kernels.

\subsection{Endpoint-transversality lemma}
\label{sec:endpoint-lemma}

Consider a one-particle transition
$(p,\sigma)\rightarrow(p',\sigma')$ and assign the semiclassical endpoint
tensors
\begin{align}
\Theta_{p,\sigma}^{\mu\nu}
&=-\frac{\hbar}{m^2c^2}S_{p,\sigma}^{\mu\nu},
&
\Theta_{p,\sigma}^{\mu\nu}p_\nu&=0,
\label{eq:Theta-in-endpoint}\\
\Theta_{p',\sigma'}^{\mu\nu}
&=-\frac{\hbar}{m^2c^2}S_{p',\sigma'}^{\mu\nu},
&
\Theta_{p',\sigma'}^{\mu\nu}p'_\nu&=0.
\label{eq:Theta-out-endpoint}
\end{align}
Each tensor is constructed from the momentum and spin data of its own
physical state, and neither is a value of one universal background.  A
single frozen tensor could not make both fermion legs transverse, since the
kernel of $\Theta_{p,\sigma}$ is the two-plane spanned by $p$ and the spin
axis, which contains $p'$ only for special kinematics, and the endpoint
assignment removes that obstruction.  For any numerical coefficients $a$
and $b$, define
\begin{equation}
\Theta_V^{\mu\nu}
=a\Theta_{p',\sigma'}^{\mu\nu}
+b\Theta_{p,\sigma}^{\mu\nu}.
\label{eq:Theta-V-general}
\end{equation}
Antisymmetry and endpoint transversality then imply
\begin{equation}
p'_\mu\Theta_V^{\mu\nu}p_\nu=0,
\label{eq:endpoint-identity}
\end{equation}
because $p'_\mu\Theta_{p',\sigma'}^{\mu\nu}=0$ follows from outgoing
transversality together with antisymmetry, while
$\Theta_{p,\sigma}^{\mu\nu}p_\nu=0$ is incoming transversality.  A
postulated scalar phase dressing,
\begin{equation}
\Gamma_{\rm ph}^{\alpha}(p',p)
=-ie\gamma^\alpha
\exp\left[
\frac{i}{2\hbar^2}
p'_\mu\Theta_V^{\mu\nu}p_\nu
\right],
\label{eq:endpoint-phase-vertex}
\end{equation}
therefore reduces to the ordinary vertex, and the same conclusion survives
the addition of any $\Theta_\perp^{\mu\nu}$ obeying
$p'_\mu\Theta_\perp^{\mu\nu}p_\nu=0$.  This is an exact kinematical lemma
about one specified bilinear phase, and its scope must be stated narrowly,
since it removes the endpoint-adapted external Moyal phase as a candidate
signal and nothing further.

The lemma leaves the most general vertex undetermined.  With $q=p'-p$,
\begin{equation}
\Gamma^\alpha(p',p)
=\gamma^\alpha F_1(q^2)
+\frac{i\sigma^{\alpha\beta}q_\beta}{2mc}F_2(q^2)
+\delta\Gamma_\perp^\alpha,
\qquad
q_\alpha\delta\Gamma_\perp^\alpha=0,
\label{eq:general-vertex}
\end{equation}
and structures such as $G_\Theta(q^2)\Theta_V^{\alpha\beta}q_\beta$ are in
no way excluded by \eqref{eq:endpoint-identity}.  The endpoint average
$a=b=1/2$ is a symmetric midpoint prescription, obtained here from no
action and by no means unique.  Null results of this structural kind have
precedents in other covariant quantizations, since for a class of
T-Minkowski noncommutative spacetimes the Wightman functions coincide with
their commutative counterparts \cite{Fabiano:2026TMinkowski}, although the
mechanism there involves the spacetime symmetry twist and not the
spin-dependent endpoint transversality at work in
\eqref{eq:endpoint-identity}.

\subsection{Straight-line transport, Ward defect, and transverse matching}
\label{sec:Wilson-matching}

For a straight path $p(t)=p+tq$ from $p$ to $p'=p+q$, antisymmetry removes
the path parameter from the numerator of the line integral and gives the
closed form
\begin{equation}
\int_0^1\mathcal A_\mu(p+tq)q^\mu dt
=\Sigma^{\mu\nu}q_\mu p_\nu J(p,q),
\qquad
J(p,q)\equiv\int_0^1\frac{dt}{(p+tq)^2}.
\label{eq:line-integral-closed}
\end{equation}
The scalar $J$ is regular whenever the complete segment remains inside the
timelike domain.  In the Dirac representation,
$\sigma^{\mu\nu}q_\mu p_\nu=\tfrac{i}{2}[\slashed q,\slashed p]$.
Setting $c=\hbar=1$ for the next identity, an on-shell external pair obeys
\begin{equation}
(\slashed p'+m)[\slashed q,\slashed p]\slashed q(\slashed p+m)
=-2mq^2(\slashed p'+m)(\slashed p+m).
\label{eq:ward-defect-identity}
\end{equation}
Consequently, the Ward defect of a vertex multiplied by the linearized
transporter is regular and proportional to $q^2J(p,q)\bar u(p')u(p)$.  Its
longitudinal part can be repaired by a local term proportional to
$q^\alpha$.  At the two-photon level both identities
\begin{equation}
k_\mu\mathcal M_{\rm comp}^{\mu\nu}=0,
\qquad
k'_\nu\mathcal M_{\rm comp}^{\mu\nu}=0
\label{eq:two-photon-Ward}
\end{equation}
must hold.  A useful decomposition is
\begin{equation}
\mathcal M_{\rm comp}^{\mu\nu}
=\mathcal M_s^{\mu\nu}+\mathcal M_u^{\mu\nu}
+\mathcal M_{{\rm ct},L}^{\mu\nu}
+\mathcal T_\perp^{\mu\nu},
\qquad
k_\mu\mathcal T_\perp^{\mu\nu}
=k'_\nu\mathcal T_\perp^{\mu\nu}=0.
\label{eq:gauge-complete-decomposition}
\end{equation}
The Ward identities fix the required longitudinal contact structure but do
not determine the doubly transverse amplitude $\mathcal T_\perp^{\mu\nu}$.
Appendix~\ref{app:straight-line} records the algebra behind
\eqref{eq:line-integral-closed} and \eqref{eq:ward-defect-identity}.

The scalar phase \eqref{eq:endpoint-phase-vertex} retains only an abelian
number, whereas the exact centroid geometry supplies the matrix-valued
transporter \eqref{eq:Wilson-line}.  An earlier form of this work therefore
introduced the following matching hypothesis, which we retain as the
definition of a possible dynamical excess rather than as an identity of QED.

\begin{quote}
An interaction that changes the momentum of a spinning particle may carry,
in addition to the standard change of spin frame and its local field-theory
kernel, an independently matched transport along the momentum routing of
that kernel.
\end{quote}

To keep the fiber indices explicit, let
$\widetilde{\mathcal K}_0(p',p):V_p\to V_p$ denote the ordinary local
kernel written in the initial spin frame.  Its transported counterpart is
the map $\mathcal K_\gamma(p',p):V_p\to V_{p'}$ defined by
\begin{equation}
\mathcal K_\gamma(p',p)
=U_\gamma(p',p)\circ\widetilde{\mathcal K}_0(p',p)
+\mathcal K_{\rm ct,\gamma}(p',p),
\label{eq:kernel-matching}
\end{equation}
where $\mathcal K_{\rm ct,\gamma}:V_p\to V_{p'}$ collects the local and
contact operators required by spacetime gauge symmetry.  In fiber indices
this reads
$(\mathcal K_\gamma)^{a'}{}_{b}
=(U_\gamma)^{a'}{}_{a}
(\widetilde{\mathcal K}_0)^{a}{}_{b}
+(\mathcal K_{\rm ct,\gamma})^{a'}{}_{b}$, and
\eqref{eq:kernel-matching} remains an ansatz for additional dynamical
matching rather than a consequence of the coordinate commutator.  Standard
Dirac and QED amplitudes already contain changes of spin basis and Wigner
rotations.  Multiplying them again by the same geometric transport would
double count that kinematics unless a gauge-complete action produces an
independent coefficient.

The advantage of the hypothesis over an isolated phase is structural.  An
open Wilson line changes with the spin frames at $p$ and $p'$, so a single
open factor need not be observable, whereas two paths $\gamma_1$ and
$\gamma_2$ sharing endpoints define
\begin{equation}
\mathcal H_{12}
\equiv U_{\gamma_1}(p',p)U_{\gamma_2}^{-1}(p',p),
\label{eq:relative-holonomy}
\end{equation}
which transforms by conjugation.  For a small loop
$\gamma_1\circ\gamma_2^{-1}$ and an oriented surface $\Sigma_{12}$ with
area bivector
\begin{equation}
\mathcal S_{12}^{\mu\nu}
\equiv\int_{\Sigma_{12}}dq^\mu\wedge dq^\nu,
\label{eq:momentum-area}
\end{equation}
the nonabelian Stokes expansion gives
\begin{equation}
\mathcal H_{12}
=\bm1+\frac{i}{2\hbar}
\mathcal F_{p\,\mu\nu}(p_\ast)
\mathcal S_{12}^{\mu\nu}
+O(\mathcal S^{3/2}),
\label{eq:small-loop-holonomy}
\end{equation}
with $p_\ast$ inside the loop.  The first path-dependent invariant is
therefore controlled by the same curvature that appears in
$[X^\mu,X^\nu]$.

A consistent electromagnetic completion must still satisfy the
Ward-Takahashi identity
\begin{equation}
q_\alpha\Gamma_{\rm off}^{\alpha}(r+q,r)
=G_{\rm off}^{-1}(r+q)-G_{\rm off}^{-1}(r),
\label{eq:Ward-Takahashi}
\end{equation}
which neither $U_\gamma$ alone nor an arbitrary contact term guarantees.
Slavnov-Taylor analyses in noncommutative gauge theories make the same
point at one loop and beyond, tying two-, three- and four-point functions
together so tightly that isolated channel phases cannot be consistent
\cite{Blaschke:2013SlavnovTaylor}.  The test that settles the matching
hypothesis is accordingly to obtain \eqref{eq:kernel-matching} from an
action or worldline functional and to verify all Ward identities.

\section{Semiclassical dynamics and observable discriminants}
\label{sec:phenomenology}

Let a semiclassical packet be described by a Hamiltonian $H(X,p,S)$, with
spin following a specified adiabatic or BMT-type transport law.  The
centroid brackets give
\begin{align}
\dot X^\mu
&=\{X^\mu,H\}
\notag\\
&=\frac{\partial H}{\partial p_\mu}
-\frac{S^{\mu\nu}}{p^2}
\frac{\partial H}{\partial X^\nu}
+\text{spin-transport terms}
\notag\\
&=\frac{\partial H}{\partial p_\mu}
+\frac{S^{\mu\nu}}{p^2}\dot p_\nu
+\text{spin-transport terms},
\label{eq:anomalous-velocity}
\end{align}
and the curvature contribution integrates to
\begin{equation}
\Delta X_{\rm anom}^{\mu}
=\int_\gamma\frac{S^{\mu\nu}(p)}{p^2}\,dp_\nu.
\label{eq:side-jump-integral}
\end{equation}
This is the massive covariant counterpart of the anomalous velocities and
spin-dependent side-jumps familiar from Berry-curvature descriptions of
chiral and Dirac particles
\cite{StoneDwivediZhou:2015,ChenSonStephanov:2014,DuvalHorvathy:2003}.

The same scale follows from recentering after a small impulse, with no
reference to a Hamiltonian at all.  Suppose initially
$\Sigma^{\mu\nu}p_\nu=0$, and during the impulse take $\delta z^\mu=0$,
$\delta\Sigma^{\mu\nu}=0$ and $p\cdot\delta p=0$ to leading order.  Varying
\eqref{eq:X-definition} then gives
\begin{equation}
\delta X^\mu
=\frac{S^{\mu\nu}\delta p_\nu}{m^2c^2}
+O(\delta p^2),
\label{eq:recentring-jump}
\end{equation}
After the laboratory conversion
\eqref{eq:laboratory-dictionary}, the same fixed-$(z,\Sigma)$ variation at
the rest point becomes
\begin{equation}
\delta\bm q_{\rm map}
=-\frac{\bm s\times\delta\bm p}{m^2c^2},
\qquad
|\delta\bm q_{\rm map}|
\sim\lambda_{\rm C}
\frac{|\delta\bm p|}{mc}
\frac{|\langle\bm s\rangle|}{\hbar}.
\label{eq:side-jump-rest}
\end{equation}
This displacement is transverse, reverses with the polarization, and is of
Compton scale.  It is a variation of the composite map at fixed reference
variables, not yet the Hamiltonian trajectory of the projected TD centroid.
At rest its numerical value coincides with the total charge-centroid jump
obtained below.  The distinction resolves the apparent conflict between the
covariant recentering formula and the inertial motion of the TD mass
centroid.

\subsection{Minimal coupling, spin-orbit interaction, and BMT precession}
\label{sec:BMT-side-jump}

Let $q_\ell$ denote the signed lepton charge.  Minimal coupling is coupling
at the charge position, so reducing
$q_\ell\phi(\bm q_D)$ to the positive-energy block and expanding to first
order in a slowly varying electric field gives
\begin{equation}
H_{++}=W(\bm p)+q_\ell\phi(\bm x_{\rm NW})+h_{\rm SO},
\qquad
h_{\rm SO}
=\frac{q_\ell c^2}{W(W+M)}
\bm E\cdot(\bm s\times\bm p).
\label{eq:spin-orbit-Hamiltonian}
\end{equation}
Equivalently,
$h_{\rm SO}=-q_\ell c^2\bm s\cdot(\bm E\times\bm p)/[W(W+M)]$.
The nonrelativistic limit is
\begin{equation}
h_{\rm SO}\longrightarrow
-\frac{q_\ell}{2m^2c^2}
\bm s\cdot(\bm E\times\bm p),
\label{eq:Thomas-spin-orbit}
\end{equation}
which contains the Thomas factor $1/2$.  With
$\{s_i,s_j\}=\epsilon_{ijk}s_k$, the corresponding precession is
\begin{equation}
\dot{\bm s}=\bm\Omega\times\bm s,
\qquad
\bm\Omega
=\frac{q_\ell c^2}{W(W+M)}\bm p\times\bm E
=\frac{q_\ell}{(\gamma+1)mc^2}\bm v\times\bm E.
\label{eq:BMT-electric}
\end{equation}
This is the BMT result for $g=2$ in a purely electric field.  It is exact in
$\gamma$ within first order in the field and in $\hbar$, after the free FW
transformation and the slowly varying-field expansion.  Magnetic fields,
an anomalous moment, Darwin and Stern-Gerlach terms, and higher field
gradients lie outside this statement \cite{Bargmann:1959gz}.

\subsection{Centroid-resolved impulse budget}
\label{sec:impulse-budget}

Apply a short transverse impulse $\delta\bm p$ to a particle initially at
rest and measure displacements in units of
\begin{equation}
\bm u\equiv\frac{\bm s\times\delta\bm p}{m^2c^2}.
\label{eq:side-jump-unit}
\end{equation}
The momentum dependence of \eqref{eq:spin-orbit-Hamiltonian} supplies the
dynamical Newton-Wigner displacement $-\bm u/2$.  The changes of the
kinematic shifts are
\begin{equation}
\delta\mathcal A_B=-\frac12\bm u,
\qquad
\delta\bm\Delta_{\rm TD}=+\frac12\bm u.
\label{eq:kinematic-impulse-shifts}
\end{equation}
Adding the appropriate shift to the same dynamical term gives
table~\ref{tab:side-jump-budget}.

\begin{table}[t]
\centering
\caption{Transverse displacement under an impulse from rest, in units of
$\bm u$ from \eqref{eq:side-jump-unit}, to first order in the field and in
$\hbar$.  ``Recentering'' denotes the change of the position-dependent
shift relative to Newton-Wigner.}
\label{tab:side-jump-budget}
\begin{tabular}{@{}lccc@{}}
\toprule
position & dynamical part & recentering & total side jump \\
\midrule
Newton-Wigner & $-\tfrac12$ & $0$ & $-\tfrac12$ \\
charge/Dirac & $-\tfrac12$ & $-\tfrac12$ & $-1$ \\
TD projected $=$ TD irrep & $-\tfrac12$ & $+\tfrac12$ & $0$ \\
\bottomrule
\end{tabular}
\end{table}

An electromagnetic detector couples at the charge position and therefore
reconstructs the second line unless its analysis explicitly converts to a
different centroid.  The TD mass centroid has no total transverse jump in
this rest-impulse protocol: its dynamical and recentering halves cancel.
The projected TD centroid is not a fourth line, because
\eqref{eq:projected-TD-position} already identifies it with the TD irrep up
to the radial, spin-independent remnant.  Agreement with this table tests
the localization dictionary of the standard Dirac theory; new dynamics
would be a calibrated deviation from these values.

\subsection{Compton scattering: geometry, QED trace, and conditional response}
\label{sec:Compton}

Compton scattering supplies two fermion routings with common external
endpoints, but three logically different comparisons must be kept apart.
The centroid connection defines a geometric holonomy for prescribed paths;
canonical boosts define a Thomas-Wigner comparator on the unit velocity
hyperboloid; and the QED amplitude contains off-shell propagators and local
photon vertices in addition to changes of spin basis.  A fourth object, the
polarized channel trace, can be evaluated exactly without introducing new
dynamics.  A rate modulation appears only in a fifth step, after an extra
relative phase or a gauge-complete transverse operator has been supplied.

\subsubsection{Timelike routings and their off-shell content}
\label{sec:Compton-routing}

For
\begin{equation}
e^-(p)+\gamma(k)\longrightarrow e^-(p')+\gamma(k'),
\qquad
p+k=p'+k',
\label{eq:Compton-kinematics}
\end{equation}
the two fermion routings are
\begin{align}
\gamma_s:&\quad p\longrightarrow r_s=p+k\longrightarrow p',
\label{eq:Compton-s-path}\\
\gamma_u:&\quad p\longrightarrow r_u=p-k'\longrightarrow p'.
\label{eq:Compton-u-path}
\end{align}
Endpoints alone do not determine a Wilson line, so we take each arrow in
\eqref{eq:Compton-s-path}-\eqref{eq:Compton-u-path} to be an affine
straight segment in momentum space, a choice that belongs to the matching
prescription and not to the Feynman graph, and a different curved
interpolation would change the enclosed area already at quadratic order.
The connection moreover lives on the timelike domain, so the two
intermediate momenta and the segments joining them must remain there.  In
the initial rest frame this requires
\begin{equation}
r_u^2=m^2c^2-2mE'_\gamma>0,
\qquad\text{hence}\qquad
E'_\gamma<\frac12mc^2,
\label{eq:Compton-timelike-domain}
\end{equation}
so that the small-loop result obtained next is a low-recoil statement.

The timelike condition permits a radial projection of each off-shell point
to the unit velocity hyperboloid, but it does not put the internal electron
on the physical mass shell.  In natural units for the remainder of this
paragraph, define
\begin{equation}
M_r\equiv\sqrt{r^2},
\qquad
u_r^\mu\equiv\frac{r^\mu}{M_r},
\qquad
M_s^2=m^2+2p\cdot k,
\qquad
M_u^2=m^2-2p\cdot k'.
\label{eq:Compton-effective-masses}
\end{equation}
Except at zero photon energy, $M_s\neq m$ and $M_u\neq m$.  Auxiliary
spinors with mass $M_r$ satisfy
\begin{equation}
\sum_\lambda u_\lambda(r;M_r)\bar u_\lambda(r;M_r)=\slashed r+M_r,
\qquad
\sum_\lambda v_\lambda(r;M_r)\bar v_\lambda(r;M_r)=\slashed r-M_r.
\label{eq:auxiliary-spin-sums}
\end{equation}
They give the exact algebraic decomposition
\begin{equation}
\slashed r+m
=\frac{M_r+m}{2M_r}(\slashed r+M_r)
+\frac{M_r-m}{2M_r}(\slashed r-M_r) 
\label{eq:offshell-numerator-decomposition}
\end{equation}
and hence
\begin{equation}
\frac{\slashed r+m}{r^2-m^2}
=\frac{1}{2M_r}\left[
\frac{\slashed r+M_r}{M_r-m}
+\frac{\slashed r-M_r}{M_r+m}
\right]. 
\label{eq:offshell-propagator-decomposition}
\end{equation}
The first term admits a positive-energy auxiliary spin sum whose canonical
bases carry Wigner rotations; the second term is required by the exact
identity and belongs to the complementary auxiliary sector.  Neither set is
a physical intermediate electron state, because its mass is $M_r$ rather
than $m$, and the photon vertices remain nontrivial spin operators.  At low
recoil,
\begin{equation}
M_s-m=\omega+O(\omega^2/m),
\qquad
M_u-m=-\omega'+O(\omega'^2/m),
\label{eq:offshell-low-energy-masses}
\end{equation}
whereas $M_r+m$ stays finite.  The positive auxiliary sector therefore
provides the dominant geometric comparator, but the QED amplitude is not a
pure holonomy factor.  Equation \eqref{eq:radial-projection} explains why
the numerical normalization $r\mapsto r/\sqrt{r^2}$ is geometrically
legitimate: it projects the affine route to velocity space without erasing
the off-shell terms in \eqref{eq:offshell-propagator-decomposition}.

\subsubsection{Centroid holonomy and the Thomas-Wigner comparator}
\label{sec:Compton-holonomy}

The transports along the two routings are
\begin{align}
U_s&=U(p',r_s)U(r_s,p),
&
U_u&=U(p',r_u)U(r_u,p),
\label{eq:Compton-transports}
\end{align}
and the relative object
\begin{equation}
\mathcal H_{su}=U_s U_u^{-1}
\label{eq:Compton-holonomy}
\end{equation}
is a closed momentum-space holonomy.  For $|k|,|k'|\ll|p|$ and the
orientation chosen in
\eqref{eq:Compton-s-path}-\eqref{eq:Compton-u-path},
\begin{align}
\mathcal H_{su}
&=\exp\left[
-\frac{i}{\hbar}
\mathcal F_p^{\mu\nu}(p)k_\mu k'_\nu
+O(k^3/p^3)
\right]
\notag\\
&=\exp\left[
\frac{i}{\hbar}
\frac{S^{\mu\nu}k_\mu k'_\nu}{p^2}
+O(k^3/p^3)
\right],
\label{eq:Compton-small-holonomy}
\end{align}
an expansion detailed in appendix~\ref{app:Compton-loop}.  Traversing the
loop in the opposite order reverses the overall orientation, so observables
depend on the corresponding interference convention.

The radial identity \eqref{eq:radial-projection} maps the four timelike
vertices to
\begin{equation}
u_p=\frac{p}{m},
\qquad
u_s=\frac{r_s}{M_s},
\qquad
u_{p'}=\frac{p'}{m},
\qquad
u_u=\frac{r_u}{M_u}.
\label{eq:Compton-unit-velocities}
\end{equation}
Let $R_W(u_b,u_a)$ denote the Wigner rotation associated with canonical
boosts between two such velocities.  The closed comparator is
\begin{equation}
\mathcal H_W
=\left[R_W(u_{p'},u_s)R_W(u_s,u_p)\right]
\left[R_W(u_{p'},u_u)R_W(u_u,u_p)\right]^{-1}.
\label{eq:Wigner-comparator}
\end{equation}
It is exactly a little-group rotation on the unit hyperboloid and is unitary
in the spin-$1/2$ Hilbert representation.  The radially projected centroid
connection defines the same geometric comparison of canonical spin frames
\cite{ThomasWignerHolonomy:2024}.  For an electron initially at rest and at
low recoil, both give
\begin{equation}
\theta_W
=\frac{E_\gamma E'_\gamma}{m^2c^4}
\sin\theta_\gamma
+O\!\left((E_\gamma/mc^2)^3\right).
\label{eq:Wigner-angle}
\end{equation}
This equality concerns the geometry of timelike paths after radial
projection.  Equations \eqref{eq:offshell-propagator-decomposition} and
\eqref{eq:Wigner-comparator} show why it cannot be promoted to an exact
factorization of the QED amplitude.

For an initially resting electron, define
$\langle\bm s\rangle=(\hbar/2)P_e\hat{\bm s}$ and let $\hat{\bm n}$ be
normal to the photon scattering plane.  The leading dimensionless invariant
is
\begin{equation}
\chi_{\rm hol}
\equiv
\frac{S^{\mu\nu}k_\mu k'_\nu}{\hbar p^2}
\simeq
\frac{P_e}{2}
\frac{E_\gamma E'_\gamma}{m^2c^4}
\sin\theta_\gamma\,
(\hat{\bm s}\cdot\hat{\bm n}),
\label{eq:Compton-chi}
\end{equation}
up to the orientation sign.  It vanishes for forward or backward
scattering and for spin lying in the scattering plane, grows largest for
spin normal to that plane, and reverses under
$\hat{\bm s}\rightarrow-\hat{\bm s}$.  All of these properties follow from
the curvature alone, whereas whether a rate depends linearly or
quadratically on $\chi_{\rm hol}$ is settled by the spin analysis and by
the gauge-complete amplitude.

\subsubsection{Exact tree-level channel trace}
\label{sec:Compton-trace}

The QED baseline can be computed independently of any transport dressing.
Set $c=\hbar=1$ in this calculation and use the initial-rest-frame
kinematics
\begin{equation}
p=(m,\bm0),
\qquad
k=\omega(1,0,0,1),
\qquad
k'=\omega'(1,\sin\theta_\gamma,0,\cos\theta_\gamma),
\label{eq:Compton-rest-kinematics}
\end{equation}
with
$\omega'=\omega/[1+(\omega/m)(1-\cos\theta_\gamma)]$.  The polarized
initial density matrix is
\begin{equation}
\rho_e(p;P_e,a)
=\frac12(\slashed p+m)(1+P_e\gamma_5\slashed a),
\qquad
a=(0,\hat{\bm s}),
\qquad |P_e|\leq1.
\label{eq:electron-density-matrix}
\end{equation}
Summing the final electron spin and physical transverse photon
polarizations, while summing rather than averaging the two initial photon
polarizations, gives
\begin{equation}
X_{\rm QED}^{\Sigma_\gamma}
=4e^4\left(
\frac{\omega'}{\omega}+\frac{\omega}{\omega'}
-\sin^2\theta_\gamma\right).
\label{eq:KN-summed}
\end{equation}
The same convention gives the exact imaginary channel correlator
\begin{equation}
\operatorname{Im}X_{su}^{\Sigma_\gamma}
=2e^4P_e(\hat{\bm s}\cdot\hat{\bm n})\sin\theta_\gamma
\left[
\frac{\omega\omega'}{m^2}(1-\cos\theta_\gamma)^2
-\cos\theta_\gamma
\right].
\label{eq:Im-channel-correlator}
\end{equation}
For the conventional unpolarized cross-section normalization, which averages
the initial photon polarization, both right-hand sides are divided by two:
\begin{equation}
\overline X_{\rm QED}=\frac12X_{\rm QED}^{\Sigma_\gamma},
\qquad
\operatorname{Im}\overline X_{su}
=\frac12\operatorname{Im}X_{su}^{\Sigma_\gamma}.
\label{eq:photon-average-convention}
\end{equation}
This resolves the factor-of-two difference between the printed
Klein-Nishina normalization and the supporting scripts; every relative
ratio below is unchanged.

The trace calculation passes four independent algebraic checks.  The
diagonal channel terms are real; the full tree-level rate is independent of
$P_e$; the imaginary correlator vanishes for spin in the scattering plane;
and \eqref{eq:KN-summed} reduces exactly to the Klein-Nishina combination.
The term proportional to $-\cos\theta_\gamma$ in
\eqref{eq:Im-channel-correlator} is not energy suppressed, because the
low-energy theorem constrains the sum of channels while the channel
correlator retains pieces that cancel in that sum.

Equation \eqref{eq:Im-channel-correlator} is an exact trace in the declared
physical-polarization basis, but it is not an observable by itself.  The
$s$- and $u$-channel split is not separately gauge invariant: replacing the
physical transverse sums by $-g_{\mu\nu}$ changes the summed result to
\begin{equation}
\operatorname{Im}X_{su}^{(-g)}
=-6e^4P_e(\hat{\bm s}\cdot\hat{\bm n})
\sin\theta_\gamma,
\label{eq:Im-channel-covariant-sum}
\end{equation}
an order-one change in both magnitude and angular dependence.  We therefore
refer to \eqref{eq:Im-channel-correlator} as a polarized channel correlator,
not as a QED asymmetry.  Appendix~\ref{app:Compton-trace} gives the trace
setup and its validation sequence.

\subsubsection{Minimal relative-phase benchmark}
\label{sec:Compton-benchmark}

Comparing this invariant with the older frozen-line ansatz is instructive.
Assigning one constant tensor to the internal fermion and removing a common
phase, the two QED channels take the diagnostic form
\begin{equation}
\mathcal M_{\rm toy}
=e^{+i\Phi_{\mathcal C}}\mathcal M_s^{(0)}
+e^{-i\Phi_{\mathcal C}}\mathcal M_u^{(0)},
\qquad
\Phi_{\mathcal C}
=\frac{1}{2\hbar^2}
k_\mu\overline\Theta_{\mathcal C}^{\mu\nu}k'_\nu,
\label{eq:Compton-toy-amplitude}
\end{equation}
Writing
$X_{su}=\sum\mathcal M_s^{(0)}\mathcal M_u^{(0)*}=R+iI$, the exact change
of the squared amplitude caused by this ansatz is
\begin{equation}
\Delta|\mathcal M|^2
=-2\sin(2\Phi_{\mathcal C})I
-4\sin^2\!\Phi_{\mathcal C}\,R.
\label{eq:toy-amplitude-change}
\end{equation}
and in the same rest-frame polarization
\begin{equation}
\Phi_{\mathcal C}
=-\frac{P_e}{4}
\frac{E_\gamma E'_\gamma}{m^2c^4}
\sin\theta_\gamma\,
(\hat{\bm s}\cdot\hat{\bm n}),
\label{eq:Compton-frozen-phase}
\end{equation}
so the frozen phase reproduces the spin, energy and angular scales of the
geometric invariant, within a c-number expectation-value replacement.  The
reweighted pair of diagrams is nevertheless
not gauge invariant, since ordinary QED satisfies the Ward identity only
after the two channels are combined.  Defining
\begin{equation}
A^\nu\equiv
k_\mu\mathcal M_s^{(0)\,\mu\nu}
=-k_\mu\mathcal M_u^{(0)\,\mu\nu},
\label{eq:Compton-A}
\end{equation}
one finds
\begin{equation}
k_\mu\mathcal M_{\rm toy}^{\mu\nu}
=2i\sin\Phi_{\mathcal C}\,A^\nu\neq0.
\label{eq:Compton-Ward-violation}
\end{equation}
Contact terms and, in a frozen NCQED benchmark, additional gauge-sector
diagrams must be added before the Ward identity holds.  A star-gauge
completion generates them automatically, which is one more reason why the
matching has to come from an action.

\begin{figure}[t]
\centering
\includegraphics[width=0.72\linewidth]{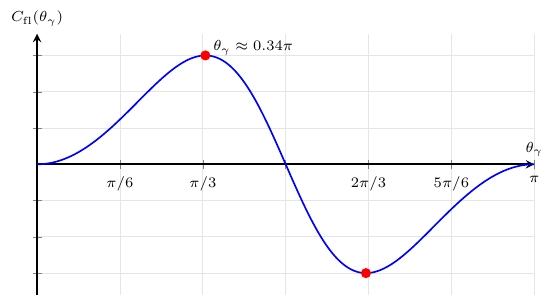}
\caption{Leading angular response
$C_{\rm fl}(\theta_\gamma)$ of the minimal relative-phase benchmark in
eq.~\eqref{eq:Compton-template}.  The curve is not a QED cross section; a
gauge-complete transverse matching may alter its coefficient and shape.}
\label{fig:Compton-template}
\end{figure}

For continuity with the exploratory calculation, define the dimensionless
angular template
\begin{equation}
\mathcal T_{\rm fl}(\theta_\gamma)
\equiv
C_{\rm fl}(\theta_\gamma)
\left(\frac{E_\gamma}{mc^2}\right)^2P_e^2,
\qquad
C_{\rm fl}(\theta_\gamma)
=\frac{\sin^2\theta_\gamma\cos\theta_\gamma}
{2(1+\cos^2\theta_\gamma)}.
\label{eq:Compton-template}
\end{equation}
The factor $P_e^2$ has a restricted origin: the c-number fiber replacement
sets $\Phi_{\mathcal C}\to P_e\Phi_0$, while the density-matrix trace sets
$I\to P_e I_0$.  It is therefore the prediction of a mean-spin or
mean-field benchmark, not a general law for a partially polarized ensemble.
For $P_e=\pm1$ the calculation is a pure-state result.  For $|P_e|<1$, a
matrix holonomy must act on the density matrix before the channel kernels
are traced, and that calculation need not scale as $P_e^2$.  Within the
mean-spin benchmark, the response is a spin-alignment effect, even under
$\hat{\bm s}\to-\hat{\bm s}$, and not a vector single-spin asymmetry.
The function $C_{\rm fl}$ is odd about $\theta_\gamma=\pi/2$, vanishes at
$0$, $\pi/2$ and $\pi$, and reaches $|C_{\rm fl}|_{\max}=0.150$ at
$\cos^2\theta_\gamma=\sqrt5-2$, that is, at
$\theta_\gamma\simeq61^\circ$.  Figure~\ref{fig:Compton-template} displays
this benchmark shape, which is not a cross-section prediction, since
gauge-restoring contact terms and transverse matching operators can change
both its coefficient and its angular dependence.

For example, an operatorial spin holonomy about $\hat{\bm n}$ gives
\begin{equation}
\rho_n(P_e)=\frac12(\bm1+P_e\sigma_n),
\qquad
H_W=\exp\!\left(-\frac{i}{2}\theta_W\sigma_n\right),
\qquad
\operatorname{tr}(\rho_nH_W)
=\cos\frac{\theta_W}{2}-iP_e\sin\frac{\theta_W}{2}.
\label{eq:density-matrix-holonomy}
\end{equation}
The physical interference must then be formed as
$2\operatorname{Re}\operatorname{tr}(K_s\rho_nK_u^\dagger)$, with the
channel kernels included; its polarization dependence cannot be inferred by
replacing $\sigma_n$ with its expectation value at the outset.

The distinction between polarization treatments is summarized in
table~\ref{tab:polarization-status}.

\begin{table}[t]
\centering
\footnotesize
\caption{Polarization dependence of the relative-phase calculation.}
\label{tab:polarization-status}
\begin{tabular}{@{}L{0.34\linewidth}L{0.20\linewidth}L{0.36\linewidth}@{}}
\toprule
treatment & $P_e$ dependence & status \\
\midrule
pure eigenstate, $P_e=\pm1$ & numerical values below & supported by the exact trace \\
mean-spin fiber, $\overline\Theta\propto\langle S\rangle$ & $P_e^2$ & semiclassical c-number benchmark \\
matrix holonomy acting on $\rho_e$ & to be calculated & not fixed by the present scripts \\
\bottomrule
\end{tabular}
\end{table}

The kinematic size of the benchmark is worth recording together with its
status.  For an electron, $(E_\gamma/mc^2)^2\simeq3.8\times10^{-4}$ at
$E_\gamma=10\,$keV and $3.8\times10^{-2}$ at $100\,$keV, so with
$|C_{\rm fl}|_{\max}=0.150$ and full polarization, evaluation of the exact
expressions gives $5.690\times10^{-5}$ at $10\,$keV and
$5.048\times10^{-3}$ at $100\,$keV.  The $P_e^2$ extrapolation belongs only
to the mean-spin line of table~\ref{tab:polarization-status}.  These numbers
measure the size of the minimal phase benchmark and do not predict a
gauge-complete rate.  Near
$E_\gamma\sim mc^2$, where the nominal effect grows, recoil reorients the
spin tensor and the frozen-line assumption is the first thing to break
down, so a reliable statement in that regime requires the gauge-complete
theory.  At low recoil the completion-independent quantity is the holonomy
scale $|\chi_{\rm hol}|\sim(|P_e|/2)E_\gamma E'_\gamma/(m^2c^4)$ instead of
a fixed rate correction.  Polarimetry of the final photon and comparison of
opposite electron polarizations preserve interference information that an
unpolarized total rate erases, and a genuine signal has to rotate rigidly
with the electron spin, whereas a fixed laboratory anisotropy would point
to a background tensor or to an instrumental systematic.  Sources of
quasi-monochromatic polarized photons produced by inverse Compton
backscattering, such as HI$\gamma$S and the ELI-NP gamma system, match this
kinematics \cite{Weller:2009HIGS,ELINP:TDR}.  Laser-based probes of
noncommutative QED with lepton beams \cite{Tizchang:2017NCQEDLaser} test a
universal background tensor inside a common star-gauge theory and therefore
constrain a different object, so their bounds do not transfer to the
state-borne tensor without an observable-level matching.

An amplitude-level EFT makes the absence of double counting explicit.  The
pure geometric branch is
\begin{equation}
\mathcal M_{\rm physical}=\mathcal M_{\rm QED}.
\label{eq:pure-geometric-amplitude}
\end{equation}
Additional dynamics may be written as
\begin{equation}
\mathcal M^{\mu\nu}
=\mathcal M_{\rm QED}^{\mu\nu}
+\chi_{\rm hol}\sum_i c_i(s,t)T_i^{\mu\nu},
\qquad
k_\mu T_i^{\mu\nu}=k'_\nu T_i^{\mu\nu}=0.
\label{eq:transverse-EFT-excess}
\end{equation}
The coefficients $c_i=0$ define the standard geometric interpretation.  A
nonzero $c_i$ is new physics and must be obtained from an action or bounded
experimentally.  The minimal phase curve corresponds to one
non-gauge-complete way of imitating such an excess; it is not a value of the
coefficients in \eqref{eq:transverse-EFT-excess}.

\subsection{The anomalous magnetic moment as a matching test}
\label{sec:gminus2}

For the direct on-shell 1PI endpoint dressing in
\eqref{eq:endpoint-phase-vertex}, the endpoint lemma gives
\begin{equation}
\delta F_2(0)
\big|_{\text{direct scalar endpoint dressing}}=0.
\label{eq:gminus2-scalar-null}
\end{equation}
This null result constrains neither the loop contributions of a theory
whose elementary vertices carry such phases nor the value of $F_2(0)$ in a
completed theory, and two logically consistent branches remain open.

In the \emph{geometric pullback branch}, the change from $z$ to $X$ is
implemented as an exact redefinition of all fields, sources, observables
and the functional measure.  Should that pullback be unitary and anomaly
free, the equivalence theorem leaves the $S$-matrix and with it $g-2$
unchanged, so the null result would follow from dynamical equivalence and
not from \eqref{eq:endpoint-identity}.  This conclusion is conditional:
the centroid map is many-to-one on the ambient data and is momentum
dependent, so invertibility, the functional measure and the treatment of
sources must be established before the equivalence theorem can be invoked.

In the \emph{dynamical branch}, the connection or the fibered twist enters
the interaction action as additional physics, which permits a Pauli
matching coefficient,
\begin{equation}
\Delta\Gamma^\alpha
=\frac{i\sigma^{\alpha\beta}q_\beta}{2mc}
c_{\rm NC}
+\cdots,
\qquad
\delta a=\delta F_2(0)=c_{\rm NC}.
\label{eq:Pauli-matching}
\end{equation}
That coefficient must then be calculated in a specified completion or
constrained by the precision data on lepton moments
\cite{Fan:2022eto,Aoyama:2020ynm,Aliberti:2025beg,Muong-2:2023cdq}.  An
order-one matching coefficient at the Compton scale would be severely
incompatible with those measurements; a frequently quoted estimate
$\delta a\sim O(\alpha)$ is therefore a naturalness warning conditional on
the completion producing an unsuppressed Pauli operator, not a theorem
about every transverse two-photon structure.  Lepton moments directly
constrain the operators that match onto $F_2(0)$ and do not, without an EFT
calculation, bound every doubly transverse coefficient in
\eqref{eq:transverse-EFT-excess}.  An
economical hypothesis lying between the two branches holds that only closed
momentum-space holonomies survive as observables, which predicts no
independent effect for a single open vertex while allowing Compton or
multi-impulse interference to probe the curvature.  A gauge-complete
matching that produced an unavoidable local Pauli operator would falsify
it.

\subsection{Experimental logic}
\label{sec:experimental-logic}

Before the dynamical branch is selected the geometry predicts no universal
number, and what it does predict are correlated transformation properties.
Table~\ref{tab:observables} states the discriminants without converting
templates into bounds.  It should be read against the universal-background
searches of
refs.~\cite{Hewett:2001,Hinchliffe:2004,Carlson:2001,Colladay:1998fq,
Kostelecky:2008,Mattingly:2005re}, whose bounds constrain a constant
Lorentz-violating tensor and therefore do not apply directly to a
state-borne, transversality-constrained one.

\begin{table}[t]
\centering
\caption{Standard or geometric baselines and the assumptions required for a
discriminator of additional dynamics.}
\label{tab:observables}
\footnotesize
\begin{tabular}{@{}L{0.17\linewidth}L{0.27\linewidth}L{0.25\linewidth}L{0.22\linewidth}@{}}
\toprule
Probe & baseline & additional assumption & genuine discriminator \\
\midrule
Packet interference &
same-fiber phase quadratic in packet width &
common conditioned fiber and coherent transport &
departure from width and spin-reversal scaling \\
Side-jump &
the three centroid values in table~\ref{tab:side-jump-budget} &
detector-to-centroid calibration &
deviation from the calibrated position value \\
Compton angle &
Thomas-Wigner value \eqref{eq:Wigner-angle} &
none for the geometric/QED comparator &
doubly transverse excess beyond the baseline \\
Compton rate response &
Klein-Nishina and the channel trace &
gauge-complete contact and transverse operator &
nonzero fitted $c_i(s,t)$ in \eqref{eq:transverse-EFT-excess} \\
$g-2$ &
dynamical Pauli matching beyond the geometric pullback &
completion that generates $F_2(0)$ &
nonzero $c_{\rm NC}$ after standard contributions \\
Three-impulse protocol &
leakage identity \eqref{eq:projected-associator} &
intermediate reprojection as part of the protocol &
bracketing dependence with regulator control \\
\bottomrule
\end{tabular}
\end{table}
\section{Projected products and a controlled nonassociative branch}
\label{sec:projected}

The exact centroid algebra is associative, and relaxing the supplementary
condition at the level of states alone does nothing to change that.  The
simultaneous assumptions
\begin{equation}
S^{\mu\nu}p_\nu|\mathrm{phys}\rangle=0,
\qquad
[S^{\mu\nu},X^\rho]=0,
\qquad
[X^\rho,p_\nu]=i\hbar\delta^\rho_\nu
\label{eq:naive-weak-assumptions}
\end{equation}
are in fact unstable under the position operation, since
\begin{equation}
[S^{\mu\nu}p_\nu,X^\rho]
=-i\hbar S^{\mu\rho}.
\label{eq:weak-obstruction}
\end{equation}
An ordinary commutator in any associative operator algebra obeys the Jacobi
identity, so a nonvanishing coordinate Jacobiator signals either an
inconsistent truncation or a genuinely different product.  Rather than
infer nonassociativity from a failed identity, we exhibit such a product
explicitly.

The truncated calculation nevertheless leaves behind a useful diagnostic.
Treating $S$ as an external frozen label and defining the formal bracket
\begin{equation}
\{X^\mu,X^\nu\}_{\rm tr}
=-\frac{S^{\mu\nu}}{p^2},
\qquad
\{X^\mu,p^\nu\}_{\rm tr}=\eta^{\mu\nu},
\qquad
\{S^{\mu\nu},X^\rho\}_{\rm tr}=0,
\label{eq:truncated-bracket}
\end{equation}
one finds the Jacobiator
\begin{align}
\mathfrak J_{\rm tr}^{\mu\nu\rho}
&\equiv
\{\{X^\mu,X^\nu\}_{\rm tr},X^\rho\}_{\rm tr}
+\text{cyclic}
\notag\\
&=-\frac{2}{(p^2)^2}
\left(
S^{\mu\nu}p^\rho
+S^{\nu\rho}p^\mu
+S^{\rho\mu}p^\nu
\right)
\notag\\
&=\frac{2}{(p^2)^2}
\epsilon^{\mu\nu\rho\sigma}W_\sigma,
\label{eq:truncated-Jacobiator}
\end{align}
where the last step uses $T^{\mu\nu\rho}=-\epsilon^{\mu\nu\rho\sigma}
W_\sigma$, the dual form of \eqref{eq:T-def}.  This is precisely the
contribution that the mixed bracket \eqref{eq:reduced-SX} cancels in the
full algebra, so \eqref{eq:truncated-Jacobiator} is a target that a new
effective quantization would have to reproduce and not the Jacobiator of
the exact centroid.  Three-cocycles of the same $\epsilon\,W$ form occur in
nongeometric $R$-flux backgrounds and in the phase-space description of
magnetic sources
\cite{Bakas:2014threeCocycles,Mylonas:2012membrane,Jackiw:2001yj}, with the
Pauli-Lubanski pseudovector playing here the part of the flux density.

\subsection{Composition with intermediate projection}
\label{sec:projected-product}

Let $P$ project the ambient kinematical space onto a chosen physical
sector, for instance a regulated mass band together with a
supplementary-condition sector, and set $Q=1-P$.  Reading a spin
supplementary condition as a physical-state projection is familiar from
field theory \cite{Kim:2023covariantSSC}, and the composition defined here
makes that projection part of the product itself.  For ambient
representatives $A$ and $B$, set
\begin{equation}
A\circ_P B\equiv PABP.
\label{eq:projected-product}
\end{equation}
For differential operators such as
$X^\mu=i\hbar\partial_p^\mu+\mathcal A^\mu$, multiplying by the
characteristic function of a sharp momentum shell produces boundary
distributions and fails by itself to define a common operator domain.  We
therefore adopt a finite-volume regulator together with an orthogonal
finite-rank projection
\begin{equation}
P_N=\sum_{a=1}^{N}|u_a\rangle\langle u_a|,
\qquad P_N^2=P_N,
\label{eq:regulated-projector}
\end{equation}
whose smooth packet modes $u_a$ lie in a common invariant core and span a
regulated mass band, leaving a sharp-shell limit to be studied separately.
Suppressing the label $N$ and letting $A,B,C$ denote fixed ambient lifts,
we note that different lifts sharing the same compression $PAP$ can produce
different projected products, a dependence that belongs to the effective
protocol and is no invariant of the compressed operator alone.  Repeated
use of $P^2=P$ gives the exact associator
\begin{align}
(A,B,C)_P
&\equiv
(A\circ_P B)\circ_P C
-A\circ_P(B\circ_P C)
\notag\\
&=PABPCP-PAPBCP
\notag\\
&=P A Q B C P-P A B Q C P,
\label{eq:projected-associator}
\end{align}
with the algebraic identity spelled out in
appendix~\ref{app:projected-identity}.  Failure of associativity is
controlled entirely by intermediate leakage into $Q\mathcal H$, and for the
covariant position such leakage is generic, since
\eqref{eq:p2-not-preserved} forces $QX^\mu P\neq0$ whenever $P$ sharply
selects a mass shell.

For three position operations,
\begin{equation}
(X^\mu,X^\nu,X^\rho)_P
=PX^\mu QX^\nu X^\rho P
-PX^\mu X^\nu QX^\rho P,
\label{eq:coordinate-projected-associator}
\end{equation}
an object with a direct operational reading, since it compares two
protocols that apply three momentum-changing or localization operations
while inserting the physical reprojection at different intermediate stages.
None of this contradicts ordinary quantum mechanics.  The ambient operator
product is associative, and so is the algebra of genuine endomorphisms
$PAP$ acting entirely within $P\mathcal H$, whereas the product
\eqref{eq:projected-product} concerns ambient generators that fail to
preserve the sector and a protocol that discards their $Q$ components.  Its
nonassociativity is effective and projection dependent.

\subsection{Spin three-form and quasi-Hopf program}
\label{sec:quasi-Hopf-program}

The totally antisymmetric part of
\eqref{eq:coordinate-projected-associator} is a rank-three tensor, and
since $W^\mu$ is an axial vector its contraction with the pseudotensor
$\epsilon^{\mu\nu\rho\sigma}$ has ordinary tensor parity.  The
finite-volume, finite-rank regulator \eqref{eq:regulated-projector} is not
Lorentz covariant in general and can generate additional structures built
from its box and mode data, so the parametrization that follows is an
ansatz for a regulator-removal limit $N\to\infty$ in which those artifacts
disappear.  At leading order in spin, covariance of that limit allows
\begin{equation}
\mathfrak J_P^{\mu\nu\rho}
=\zeta_P\frac{\hbar^2}{(p^2)^2}
\epsilon^{\mu\nu\rho\sigma}W_\sigma
+\text{derivative and higher-spin terms},
\label{eq:projected-W-ansatz}
\end{equation}
where $\zeta_P$ depends on the projector and on the precise reprojection
protocol, the prefactor becoming $\hbar^2/(m^4c^4)$ on a massive shell.
Equation~\eqref{eq:projected-W-ansatz} carries the dimension of a
coordinate three-volume and reproduces the covariant structure suggested by
\eqref{eq:truncated-Jacobiator}, while remaining a matching ansatz and not
a universal coefficient.

Two tests must be passed before familiar algebraic names may be attached to
this branch.  The commutator algebra would be of Malcev type only if
\begin{equation}
J(a,b,[a,c])=[J(a,b,c),a]
\label{eq:Malcev-identity}
\end{equation}
holds, so that a nonvanishing Jacobiator by itself settles nothing.  A
quasi-Hopf description requires in addition an invertible cochain
$\mathcal F_{\rm op}$ and the associator
\begin{align}
\Phi_{\mathcal F}
={}&(1\otimes\mathcal F_{\rm op})
({\rm id}\otimes\Delta)\mathcal F_{\rm op}\times
(\Delta\otimes{\rm id})\mathcal F_{\rm op}^{-1}
(\mathcal F_{\rm op}^{-1}\otimes1),
\label{eq:quasi-Hopf-associator}
\end{align}
followed by the pentagon identity.  A comparison framework is available,
since the known $R$-flux constructions relate families of associative
momentum-surface products to a nonassociative phase-space product
\cite{MylonasSchuppSzabo:2014,Mylonas:2012membrane}, and nonassociative
differential geometry based on quasi-Hopf twists has been developed for
nongeometric gravitational backgrounds
\cite{Blumenhagen:2016nonassociativeGravity,
Aschieri:2018nonassociativeGeometry}.  Here the source of the candidate
associator would instead be spin-controlled sector leakage.  Since only
$W^2$ is a Casimir invariant, $W^\mu$ in \eqref{eq:projected-W-ansatz}
remains an operator unless a further polarized semiclassical expectation
value is taken.  Explicit calculation of $P$, of $\zeta_P$, of hermiticity,
of the Malcev identity and of the pentagon identity is the minimal program
needed to promote this branch from an effective proposal to a
nonassociative quantum geometry.

\section{Discussion}
\label{sec:discussion}

Spin-dependent noncommutativity of relativistic position is part of an established line of work on covariant centroids, spinning-particle phase spaces, and spin-dependent coordinate algebras \cite{Souriau:1970,CostaNatario:2015,Colatto_2006,Gomes:2010,DeriglazovPupasov:2014,DeriglazovTereza:2019}. Momentum-space Berry connections also provide a known mechanism by which projected Dirac positions acquire noncommuting components \cite{Berard:2004monopole,Berard:2006spinhall}. Equation~\eqref{eq:XX-PB} should therefore not be read as the introduction of spin-induced position noncommutativity itself. The narrower contribution of the present construction is to place the covariant Tulczyjew-Dixon centroid, its induced operator algebra, the momentum-space curvature calculation, the free positive-energy projection, the conditioned twist description, and the projected-product extension in a single conventionally fixed framework in which their logical status can be compared without identifying kinematics with additional dynamics.

Within the declared linear covariant ansatz, transversality fixes the transverse displacement that leads to \eqref{eq:X-definition}, modulo the longitudinal freedom $g(p^2)p^\mu$. The idempotent centroid map used throughout the subsequent analysis is the $g=0$ representative. For that representative, the composite-variable calculation yields \eqref{eq:XX-PB} together with the mixed spin-position bracket and the projected spin algebra. The explicit Jacobi cancellation is best interpreted as a consistency check on this induced algebra and, more specifically, as a diagnostic showing why $p^{-2}$ cannot be replaced by a fixed shell value before the commutators have been evaluated. It does not supply an independent interaction or an additional physical postulate.

The momentum representation sharpens this geometric reading. In the fixed spin frame used in the calculation, the multiplicative term $\mathcal A^\mu=\Sigma^{\mu\nu}p_\nu/p^2$ has the algebraic form of an $\mathfrak{so}(1,3)$-valued momentum-space connection, and the corresponding nonabelian curvature combination reproduces $-S^{\mu\nu}/p^2$. The split between the derivative and commutator pieces, $-2:+1$, is required for that result. A fully frame-covariant bundle interpretation would additionally require an explicit transformation law for $\mathcal A^\mu$ under momentum-dependent changes of spin frame. The present calculation establishes the curvature identity in the chosen realization, which is sufficient for the local transport statements used here but should not be confused with a completed global gauge construction.

\subsection{Positive-energy projection and the localization dictionary}

The free positive-energy calculation gives the sharpest internal distinction between the ambient covariant centroid and the band-projected Dirac position. The projected spin shift contains the negative of the Berry connection carried by the projected charge position, so the two terms cancel in the total projected covariant centroid. What remains is the transverse shift displayed in \eqref{eq:projected-TD-position}, together with the longitudinal anti-hermitian term whose curl does not contribute to the antisymmetrized position commutator. The resulting commutator retains the longitudinal ambient value while its transverse component is reduced by $\gamma^2$. This cancellation prevents the Berry contribution and the centroid shift from being counted as independent transverse effects.

These identities are exact statements about the free Dirac block used in the calculation, but their operational interpretation requires more than block algebra. In particular, the projected operator in \eqref{eq:projected-TD-position} has not been established here as a new self-adjoint position observable with a specified measurement protocol. The position dictionary and the impulse bookkeeping should consequently be read as relations among localization prescriptions within the stated semiclassical setup. The recovery of the Thomas spin-orbit factor and the electric-field BMT equation for $g=2$ belongs to standard relativistic spin dynamics, in agreement with earlier analyses in which spin-dependent position brackets account for the Pauli and Thomas structure \cite{DeriglazovTereza:2019}. Their role here is to test the consistency of the centroid dictionary, not to provide a beyond-QED signal.

\subsection{Conditioned twist fibers and their domain of validity}

The constant-tensor deformation used in the frozen fibers is a separate approximation from the exact operator algebra. Once a packet and polarization sector are fixed so that $\widehat\Theta^{\mu\nu}$ is replaced by the c-number $\overline\Theta_{\mathcal C}^{\mu\nu}$, the Moyal product, abelian Drinfel'd twist, braided exchange rules, deformed coproduct, and Clebsch-Gordan maps are associative within that fiber. Those statements are algebraically well defined under the conditioning assumptions and provide a local comparison with constant-$\Theta$ noncommutative theories.

For a spin-one-half particle, however, the estimate $\epsilon_S^2=2/3$ removes a parametrically controlled spin-freezing limit. The fiber construction therefore cannot serve, for the electron, as a semiclassical reduction whose error can be made arbitrarily small by narrowing the packet. It is better interpreted as a conditioned c-number chart, or as a matching device for matrix elements prepared in a specified polarization sector. The impossibility of choosing one nonzero constant bivector transverse to every momentum in an open set further prevents these fibers from being assembled into a universal constant-background theory. Momentum-space transport can relate different local charts, but a global deformation with a momentum-dependent product or twist remains an additional construction whose gauge covariance and scattering rules have not been obtained in this work.

\subsection{Compton transport, QED, and the boundary between geometry and dynamics}

The Compton analysis contains several statements with different evidential status. The closed centroid routing defines a geometric holonomy, and its small-loop expansion produces the invariant $S^{\mu\nu}k_\mu k'_\nu/(\hbar p^2)$. The associated timelike momentum polygon supplies a Thomas-Wigner comparator at leading low recoil. The displayed numerical comparison illustrates this correspondence at the kinematics that were evaluated, but it does not by itself establish a scaling law over energy and scattering angle. Independently, the internal fermion momenta in the physical QED diagrams are off shell at the electron mass, so the propagator numerators cannot be factorized exactly into transport of physical on-shell spin states along the two channels.

The polarized $s$-$u$ channel trace and the geometric holonomy are therefore diagnostic quantities, not a factorization theorem for the QED amplitude. The relative-phase dressing makes this separation especially explicit because its Ward-identity defect is nonzero and the result depends on the polarization prescription. Its angular curve can be retained as a benchmark for the symmetry pattern that a completed model might reproduce, but its magnitude is not a prediction for a measurable Compton rate. The pure geometric reading is instead summarized by \eqref{eq:pure-geometric-amplitude},
\begin{equation}
\mathcal M_{\rm physical}=\mathcal M_{\rm QED}.
\end{equation}
Any deviation from QED must consequently enter through additional, gauge-complete operators such as the doubly transverse structures in \eqref{eq:transverse-EFT-excess}, or through a separately specified interaction that generates a Pauli coefficient such as $c_{\rm NC}$.

This separation changes the status of the experimental discussion. The observables collected in table~\ref{tab:observables} are presently a map of possible matching targets, not a set of predicted deviations. The coefficients $c_i(s,t)$, $c_{\rm NC}$, and the projected coefficient $\zeta_P$ have not been calculated or bounded by the present construction. Existing measurements can constrain them only after a concrete completion maps the geometric variables into gauge-complete amplitudes or form factors. The next discriminating calculation on the dynamical branch is therefore not another phase estimate but the construction of an interaction for which both photon Ward identities hold and at least one transverse matching coefficient can be obtained. If all such coefficients vanish, the dynamical branch is removed while the covariant-centroid geometry and its localization dictionary remain intact.

\subsection{Projected products and the quasi-Hopf question}

The projected product introduces a logically different extension. The ambient operator algebra remains associative, whereas $A\circ_P B=PABP$ inserts reprojection into the composition rule. Its associator, \eqref{eq:projected-associator}, is an exact algebraic identity for an idempotent projector and shows that any failure of associativity is controlled by excursions through the complementary sector $Q\mathcal H$. This result gives a concrete meaning to projection-induced nonassociativity without modifying the ordinary operator product.

The subsequent identification with a spin-controlled three-form is more tentative. The truncated Jacobiator fixes a covariant tensor structure proportional to $\epsilon^{\mu\nu\rho\sigma}W_\sigma$, but it does not determine the coefficient of the projected associator. For the admissible finite-rank projectors $P_N$, generic nonzero leakage has not yet been demonstrated, and the coefficient $\zeta_P$ in \eqref{eq:projected-W-ansatz} remains an ansatz that can depend on the regulator and on the reprojection protocol. The comparison with nongeometric $R$-flux models is therefore structural. In those constructions the quasi-Hopf associator is tied to a three-cocycle generated by flux \cite{MylonasSchuppSzabo:2014,Aschieri:2018nonassociativeGeometry}, whereas the candidate source here would be spin-dependent sector leakage.

For this branch, the decisive program is well defined. One must construct a regulator sequence with a controlled limit, verify nonvanishing and regulator stability of the projected associator, establish the relevant adjointness properties, determine whether the Malcev identity is satisfied, and, if a cochain twist exists, verify the pentagon identity. Failure of any of these tests narrows the algebraic interpretation without affecting the exact centroid commutator. Conversely, success would supply the missing bridge between the projection identity and a quasi-Hopf effective geometry. The twist and quasi-Hopf sectors are therefore retained here as mathematically testable extensions, not as consequences already implied by spin-induced coordinate noncommutativity.

\section{Conclusions}
\label{sec:conclusions}

The covariant Tulczyjew-Dixon localization of a massive spinning particle carries a noncommuting coordinate algebra. Starting from canonical Poincar\'e variables and choosing the idempotent $g=0$ centroid representative, the composite construction gives
\begin{equation}
\{X^\mu,X^\nu\}=-\frac{S^{\mu\nu}}{p^2},
\qquad
S^{\mu\nu}p_\nu=0,
\label{eq:conclusions-bracket}
\end{equation}
together with the mixed spin-position and spin-spin brackets, derived in appendix~\ref{app:brackets}, that a consistent induced algebra requires. The bracket itself is not new, and we attribute it to \cite{DeriglazovPupasov:2014}. What the composite route contributes is the domain statement of section~\ref{sec:shell-restriction}: the replacement $p^2\to m^2c^2$ inside the centroid is exact on shell through two insertions of $\hat X$ and fails from three onward, with the obstruction in the antisymmetric channel given by the truncated Jacobiator. That criterion, and not the bracket, is the result we would ask a reader to carry away. In the momentum realization, the same structure is encoded by the $\mathfrak{so}(1,3)$-valued potential $\mathcal A^\mu=\Sigma^{\mu\nu}p_\nu/p^2$, whose nonabelian curvature combination reproduces $-S^{\mu\nu}/p^2$ with the derivative and commutator contributions in the ratio $-2:+1$. These results establish a covariant momentum-space geometry of the chosen centroid. They do not, by themselves, define a new interaction.

The free positive-energy projection supplies a second set of exact identities. The Berry term of the projected Dirac position cancels against the corresponding term in the projected spin shift, leaving the transverse displacement in \eqref{eq:projected-TD-position} and a longitudinal anti-hermitian remnant. The commutator obtained from this block calculation is anisotropic, with the transverse ambient value reduced by $\gamma^2$. The associated side-jump and $g=2$ spin-orbit relations are consistency statements within standard relativistic spin dynamics. Turning these operator relations into detector-level position observables requires a separate measurement model and, where self-adjointness is needed, a corresponding operator construction.

The local twist description has a narrower status. A fixed polarized packet defines an associative c-number fiber in which the Moyal product and the abelian Drinfel'd machinery are well defined, but $\epsilon_S^2=2/3$ prevents this replacement from becoming a controlled classical-spin limit for an electron. The fiber is consequently a conditioned chart and not a universal noncommutative background. Its most useful role is to provide a local algebraic benchmark and a consistency condition for any future momentum-dependent deformation that connects different fibers.

The Compton analysis likewise separates geometric transport from new dynamics. The closed centroid loop has a low-recoil Thomas-Wigner comparator, whereas the physical QED channels contain off-shell fermion propagation and do not factorize into on-shell holonomies. The relative-phase curve is gauge incomplete and cannot be promoted to a rate prediction. In the geometric branch, the amplitude remains the QED amplitude. A measurable excess requires an independently specified, gauge-complete interaction that produces nonzero transverse coefficients such as $c_i(s,t)$ or a Pauli matching coefficient $c_{\rm NC}$. No such coefficient is fixed by the present work, so the manuscript does not yet predict a beyond-QED deviation.

Projection-induced nonassociativity provides a separate mathematical direction. The identity \eqref{eq:projected-associator} shows exactly how reprojection can generate an effective associator through leakage into $Q\mathcal H$, while the proposed $\epsilon^{\mu\nu\rho\sigma}W_\sigma$ structure and the coefficient $\zeta_P$ remain to be established for admissible regulated projectors. A quasi-Hopf interpretation therefore depends on calculations that have not yet been performed, including regulator control, adjointness, Malcev closure, construction of the relevant cochain, and the pentagon identity.

The established content of the paper is thus a covariant noncommutative localization structure for relativistic spin, its momentum-space curvature representation, and its relation to the positive-energy Dirac position algebra. The conditioned twist fibers and the projected associator provide two mathematically distinct extensions of that structure. Their physical promotion requires additional input. For the dynamical branch, the decisive missing step is a gauge-complete matching calculation that yields at least one nonzero or demonstrably vanishing transverse coefficient. For the projected branch, it is an explicit regulated computation of the associator and its algebraic identities. Keeping these requirements separate preserves the exact geometric results while making the claims of additional dynamics and quasi-Hopf structure falsifiable.

\appendix

\section{Composite-variable calculation of the centroid brackets}
\label{app:brackets}

Write $X^\mu=z^\mu+a^\mu$ with $A^\mu=\Sigma^{\mu\nu}p_\nu$ and
$a^\mu=A^\mu/p^2$.  The elementary relations imply
\begin{align}
\{z^\mu,p^2\}&=2p^\mu,
&
\left\{z^\mu,\frac1{p^2}\right\}
&=-\frac{2p^\mu}{p^4},
\label{eq:app-zp2}\\
\{z^\mu,A^\nu\}&=-\Sigma^{\mu\nu},
&
\{A^\mu,A^\nu\}
&=p^2\Sigma^{\mu\nu}
+p^\mu A^\nu-p^\nu A^\mu,
\label{eq:app-AA}
\end{align}
the last identity following by contraction of the Lorentz algebra
\eqref{eq:canonical-spin} with $p_\alpha p_\beta$.  Since
$\{A^\mu,p^2\}=0$, the Leibniz rule applied to $a^\mu=A^\mu/p^2$ gives
\begin{align}
\{z^\mu,a^\nu\}
&=-\frac{\Sigma^{\mu\nu}}{p^2}
-\frac{2p^\mu A^\nu}{p^4},
\label{eq:app-za}\\
\{a^\mu,z^\nu\}
&=-\frac{\Sigma^{\mu\nu}}{p^2}
+\frac{2A^\mu p^\nu}{p^4},
\label{eq:app-az}\\
\{a^\mu,a^\nu\}
&=\frac{\Sigma^{\mu\nu}}{p^2}
-\frac{A^\mu p^\nu-A^\nu p^\mu}{p^4}.
\label{eq:app-shift-bracket}
\end{align}
Adding the three contributions, the coefficients of
$\Sigma^{\mu\nu}/p^2$ combine as $-2+1=-1$ and those of the
momentum-bilinear term as $+2-1=+1$, so that
\begin{align}
\{X^\mu,X^\nu\}
&=-\frac{\Sigma^{\mu\nu}}{p^2}
+\frac{A^\mu p^\nu-A^\nu p^\mu}{p^4}=-\frac{S^{\mu\nu}}{p^2}.
\label{eq:app-XX}
\end{align}
The terms generated by the momentum dependence of $p^{-2}$ are exactly what
replaces the unreduced tensor $\Sigma^{\mu\nu}$ by the transverse tensor
$S^{\mu\nu}$.  The two remaining reduced brackets follow from the same
composite variables, and we record them in full, because the Jacobi
cancellation of section~\ref{sec:algebra} turns on their relative signs and
an asserted result would leave the main statement of this paper without its
input.

Write $S^{\mu\nu}=\Sigma^{\mu\nu}-B^{\mu\nu}$ with
$B^{\mu\nu}\equiv a^\mu p^\nu-a^\nu p^\mu$.  Contracting
\eqref{eq:canonical-spin} with a single momentum gives
\begin{equation}
\{\Sigma^{\mu\nu},A^\rho\}
=\eta^{\mu\rho}A^\nu-\eta^{\nu\rho}A^\mu
-p^\mu\Sigma^{\nu\rho}+p^\nu\Sigma^{\mu\rho},
\label{eq:app-SigmaA}
\end{equation}
and since $\{\Sigma^{\mu\nu},p^\rho\}=0$ and $\{\Sigma^{\mu\nu},z^\rho\}=0$,
the unreduced tensor contributes
$\{\Sigma^{\mu\nu},X^\rho\}=\{\Sigma^{\mu\nu},A^\rho\}/p^2$.  For the
subtracted piece, \eqref{eq:app-za} and \eqref{eq:app-shift-bracket} give
\begin{equation}
\{a^\mu p^\nu,X^\rho\}
=\frac{A^\mu p^\nu p^\rho+A^\rho p^\mu p^\nu}{p^4}
-a^\mu\eta^{\nu\rho},
\label{eq:app-apX}
\end{equation}
in which the two contributions proportional to $\Sigma^{\mu\rho}p^\nu/p^2$,
one from $\{a^\mu,z^\rho\}$ and one from $\{a^\mu,a^\rho\}$, cancel against
each other.  Antisymmetrization in $\mu\nu$ removes the $A^\rho$ term of
\eqref{eq:app-apX}, so that
\begin{equation}
\{B^{\mu\nu},X^\rho\}
=\frac{(A^\mu p^\nu-A^\nu p^\mu)p^\rho}{p^4}
-a^\mu\eta^{\nu\rho}+a^\nu\eta^{\mu\rho}.
\label{eq:app-BX}
\end{equation}
Subtracting \eqref{eq:app-BX} from $\{\Sigma^{\mu\nu},A^\rho\}/p^2$, the
four terms carrying a free metric cancel identically because
$a^\mu=A^\mu/p^2$, and eliminating $\Sigma^{\mu\rho}$ in favor of
$S^{\mu\rho}$ through \eqref{eq:S-definition} cancels the remaining
momentum-bilinear term, which leaves \eqref{eq:reduced-SX}.  The result
keeps the supplementary condition stable under the position operation,
since
\begin{equation}
\{S^{\mu\nu},X^\rho\}p_\nu+S^{\mu\nu}\{p_\nu,X^\rho\}
=-\frac{(S^{\nu\rho}p_\nu)p^\mu}{p^2}=0
\label{eq:app-TD-stability}
\end{equation}
by \eqref{eq:TD-identity}, which is the check announced in
\eqref{eq:TD-preserved}.

The spin-spin bracket is shorter once one notices that the centroid map
acts on the spin tensor as a transverse projection,
\begin{equation}
S^{\mu\nu}
=\Pi^\mu{}_\alpha\Pi^\nu{}_\beta\Sigma^{\alpha\beta},
\label{eq:S-projection}
\end{equation}
which follows from \eqref{eq:S-definition} together with
$\Sigma^{\alpha\nu}p_\alpha=-A^\nu$.  The projector
\eqref{eq:Pi-definition} depends on the momentum alone, and the momenta
have vanishing brackets both among themselves and with $\Sigma^{\mu\nu}$,
so every factor of $\Pi$ passes through the bracket as a constant,
\begin{equation}
\{S^{\mu\nu},S^{\rho\sigma}\}
=\Pi^\mu{}_\alpha\Pi^\nu{}_\beta
\Pi^\rho{}_\gamma\Pi^\sigma{}_\delta
\{\Sigma^{\alpha\beta},\Sigma^{\gamma\delta}\}.
\label{eq:app-SS-projected}
\end{equation}
Each metric of \eqref{eq:canonical-spin} is then contracted between two
projectors and returns $\Pi^{\mu\rho}$ by idempotence and transversality,
while the surviving pair of projectors converts each $\Sigma$ into $S$
through \eqref{eq:S-projection}, which is \eqref{eq:reduced-SS}.  The
reduced spin algebra is therefore the Lorentz algebra of the hyperplane
orthogonal to $p^\mu$, obtained without any separate computation of the
momentum-dependent corrections.

\section{Jacobi classes, spectral domain, and constraint rank}
\label{app:jacobi}

The realization \eqref{eq:z-momentum}-\eqref{eq:S-momentum} establishes
\begin{equation}
\mathcal J(X,X,X)
=\mathcal J(S,X,X)
=\mathcal J(S,S,X)
=\mathcal J(S,S,S)=0
\label{eq:app-Jacobi-classes}
\end{equation}
inside one associative operator algebra, while the classes containing two
momenta vanish from $[p,p]=[S,p]=0$.  The coordinate cancellation was
displayed in \eqref{eq:coordinate-double-commutator}, and the mixed classes
additionally fix the relative signs of the coordinate, spin-position and
spin brackets.

For completeness, the four commuting self-adjoint momenta admit a joint
spectral measure $E$.  On
\begin{equation}
\mathcal H_\mu=E(\Omega_\mu)\mathcal H,
\qquad
\Omega_\mu=\{p\mid p^2\geq\mu^2,\ p^0>0\},
\label{eq:app-Hmu}
\end{equation}
the inverse is the bounded self-adjoint operator
\begin{equation}
\hat p^{-2}
=\int_{\Omega_\mu}\frac1{k^2}\,dE(k),
\qquad
\|\hat p^{-2}\|\leq\frac1{\mu^2},
\label{eq:app-spectral-inverse}
\end{equation}
and \eqref{eq:inverse-p2-commutator} holds on a dense invariant core of
smooth wave packets supported in the interior of $\Omega_\mu$.  A state
supported on a single mass shell is distributional with respect to the
four-dimensional momentum measure and $\hat X^\rho$ fails to preserve that
support, which is why the fixed-shell replacement of $\hat p^{-2}$ is
licensed only through two insertions, by
\eqref{eq:two-insertion-neutrality}, and not for the coordinate Jacobi
identity.

The Tulczyjew constraint $\chi^\mu=\Sigma^{\mu\nu}p_\nu$ is reducible,
since $p_\mu\chi^\mu=0$, and on its surface
\begin{equation}
C^{\mu\nu}\equiv\{\chi^\mu,\chi^\nu\}
\simeq p^2S^{\mu\nu}.
\label{eq:app-constraint-matrix}
\end{equation}
For nonzero spin $C^{\mu\nu}$ has rank two, one null direction reflecting
the reducibility along $p^\mu$ while another lies along the spin axis and
generates the residual spin gauge of the covariant spinning particle
\cite{HansonRegge:1974,Deriglazov:2017,KyrianSemerak:2007,Faddeev:1987}.  A
Moore-Penrose-type inverse can invert $C^{\mu\nu}$ only on the transverse
spin two-plane and cannot obey $SS^+=\Pi$ with $\Pi$ of rank three, so the
centroid calculation of appendix~\ref{app:brackets} stands independently of
an incomplete Dirac-bracket gauge fixing.  A complete Dirac construction
would have to choose independent constraints and fix the residual spin
gauge before inverting the second-class block.

\section{Positive-energy projection and sign checks}
\label{app:projection}

The laboratory conversion
$q^i=-X^i_{\rm cov}$ and
$\Sigma_D^{\mu\nu}=-\Sigma_{\rm cov}^{\mu\nu}$ puts the spatial
realization in the form
\begin{equation}
q^i=i\hbar\frac{\partial}{\partial p_i}
+\mathscr M^i,
\qquad
\mathscr M^i=\frac{\Sigma_D^{i\nu}p_\nu}{m^2c^2}.
\label{eq:app-lab-realization}
\end{equation}
With $K_i=(\bm\sigma\times\bm p)_i$, multiplication of the FW blocks in
\eqref{eq:FW-unitary} reduces the transverse matrix products through
\begin{equation}
(\bm\sigma\cdot\bm p)K_i(\bm\sigma\cdot\bm p)
=-\bm p^2K_i.
\label{eq:app-Pauli-K}
\end{equation}
Collecting the diagonal blocks and using
$(W+M)^2-c^2\bm p^2=2M(W+M)$ then gives
\begin{equation}
(U\mathscr M^iU^\dagger)_{++}
=\frac{\hbar K_i}{2mW}
-\frac{i\hbar p_i}{2m^2c^2}\bm1_2,
\label{eq:app-Mpp}
\end{equation}
which proves \eqref{eq:Mpp-correct}.  Adding
\eqref{eq:Berry-connection-charge} and simplifying with
\eqref{eq:Berry-cancellation-identity} proves the total operator
\eqref{eq:projected-TD-position}.  The radial remnant has vanishing curl,
and its scalar character makes its commutator with the transverse shift
cancel in the antisymmetrized curvature.

For the commutator, differentiating the TD shift and adding its nonabelian
spin bracket yields \eqref{eq:projected-commutator-general}.  Taking
$\bm p=p_3\hat{\bm z}$ and substituting
$c^2p_3^2=W^2-M^2$ factorizes the parallel numerator and denominator by
$(W+M)^2$, leaving unity.  The transverse ratio reduces to $M/W=1/\gamma$.
These factorizations prove \eqref{eq:projected-commutators} exactly on shell;
direct $4\times4$ matrix evaluation at generic numerical momenta was used
as an independent check, not as a substitute for the algebraic proof.

\section{Bopp-built Lorentz generators}
\label{app:Bopp-generators}

The Bopp coordinate can be used to build
\begin{equation}
J_{\rm B}^{\mu\nu}
=X_{\rm B}^\mu p^\nu-X_{\rm B}^\nu p^\mu+S^{\mu\nu}
=J^{\mu\nu}-\frac1{2\hbar}B^{\mu\nu},
\label{eq:app-JB}
\end{equation}
with
\begin{equation}
B^{\mu\nu}
=(\overline\Theta_{\mathcal C}p)^\mu p^\nu
-(\overline\Theta_{\mathcal C}p)^\nu p^\mu.
\label{eq:app-B}
\end{equation}
Its commutator contains the anomaly
\begin{align}
\Delta^{\mu\nu\rho\sigma}
=i\big(&
\overline\Theta_{\mathcal C}^{\mu\rho}p^\nu p^\sigma
-\overline\Theta_{\mathcal C}^{\mu\sigma}p^\nu p^\rho
\notag\\
&-\overline\Theta_{\mathcal C}^{\nu\rho}p^\mu p^\sigma
+\overline\Theta_{\mathcal C}^{\nu\sigma}p^\mu p^\rho
\big),
\label{eq:app-Bopp-anomaly}
\end{align}
whose overall coefficient is $i$ and not $i/2$, because the two
cross-commutators between $J^{\mu\nu}$ and $B^{\rho\sigma}$ contribute the
same tensor structure while $[B^{\mu\nu},B^{\rho\sigma}]=0$.  The
coordinate-built generators therefore fail to close as the ordinary Lorentz
algebra, whereas the twisted-Poincar\'e formulation keeps the one-particle
generators undeformed and places the deformation in
\eqref{eq:coproduct-J}.  Related distinctions between twist symmetry and
coordinate realizations have long been discussed in canonical
noncommutative theories
\cite{Chaichian:2004,Chaichian:2004za,Wess:2003da,Aschieri:2005yw}.  The
anomaly leaves the Casimir spectrum untouched, since in the
Pauli-Lubanski pseudovector built from $J_{\rm B}$ the $B$-dependent terms
contract symmetric momentum pairs into the Levi-Civita symbol and drop out,
giving $W_{\rm B}^\mu=W^\mu$ and preserving the eigenvalues
\eqref{eq:Poincare-Casimirs}.  Neither construction should be identified
with the exact centroid generator \eqref{eq:J-quantum}.

\section{Straight-line transport and the Ward defect}
\label{app:straight-line}

Along $p(t)=p+tq$ one has
\begin{equation}
\mathcal A_\mu(p+tq)q^\mu
=\frac{\Sigma^{\mu\nu}q_\mu p_\nu}
{(p+tq)^2},
\label{eq:app-line-integrand}
\end{equation}
because the possible term
$t\Sigma^{\mu\nu}q_\mu q_\nu$ vanishes.  Integration proves
\eqref{eq:line-integral-closed}.  For the Dirac realization, write
$\sigma^{\mu\nu}q_\mu p_\nu=\tfrac{i}{2}[\slashed q,\slashed p]$ and
sandwich the contracted vertex between the on-shell projectors
$(\slashed p'+m)$ and $(\slashed p+m)$.  Substituting
$\slashed p(\slashed p+m)=m(\slashed p+m)$,
$(\slashed p'+m)\slashed p'=m(\slashed p'+m)$ and
$2p\cdot q=-q^2$ gives
\begin{equation}
(\slashed p'+m)[\slashed q,\slashed p]\slashed q(\slashed p+m)
=-2mq^2(\slashed p'+m)(\slashed p+m),
\label{eq:app-Ward-defect}
\end{equation}
which is \eqref{eq:ward-defect-identity}.  The factor $q^2J(p,q)$ has no
massless pole, so the one-photon longitudinal defect is locally reparable.
For Compton scattering, simultaneous contraction with $k_\mu$ and
$k'_\nu$ fixes the longitudinal contact amplitude but leaves the doubly
transverse part in \eqref{eq:gauge-complete-decomposition} undetermined.

\section{Polarized Compton channel trace}
\label{app:Compton-trace}

In the natural-unit kinematics of
\eqref{eq:Compton-rest-kinematics}, the two channel matrices are
\begin{align}
\Gamma_s
&=\slashed\epsilon'^{*}
\frac{\slashed p+\slashed k+m}{2p\cdot k}
\slashed\epsilon,
\label{eq:app-Gamma-s}\\
\Gamma_u
&=\slashed\epsilon
\frac{\slashed p-\slashed k'+m}{-2p\cdot k'}
\slashed\epsilon'^{*}.
\label{eq:app-Gamma-u}
\end{align}
Insert the density matrix \eqref{eq:electron-density-matrix}, sum the final
spin with $\slashed p'+m$, and sum real transverse polarization bases for
both photons.  With real momenta and polarization vectors, the imaginary
part arises solely from the $\gamma_5$ trace.  Its Levi-Civita contraction is
proportional to
\begin{equation}
\epsilon(a,p,k,k')
\propto(\hat{\bm s}\cdot\hat{\bm n})
\omega\omega'\sin\theta_\gamma,
\label{eq:app-Compton-epsilon}
\end{equation}
and the remaining contractions give
\eqref{eq:Im-channel-correlator}.  The symbolic calculation was checked by
verifying that both diagonal terms have zero imaginary part, the full rate
is independent of $P_e$, in-plane spin makes the correlator vanish, and the
summed rate reproduces \eqref{eq:KN-summed}.  Replacing the physical bases by
$-g_{\mu\nu}$ gives \eqref{eq:Im-channel-covariant-sum}, providing a direct
diagnostic that the isolated channel correlator is not gauge invariant.

\section{Small-loop expansion for the Compton routing}
\label{app:Compton-loop}

For a small closed path $q(\tau)=p+\delta q(\tau)$, expand
\begin{equation}
\mathcal A_\mu(q)
=\mathcal A_\mu(p)
+\delta q^\alpha\partial_\alpha\mathcal A_\mu(p)+\cdots .
\label{eq:app-A-expansion}
\end{equation}
Path ordering adds the commutator part of \eqref{eq:momentum-curvature},
and the result is
\begin{equation}
U_{\partial\Sigma}
=\bm1+\frac{i}{2\hbar}
\mathcal F_{p\,\mu\nu}
\int_\Sigma dq^\mu\wedge dq^\nu+\cdots .
\label{eq:app-small-loop}
\end{equation}
The Compton quadrilateral has vertices
\begin{equation}
p,\quad p+k,\quad p+k-k'=p',\quad p-k',
\label{eq:app-Compton-vertices}
\end{equation}
and oriented area proportional to $k\wedge k'$, so substituting
$\mathcal F_p^{\mu\nu}=-S^{\mu\nu}/p^2$ yields
\eqref{eq:Compton-small-holonomy} up to the orientation convention.  Higher
terms contain covariant derivatives of $\mathcal F_p$ together with
noncommuting curvature insertions, and they distinguish the full transport
from a scalar Moyal phase even when the leading matrix element happens to
abelianize.

\section{Algebraic identity for the projected associator}
\label{app:projected-identity}

Using $P^2=P$,
\begin{align}
(A\circ_P B)\circ_P C
&=PABPCP,
\label{eq:app-left-assoc}\\
A\circ_P(B\circ_P C)
&=PAPBCP.
\label{eq:app-right-assoc}
\end{align}
Inserting $P=1-Q$ once in each expression gives
\begin{align}
PABPCP&=PABCP-P A B Q C P,
\label{eq:app-left-Q}\\
PAPBCP&=PABCP-P A Q B C P,
\label{eq:app-right-Q}
\end{align}
whose difference is \eqref{eq:projected-associator}.  It vanishes when all
three operations preserve $P\mathcal H$, and need not vanish for covariant
positions acting on a sharply selected mass sector.

\bibliographystyle{JHEP}
\bibliography{refs_3}

@article{Aschieri:2005yw,
  author = {Aschieri, P. and Dimitrijevi\'c, M. and Meyer, F. and Wess, J.},
  title = {{Noncommutative geometry and gravity}},
  journal = {Class. Quant. Grav.},
  volume = {23},
  year = {2006},
  pages = {1883--1912},
  doi = {10.1088/0264-9381/23/6/005},
  eprint = {hep-th/0510059},
  archivePrefix = {arXiv}
}

@article{bahns2005unitarity,
  author = {Bahns, D. and Doplicher, S. and Fredenhagen, K. and Piacitelli, G.},
  title = {{Ultraviolet finite quantum field theory on quantum spacetime}},
  journal = {Commun. Math. Phys.},
  volume = {237},
  year = {2003},
  pages = {221--241},
  doi = {10.1007/s00220-003-0857-x},
  eprint = {hep-th/0301100},
  archivePrefix = {arXiv}
}

@article{bargmann1948unitary,
  author = {Bargmann, V. and Wigner, E. P.},
  title = {{Group Theoretical Discussion of Relativistic Wave Equations}},
  journal = {Proc. Natl. Acad. Sci. USA},
  volume = {34},
  year = {1948},
  pages = {211--223},
  doi = {10.1073/pnas.34.5.211}
}

@article{bopp1956,
  author = {Bopp, Fritz},
  title = {{La m\'ecanique quantique est-elle une m\'ecanique statistique classique particuli\`ere?}},
  journal = {Annales de l'institut Henri Poincar\'e},
  volume = {15},
  number = {2},
  year = {1956},
  pages = {81--112}
}

@article{Carlson:2001,
  author = {Carlson, C. E. and Carone, C. D. and Lebed, R. F.},
  title = {{Bounding noncommutative QCD}},
  journal = {Phys. Lett. B},
  volume = {518},
  year = {2001},
  pages = {201--206},
  doi = {10.1016/S0370-2693(01)01053-X},
  eprint = {hep-ph/0107291},
  archivePrefix = {arXiv}
}

@article{Chaichian:2004,
  author = {Chaichian, M. and Sheikh-Jabbari, M. M. and Tureanu, A.},
  title = {{Hydrogen atom spectrum and the Lamb shift in noncommutative QED}},
  journal = {Eur. Phys. J. C},
  volume = {36},
  year = {2004},
  pages = {251},
  eprint = {hep-th/0212257},
  archivePrefix = {arXiv}
}

@article{Colatto_2006,
   title={Noncommutative geometry induced by spin effects},
   volume={73},
   ISSN={1550-2368},
   url={http://dx.doi.org/10.1103/PhysRevD.73.105007},
   DOI={10.1103/physrevd.73.105007},
   number={10},
   pages={105007},
   journal={Physical Review D},
   publisher={American Physical Society (APS)},
   author={Colatto, L. P. and Penna, A. L. A. and Santos, W. C.},
   year={2006},
   month=may }

@article{Doplicher:1994zv,
  author = {Doplicher, S. and Fredenhagen, K. and Roberts, J. E.},
  title = {{Spacetime quantization induced by classical gravity}},
  journal = {Phys. Lett. B},
  volume = {331},
  year = {1994},
  pages = {39--44},
  doi = {10.1016/0370-2693(94)90940-7},
  eprint = {gr-qc/9405057},
  archivePrefix = {arXiv}
}

@article{Doplicher:1995,
  author = {Doplicher, S. and Fredenhagen, K. and Roberts, J. E.},
  title = {{The Quantum Structure of Spacetime at the Planck Scale and Quantum Fields}},
  journal = {Commun. Math. Phys.},
  volume = {172},
  number = {1},
  year = {1995},
  pages = {187--220},
  doi = {10.1007/BF02104514},
  eprint = {hep-th/0303037},
  archivePrefix = {arXiv}
}

@article{Douglas:2001,
  author = {Douglas, M. R. and Nekrasov, N. A.},
  title = {{Noncommutative field theory}},
  journal = {Rev. Mod. Phys.},
  volume = {73},
  year = {2001},
  pages = {977},
  eprint = {hep-th/0106048},
  archivePrefix = {arXiv}
}

@article{Drinfeld:1985rx,
    author = "Drinfeld, V. G.",
    title = "{Hopf algebras and the quantum Yang-Baxter equation}",
    journal = "Sov. Math. Dokl.",
    volume = "32",
    pages = "254--258",
    year = "1985"
}

@article{Faddeev:1987,
  author = {Faddeev, L. D. and Reshetikhin, N. Y. and Takhtajan, L. A.},
  title = {{Quantization of Lie groups and Lie algebras}},
  journal = {Leningrad Math. J.},
  volume = {1},
  year = {1990},
  pages = {193--225}
}

@article{Hewett:2001,
  author = {Hewett, J. L. and Petriello, F. J. and Rizzo, T. G.},
  title = {{Signals for noncommutative QED at future e+ e- colliders}},
  journal = {Phys. Rev. D},
  volume = {64},
  year = {2001},
  pages = {075012},
  eprint = {hep-ph/0010354},
  archivePrefix = {arXiv}
}

@article{Hinchliffe:2004,
  author = {Hinchliffe, I. and Kersting, N. and Ma, Y. L.},
  title = {{Review of the phenomenology of noncommutative geometry}},
  journal = {Int. J. Mod. Phys. A},
  volume = {19},
  year = {2004},
  pages = {179--204},
  doi = {10.1142/S0217751X0401662X},
  eprint = {hep-ph/0205040},
  archivePrefix = {arXiv}
}

@article{Jackiw:2001yj,
  author = {Jackiw, R.},
  title = {{Physical instances of noncommuting coordinates}},
  journal = {Nucl. Phys. Proc. Suppl.},
  volume = {108},
  year = {2002},
  pages = {30--36},
  doi = {10.1016/S0920-5632(02)01302-6},
  eprint = {hep-th/0110057},
  archivePrefix = {arXiv}
}

@article{Kostelecky:2008,
  author = {Kosteleck{\'y}, V. A. and Russell, N.},
  title = {{Data Tables for Lorentz and CPT Violation}},
  journal = {Rev. Mod. Phys.},
  volume = {83},
  number = {1},
  year = {2011},
  pages = {11--318},
  doi = {10.1103/RevModPhys.83.11},
  eprint = {0801.0287},
  archivePrefix = {arXiv},
  primaryClass = {hep-ph}
}

@book{Madore:1999,
  author = {Madore, J.},
  title = {{An Introduction to Noncommutative Differential Geometry and its Physical Applications}},
  series = {London Mathematical Society Lecture Note Series},
  volume = {257},
  publisher = {Cambridge University Press},
  edition = {2nd},
  year = {1999},
  doi = {10.1017/CBO9780511564205}
}

@article{Colladay:1998fq,
   title={Lorentz-violating extension of the standard model},
   volume={58},
   ISSN={1089-4918},
   url={http://dx.doi.org/10.1103/PhysRevD.58.116002},
   DOI={10.1103/physrevd.58.116002},
   number={11},
   journal={Physical Review D},
   publisher={American Physical Society (APS)},
   author={Colladay, D. and Kostelecký, V. Alan},
   year={1998},
   month=Oct }

@article{Mattingly:2005re,
   title={Modern Tests of Lorentz Invariance},
   volume={8},
   ISSN={1433-8351},
   url={http://dx.doi.org/10.12942/lrr-2005-5},
   DOI={10.12942/lrr-2005-5},
   number={1},
   journal={Living Reviews in Relativity},
   publisher={Springer Science and Business Media LLC},
   author={Mattingly, David},
   year={2005},
   month={September} }

@article{Seiberg:1999,
  author = {Seiberg, N. and Witten, E.},
  title = {{String theory and noncommutative geometry}},
  journal = {JHEP},
  volume = {9909},
  year = {1999},
  pages = {032},
  eprint = {hep-th/9908142},
  archivePrefix = {arXiv}
}

@article{Seiberg:2000,
  author = {Seiberg, N. and Susskind, L. and Toumbas, N.},
  title = {{Space/time non-commutativity and causality}},
  journal = {JHEP},
  volume = {0006},
  year = {2000},
  pages = {044},
  doi = {10.1088/1126-6708/2000/06/044},
  eprint = {hep-th/0005015},
  archivePrefix = {arXiv}
}

@article{Snyder:1947,
  author = {Snyder, H. S.},
  title = {{Quantized Space-Time}},
  journal = {Phys. Rev.},
  volume = {71},
  year = {1947},
  pages = {38--41},
  doi = {10.1103/PhysRev.71.38}
}

@article{Szabo:2003,
  author = {Szabo, R. J.},
  title = {{Quantum field theory on noncommutative spaces}},
  journal = {Phys. Rept.},
  volume = {378},
  year = {2003},
  pages = {207},
  eprint = {hep-th/0109162},
  archivePrefix = {arXiv}
}

@incollection{Wess:2003da,
  author = {Wess, J.},
  title = {{Deformed coordinate spaces: Derivatives}},
  booktitle = {Lect. Notes Phys.},
  volume = {616},
  year = {2003},
  pages = {122--128},
  doi = {10.1007/3-540-36539-7},
  eprint = {hep-th/0408080},
  archivePrefix = {arXiv}
}

@article{wigner1939unitary,
    author = "Wigner, Eugene P.",
    editor = "Kim, Y. S. and Zachary, W. W.",
    title = "{On Unitary Representations of the Inhomogeneous Lorentz Group}",
    doi = "10.2307/1968551",
    journal = "Annals Math.",
    volume = "40",
    pages = "149--204",
    year = "1939"
}

@article{Chaichian:2004za,
   title={Non-commutative standard model: model building},
   volume={29},
   ISSN={1434-6052},
   url={http://dx.doi.org/10.1140/epjc/s2003-01204-7},
   DOI={10.1140/epjc/s2003-01204-7},
   number={3},
   journal={The European Physical Journal C},
   publisher={Springer Science and Business Media LLC},
   author={Chaichian, M. and Presnajder, P. and Sheikh-Jabbari, M. M. and Tureanu, A.},
   year={2003},
   month={July}, pages={413--432} }

@article{Terashima:2003,
  author = {Terashima, H. and Ueda, M.},
  title = {Relativistic Einstein-Podolsky-Rosen correlation and Bell's inequality},
  journal = {Int. J. Quantum Inf.},
  volume = {1},
  pages = {93--114},
  year = {2003}
}

@article{Alsing:2012,
  author = {Alsing, P. M. and Fuentes, I.},
  title = {Observer-dependent entanglement},
  journal = {Class. Quantum Grav.},
  volume = {29},
  pages = {224001},
  year = {2012}
}

@article{Bargmann:1959gz,
  author = {Bargmann, V. and Michel, L. and Telegdi, V. L.},
  title = {Precession of the polarization of particles moving in a homogeneous electromagnetic field},
  journal = {Phys. Rev. Lett.},
  volume = {2},
  pages = {435--436},
  year = {1959}
}

@article{Muong-2:2023cdq,
  author = {{Muon $g-2$ Collaboration}},
  title = {Measurement of the Positive Muon Anomalous Magnetic Moment to 0.20 ppm},
  journal = {Phys. Rev. Lett.},
  volume = {131},
  pages = {161802},
  year = {2023}
}

@article{Fan:2022eto,
   title={Measurement of the Electron Magnetic Moment},
   volume={130},
   ISSN={1079-7114},
   url={http://dx.doi.org/10.1103/PhysRevLett.130.071801},
   DOI={10.1103/physrevlett.130.071801},
   number={7},
   journal={Physical Review Letters},
   publisher={American Physical Society (APS)},
   author={Fan, X. and Myers, T. G. and Sukra, B. A. D. and Gabrielse, G.},
   year={2023},
   month=Feb }

@article{Aoyama:2020ynm,
    author = "Aoyama, T. and others",
    title = "{The anomalous magnetic moment of the muon in the Standard Model}",
    eprint = "2006.04822",
    archivePrefix = "arXiv",
    primaryClass = "hep-ph",
    reportNumber = "FERMILAB-PUB-20-207-T, INT-PUB-20-021, KEK Preprint 2020-5,
  MITP/20-028, KEK Preprint 2020-5, MITP/20-028, CERN-TH-2020-075, IFT-UAM/CSIC-20-74, LMU-ASC 18/20, LTH 1234,
  LU TP 20-20, LTH 1234, LU TP 20-20, MAN/HEP/2020/003, PSI-PR-20-06, UWThPh 2020-14, ZU-TH 18/20",
    doi = "10.1016/j.physrep.2020.07.006",
    journal = "Phys. Rept.",
    volume = "887",
    pages = "1--166",
    year = "2020"
}

@article{Aliberti:2025beg,
    author = "Aliberti, R. and others",
    title = "{The anomalous magnetic moment of the muon in the Standard Model: an update}",
    eprint = "2505.21476",
    archivePrefix = "arXiv",
    primaryClass = "hep-ph",
    reportNumber = "CERN-TH-2025-101, FERMILAB-PUB-25-0344-T, INT-PUB-25-015, IPARCOS-UCM-25-029, KEK Preprint 2025-22, LTH 1403, MITP-25-037, UWThPh 2025-15, UWThPh
  2025-15, ZU-TH 37/25, IPARCOS-UCM-25-029",
    doi = "10.1016/j.physrep.2025.08.002",
    journal = "Phys. Rept.",
    volume = "1143",
    pages = "1--158",
    year = "2025"
}

@article{Frenkel:1926,
  author = {Frenkel, J.},
  title = {{Die Elektrodynamik des rotierenden Elektrons}},
  journal = {Z. Phys.},
  volume = {37},
  year = {1926},
  pages = {243--262},
  doi = {10.1007/BF01397099}
}

@article{Mathisson:1937,
  author = {Mathisson, M.},
  title = {{Neue Mechanik materieller Systeme}},
  journal = {Acta Phys. Pol.},
  volume = {6},
  year = {1937},
  pages = {163--200},
  note = {Republished in Gen. Relativ. Gravit. \textbf{42}, 1011 (2010), doi:10.1007/s10714-010-0939-y}
}

@article{Pirani:1956,
  author = {Pirani, F. A. E.},
  title = {{On the physical significance of the Riemann tensor}},
  journal = {Acta Phys. Pol.},
  volume = {15},
  year = {1956},
  pages = {389--405},
  note = {Republished in Gen. Relativ. Gravit. \textbf{41}, 1215 (2009), doi:10.1007/s10714-009-0787-9}
}

@article{Tulczyjew:1959,
  author = {Tulczyjew, W.},
  title = {{Motion of multipole particles in general relativity theory}},
  journal = {Acta Phys. Pol.},
  volume = {18},
  year = {1959},
  pages = {393--409}
}

@article{Dixon:1964,
  author = {Dixon, W. G.},
  title = {{A covariant multipole formalism for extended test bodies in general relativity}},
  journal = {Nuovo Cimento},
  volume = {34},
  year = {1964},
  pages = {317--339},
  doi = {10.1007/BF02734579}
}

@article{Papapetrou:1951,
  author = {Papapetrou, A.},
  title = {{Spinning test-particles in general relativity. I}},
  journal = {Proc. R. Soc. Lond. A},
  volume = {209},
  number = {1097},
  year = {1951},
  pages = {248--258},
  doi = {10.1098/rspa.1951.0200}
}

@article{Corinaldesi:1951,
  author = {Corinaldesi, E. and Papapetrou, A.},
  title = {{Spinning test-particles in general relativity. II}},
  journal = {Proc. R. Soc. Lond. A},
  volume = {209},
  number = {1097},
  year = {1951},
  pages = {259--268},
  doi = {10.1098/rspa.1951.0201}
}

@article{Pryce:1948,
  author = {Pryce, M. H. L.},
  title = {{The mass-centre in the restricted theory of relativity and its connexion with the quantum theory of elementary particles}},
  journal = {Proc. R. Soc. Lond. A},
  volume = {195},
  number = {1040},
  year = {1948},
  pages = {62--81},
  doi = {10.1098/rspa.1948.0103}
}

@article{Moller:1949,
  author = {M{\o}ller, C.},
  title = {{Sur la dynamique des syst\`emes ayant un moment angulaire interne}},
  journal = {Ann. Inst. Henri Poincar\'e},
  volume = {11},
  year = {1949},
  pages = {251--278}
}

@article{NewtonWigner:1949,
  author = {Newton, T. D. and Wigner, E. P.},
  title = {{Localized states for elementary systems}},
  journal = {Rev. Mod. Phys.},
  volume = {21},
  year = {1949},
  pages = {400--406},
  doi = {10.1103/RevModPhys.21.400}
}

@article{HansonRegge:1974,
  author = {Hanson, A. J. and Regge, T.},
  title = {{The relativistic spherical top}},
  journal = {Ann. Phys. (N.Y.)},
  volume = {87},
  year = {1974},
  pages = {498--566},
  doi = {10.1016/0003-4916(74)90046-3}
}

@misc{DuvalHorvathy:2003,
      title={Mathisson' spinning electron: noncommutative mechanics \& exotic Galilean symmetry, 66 years ago}, 
      author={P. A. Horvathy},
      year={2003},
      eprint={hep-th/0303099},
      archivePrefix={arXiv},
      primaryClass={hep-th},
      url={https://arxiv.org/abs/hep-th/0303099}, 
}

@article{Deriglazov:2017,
  author = {Deriglazov, A. A. and Guzm\'an Ram\'irez, W.},
  title = {{Recent progress on the description of relativistic spin: vector model of spinning particle and rotating body with gravimagnetic moment in General Relativity}},
  journal = {Adv. Math. Phys.},
  volume = {2017},
  year = {2017},
  pages = {7397159},
  doi = {10.1155/2017/7397159},
  eprint = {1710.07135},
  archivePrefix = {arXiv}
}

@incollection{CostaNatario:2015,
  author = {Costa, L. F. O. and Nat\'ario, J.},
  title = {{Center of mass, spin supplementary conditions, and the momentum of spinning particles}},
  booktitle = {{Equations of Motion in Relativistic Gravity}},
  series = {Fund. Theor. Phys.},
  volume = {179},
  year = {2015},
  pages = {215--258},
  doi = {10.1007/978-3-319-18335-0_6},
  eprint = {1410.6443},
  archivePrefix = {arXiv}
}

@article{KyrianSemerak:2007,
  author  = {Kyri{\'a}n, R. and Semer{\'a}k, O.},
  title   = {Spinning test particles in a {Kerr} field~-- {II}},
  journal = {Mon. Not. R. Astron. Soc.},
  volume  = {382},
  number  = {4},
  pages   = {1922--1932},
  year    = {2007},
  doi     = {10.1111/j.1365-2966.2007.12502.x}
}

@article{Weller:2009HIGS,
  author  = {Weller, H. R. and Ahmed, M. W. and Gao, H. and Tornow, W. and Wu, Y. K. and Gai, M. and Miskimen, R.},   
  title   = {Research opportunities at the upgraded {HI$\gamma$S} facility},
  journal = {Prog. Part. Nucl. Phys.},
  volume  = {62},
  number  = {1},
  pages   = {257--303},
  year    = {2009},
  doi     = {10.1016/j.ppnp.2008.07.001}
}

@article{ELINP:TDR,
    author = "Adriani, O. and others",
    title = "{Technical Design Report EuroGammaS proposal for the ELI-NP Gamma beam System}",
    eprint = "1407.3669",
    archivePrefix = "arXiv",
    primaryClass = "physics.acc-ph",
    month = "7",
    year = "2014"
}

@article{Zakrzewski:1994,
  author = {Zakrzewski, S.},
  title = {{Extended phase space for a spinning particle}},
  year = {1994},
  eprint = {hep-th/9412100},
  archivePrefix = {arXiv},
  doi = {10.1088/0305-4470/28/24/028}
}

@article{Chryssomalakos:2009,
  author = {Chryssomalakos, C. and Hern{\'a}ndez-Coronado, H. and Okon, E.},
  title = {{Center of mass in special and general relativity and its role in an effective description of spacetime}},
  journal = {J. Phys. Conf. Ser.},
  volume = {174},
  pages = {012026},
  year = {2009},
  eprint = {0901.3349},
  archivePrefix = {arXiv},
  primaryClass = {gr-qc},
  doi = {10.1088/1742-6596/174/1/012026}
}

@article{Aguilar:2013,
  author = {Aguilar, P. and Chryssomalakos, C. and Hern{\'a}ndez-Coronado, H. and Okon, E.},
  title = {{Position operators and center of mass: new perspectives}},
  year = {2013},
  eprint = {1306.0504},
  archivePrefix = {arXiv},
  primaryClass = {physics.gen-ph},
  doi = {10.1142/S0217751X13501467}
}

@article{Gomes:2010,
  author = {Gomes, M. and Kupriyanov, V. G. and da Silva, A. J.},
  title = {{Noncommutativity due to spin}},
  journal = {Phys. Rev. D},
  volume = {81},
  pages = {085024},
  year = {2010},
  eprint = {1002.4173},
  archivePrefix = {arXiv},
  primaryClass = {hep-th},
  doi = {10.1103/PhysRevD.81.085024}
}

@article{DeriglazovPupasov:2014,
  author = {Deriglazov, A. A. and Pupasov-Maksimov, A. M.},
  title = {{Lagrangian for Frenkel electron and position's non-commutativity due to spin}},
  journal = {Eur. Phys. J. C},
  volume = {74},
  pages = {3101},
  year = {2014},
  eprint = {1312.6247},
  archivePrefix = {arXiv},
  primaryClass = {hep-th},
  doi = {10.1140/epjc/s10052-014-3101-2}
}

@article{DeriglazovTereza:2019,
  author = {Deriglazov, A. A. and Tereza, D. M.},
  title = {{Covariant version of Pauli Hamiltonian, spin-induced noncommutativity, Thomas precession and precession of spin}},
  journal = {Phys. Rev. D},
  volume = {100},
  pages = {105009},
  year = {2019},
  eprint = {1910.11140},
  archivePrefix = {arXiv},
  primaryClass = {physics.gen-ph},
  doi = {10.1103/PhysRevD.100.105009}
}

@article{MuchVergara:2017,
  author = {Much, A. and Vergara, J. D.},
  title = {{A Poincar{\'e} covariant noncommutative spacetime?}},
  year = {2017},
  eprint = {1704.07932},
  archivePrefix = {arXiv},
  primaryClass = {math-ph},
  doi = {10.1142/S0219887818501591}
}

@article{Kim:2023covariantSSC,
  author        = {Kim, Jung-Wook and Steinhoff, Jan},
  title         = {Spin supplementary condition in quantum field theory: covariant SSC and physical state projection},
  journal       = {JHEP},
  volume        = {07},
  pages         = {042},
  year          = {2023},
  doi           = {10.1007/JHEP07(2023)042},
  eprint        = {2302.01944},
  archivePrefix = {arXiv}
}

@article{Ramirez:2014frenkel,
  author        = {Guzman Ramirez, Walberto and Deriglazov, Alexei A. and Pupasov-Maksimov, Andrey M.},
  title         = {Frenkel electron and a spinning body in a curved background},
  journal       = {JHEP},
  volume        = {03},
  pages         = {109},
  year          = {2014},
  doi           = {10.1007/JHEP03(2014)109},
  eprint        = {1311.5743},
  archivePrefix = {arXiv}
}

@article{Girelli:2011snyder,
  author        = {Girelli, Florian and Livine, Etera R.},
  title         = {Scalar field theory in Snyder space-time: alternatives},
  journal       = {JHEP},
  volume        = {03},
  pages         = {132},
  year          = {2011},
  doi           = {10.1007/JHEP03(2011)132},
  eprint        = {1004.0621},
  archivePrefix = {arXiv}
}

@article{Meier:2023twistGauge,
  author        = {Meier, Tim and van Tongeren, Stijn J.},
  title         = {Gauge theory on twist-noncommutative spaces},
  journal       = {JHEP},
  volume        = {12},
  pages         = {045},
  year          = {2023},
  doi           = {10.1007/JHEP12(2023)045},
  eprint        = {2305.15470},
  archivePrefix = {arXiv}
}

@article{Borsato:2026scaleTwist,
  author        = {Borsato, Riccardo and Meier, Tim},
  title         = {Non-commutative deformations of gauge theories via Drinfel'd twists of the scale symmetry},
  journal       = {JHEP},
  volume        = {05},
  pages         = {309},
  year          = {2026},
  doi           = {10.1007/JHEP05(2026)309},
  eprint        = {2512.04162},
  archivePrefix = {arXiv}
}

@article{Kupriyanov:2020novelGauge,
  author        = {Kupriyanov, Vladislav G. and Vitale, Patrizia},
  title         = {A novel approach to non-commutative gauge theory},
  journal       = {JHEP},
  volume        = {08},
  pages         = {041},
  year          = {2020},
  doi           = {10.1007/JHEP08(2020)041},
  eprint        = {2004.14901},
  archivePrefix = {arXiv}
}

@article{Blumenhagen:2018Linfinity,
  author        = {Blumenhagen, Ralph and Brunner, Ilka and Kupriyanov, Vladislav and Lust, Dieter},
  title         = {Bootstrapping non-commutative gauge theories from $L_{\infty}$ algebras},
  journal       = {JHEP},
  volume        = {05},
  pages         = {097},
  year          = {2018},
  doi           = {10.1007/JHEP05(2018)097},
  eprint        = {1803.00732},
  archivePrefix = {arXiv}
}

@article{Kurkov:2022U1Linfinity,
  author        = {Kurkov, Maxim and Vitale, Patrizia},
  title         = {Four-dimensional noncommutative deformations of $U(1)$ gauge theory and $L_{\infty}$ bootstrap},
  journal       = {JHEP},
  volume        = {01},
  pages         = {032},
  year          = {2022},
  doi           = {10.1007/JHEP01(2022)032},
  eprint        = {2108.04856},
  archivePrefix = {arXiv}
}

@article{Blaschke:2013SlavnovTaylor,
  author        = {Blaschke, Daniel N. and Grosse, Harald and Wallet, Jean-Christophe},
  title         = {Slavnov-Taylor identities, non-commutative gauge theories and infrared divergences},
  journal       = {JHEP},
  volume        = {06},
  pages         = {038},
  year          = {2013},
  doi           = {10.1007/JHEP06(2013)038},
  eprint        = {1302.2903},
  archivePrefix = {arXiv}
}

@article{Bakas:2014threeCocycles,
  author        = {Bakas, Ioannis and Lust, Dieter},
  title         = {3-cocycles, non-associative star-products and the magnetic paradigm of $R$-flux string vacua},
  journal       = {JHEP},
  volume        = {01},
  pages         = {171},
  year          = {2014},
  doi           = {10.1007/JHEP01(2014)171},
  eprint        = {1309.3172},
  archivePrefix = {arXiv}
}

@article{Mylonas:2012membrane,
  author        = {Mylonas, Dionysios and Schupp, Peter and Szabo, Richard J.},
  title         = {Membrane sigma-models and quantization of non-geometric flux backgrounds},
  journal       = {JHEP},
  volume        = {09},
  pages         = {012},
  year          = {2012},
  doi           = {10.1007/JHEP09(2012)012},
  eprint        = {1207.0926},
  archivePrefix = {arXiv}
}

@article{Blumenhagen:2016nonassociativeGravity,
  author        = {Blumenhagen, Ralph and Fuchs, Michael},
  title         = {Towards a theory of nonassociative gravity},
  journal       = {JHEP},
  volume        = {07},
  pages         = {019},
  year          = {2016},
  doi           = {10.1007/JHEP07(2016)019},
  eprint        = {1604.03253},
  archivePrefix = {arXiv}
}

@article{Aschieri:2018nonassociativeGeometry,
  author        = {Aschieri, Paolo and Dimitrijevic Ciric, Marija and Szabo, Richard J.},
  title         = {Nonassociative differential geometry and gravity with non-geometric fluxes},
  journal       = {JHEP},
  volume        = {02},
  pages         = {036},
  year          = {2018},
  doi           = {10.1007/JHEP02(2018)036},
  eprint        = {1710.11467},
  archivePrefix = {arXiv}
}

@article{Fabiano:2026TMinkowski,
  author        = {Fabiano, Giuseppe and Mercati, Flavio},
  title         = {Covariant quantization of field theories on T-Minkowski noncommutative spacetimes},
  journal       = {JHEP},
  volume        = {06},
  pages         = {095},
  year          = {2026},
  doi           = {10.1007/JHEP06(2026)095}
}

@article{Tizchang:2017NCQEDLaser,
  author        = {Tizchang, S. and Batebi, S. and Haghighat, M. and Mohammadi, R.},
  title         = {Using an intense laser beam in interaction with muon/electron beam to probe the noncommutative QED},
  journal       = {JHEP},
  volume        = {02},
  pages         = {003},
  year          = {2017},
  doi           = {10.1007/JHEP02(2017)003}
}

@article{StoneDwivediZhou:2015,
  author = {Stone, Michael and Dwivedi, Vatsal and Zhou, Tianci},
  title = {{Berry phase, Lorentz covariance, and anomalous velocity for Dirac and Weyl particles}},
  journal = {Phys. Rev. D},
  volume = {91},
  number = {2},
  pages = {025004},
  year = {2015},
  eprint = {1406.0354},
  archivePrefix = {arXiv},
  primaryClass = {hep-th},
  doi = {10.1103/PhysRevD.91.025004}
}

@article{ChenSonStephanov:2014,
  author = {Chen, Jing-Yuan and Son, Dam T. and Stephanov, Mikhail A. and Yee, Ho-Ung and Yin, Yi},
  title = {{Lorentz invariance in chiral kinetic theory}},
  journal = {Phys. Rev. Lett.},
  volume = {113},
  number = {18},
  pages = {182302},
  year = {2014},
  eprint = {1404.5963},
  archivePrefix = {arXiv},
  primaryClass = {hep-th},
  doi = {10.1103/PhysRevLett.113.182302}
}

@article{MylonasSchuppSzabo:2014,
  author = {Mylonas, Dionysios and Schupp, Peter and Szabo, Richard J.},
  title = {{Non-geometric fluxes, quasi-Hopf twist deformations and nonassociative quantum mechanics}},
  journal = {J. Math. Phys.},
  volume = {55},
  number = {12},
  pages = {122301},
  year = {2014},
  eprint = {1312.1621},
  archivePrefix = {arXiv},
  primaryClass = {hep-th},
  doi = {10.1063/1.4902378}
}

@article{Xu:2001,
  author = {Xu, Ping},
  title = {{Quantum groupoids}},
  journal = {Commun. Math. Phys.},
  volume = {216},
  pages = {539--581},
  year = {2001},
  eprint = {math/9905192},
  archivePrefix = {arXiv},
  doi = {10.1007/s002200000334}
}

@article{Berard:2004monopole,
  author = {B\'erard, A. and Mohrbach, H.},
  title = {{Monopole and Berry phase in momentum space in noncommutative quantum mechanics}},
  journal = {Phys. Rev. D},
  volume = {69},
  year = {2004},
  pages = {127701},
  doi = {10.1103/PhysRevD.69.127701},
  eprint = {hep-th/0310167},
  archivePrefix = {arXiv}
}

@article{Berard:2006spinhall,
  author = {B\'erard, A. and Mohrbach, H.},
  title = {{Spin Hall effect and Berry phase of spinning particles}},
  journal = {Phys. Lett. A},
  volume = {352},
  year = {2006},
  pages = {190},
  eprint = {hep-th/0404165},
  archivePrefix = {arXiv}
}

@book{Souriau:1970,
  author = {Souriau, J.-M.},
  title = {{Structure des syst\`emes dynamiques}},
  publisher = {Dunod},
  address = {Paris},
  year = {1970},
  note = {English translation: Structure of Dynamical Systems, Birkh\"auser (1997)}
}

@article{Bliokh:2017,
  author = {Bliokh, K. Y. and Dennis, M. R. and Nori, F.},
  title = {{Position, spin, and orbital angular momentum of a relativistic electron}},
  journal = {Phys. Rev. A},
  volume = {96},
  pages = {023622},
  year = {2017},
  eprint = {1706.01658},
  archivePrefix = {arXiv}
}

@article{ThomasWignerHolonomy:2024,
  author = {Palge, Veiko and Pfeifer, Christian},
  title = {{Thomas--Wigner rotation as a holonomy for spin-1/2 particles}},
  journal = {Phys. Rev. A},
  volume = {109},
  pages = {032206},
  year = {2024},
  eprint = {2310.08121},
  archivePrefix = {arXiv}
}

@article{Fleming:1965,
  author = {Fleming, Gordon N.},
  title = {{Covariant Position Operators, Spin, and Locality}},
  journal = {Phys. Rev.},
  volume = {137},
  pages = {B188--B197},
  year = {1965},
  doi = {10.1103/PhysRev.137.B188}
}

\end{document}